\documentclass[
    a4paper,
    reprint,
    amsmath,
    amssymb,
    aps,
    rmp,
    unsortedaddress,
]{revtex4-2}

\usepackage[T1]{fontenc}

\usepackage[dvipsnames,table,xcdraw]{xcolor}
\definecolor{rmpblue}{HTML}{2e3092}
\usepackage[
	colorlinks=true,
	citecolor=rmpblue,
	linkcolor=rmpblue,
	filecolor=magenta,
	urlcolor=rmpblue,
]{hyperref}
\usepackage{graphicx}
\usepackage{array}
\usepackage{lipsum}
\usepackage{epigraph}
\usepackage{multirow}
\usepackage{blindtext}
\usepackage{tabularx}
\usepackage{cases}
\usepackage{comment}
\usepackage{siunitx}

\usepackage{layouts}
\usepackage{physics}

\newcommand{\qqoute}[1]{``#1''}

\newcommand{\unitvec}[1]{\bar{\mathbf{#1}}}
\newcommand{\um}{\SI{}{\micro\meter}}

\DeclareMathAlphabet\mathbfcal{OMS}{cmsy}{b}{n}

\newcommand{\iu}{{i}\mkern1mu}
\newcommand{\eu}{{e}\mkern1mu}

\newcommand{\white}[1]{\textcolor{white}{#1}}

\newcommand{\berrydot}{\cdot (\grad)}

\newcommand{\ten}[1]{\boldsymbol{\vb{\hat{#1}}}}

\newcommand{\affilRIKENQ}{RIKEN Center for Quantum Computing, RIKEN, Wako-shi, Saitama 351-0198, Japan}
\newcommand{\affilANU}{Nonlinear Physics Center, Research School of Physics, Australian National University, Canberra ACT 2601, Australia}
\newcommand{\affilMichigan}{Physics Department, University of Michigan, Ann Arbor, Michigan 48109-1040, USA}
\newcommand{\affilKing}{King’s College London, Department of Physics and London Centre for Nanotechnology, London, United Kingdom}
\newcommand{\affilDIPC}{Donostia International Physics Center (DIPC), Donostia-San Sebastián 20018, Spain}
\newcommand{\affilE}{Centre of Excellence ENSEMBLE3 Sp. z o.o., 01-919 Warsaw, Poland}
\newcommand{\affilIkerbasque}{IKERBASQUE, Basque Foundation for Science, Bilbao 48009, Spain}

\begin{document}

\title{Radiation forces and torques in optics and acoustics}

\author{Ivan Toftul}
\affiliation{\affilANU}
\affiliation{\affilRIKENQ}
\email{Contact author: Ivan.Toftul@anu.edu.au}

\author{Sebastian Golat}
\affiliation{\affilKing}

\author{Francisco J. Rodr\'\i guez-Fortu{\~n}o}
\affiliation{\affilKing}

\author{Franco Nori}
\affiliation{\affilRIKENQ}
\affiliation{\affilMichigan}

\author{Yuri Kivshar}
\affiliation{\affilANU}

\author{Konstantin Y. Bliokh}
\affiliation{\affilDIPC}
\affiliation{\affilIkerbasque}
\affiliation{\affilE}
\email{Contact author: konstantin.bliokh@dipc.org}

\date{\today}

\begin{abstract}
The mechanical action of various kinds of waves has been recognized for several centuries. The first tide of scientific interest in wave-induced forces and torques emerged at the turn of the 20th century, with the development of wave theories and the concepts of wave momentum and angular momentum. A second surge occurred in the past several decades, driven by technological breakthroughs: the invention of lasers and the controlled generation of structured wave fields. This resulted in major discoveries, including optical trapping and manipulation of small particles, from atomic to micro sizes, as well as acoustic manipulation of larger particles, including biological cells and samples. 
Nowadays, radiation forces and torques underpin numerous applications: optical and acoustic tweezers, acoustofluidic sorting of biological cells, optomechanical systems operating in both classical and quantum regimes, solar sails, quantum simulators, volumetric displays, etc. 
In this review, we present a unifying perspective on optical and acoustic forces and torques acting on various particles, addressing both their theoretical foundations and key applications. Our approach relies on the universal connection between the local energy, momentum, and spin densities of wave fields and the principal forces and torques exerted on small particles. 
Moreover, we describe important cases of nontrivial (e.g., lateral and pulling) forces and complex (e.g., chiral and anisotropic) particles. 
We also highlight significant experimental achievements involving optical and acoustic manipulation in structured wave fields. Our aim is to illuminate the common fundamental origins and close interconnections between the mechanical actions of optical and acoustic fields, thereby fostering a deeper understanding and advancing the development of optomechanical and acoustomechanical applications.     

\end{abstract}

\maketitle
\tableofcontents

\section{Introduction}
\label{sec:intro}

\begin{figure*}
\centering
\includegraphics[width=0.85\linewidth]{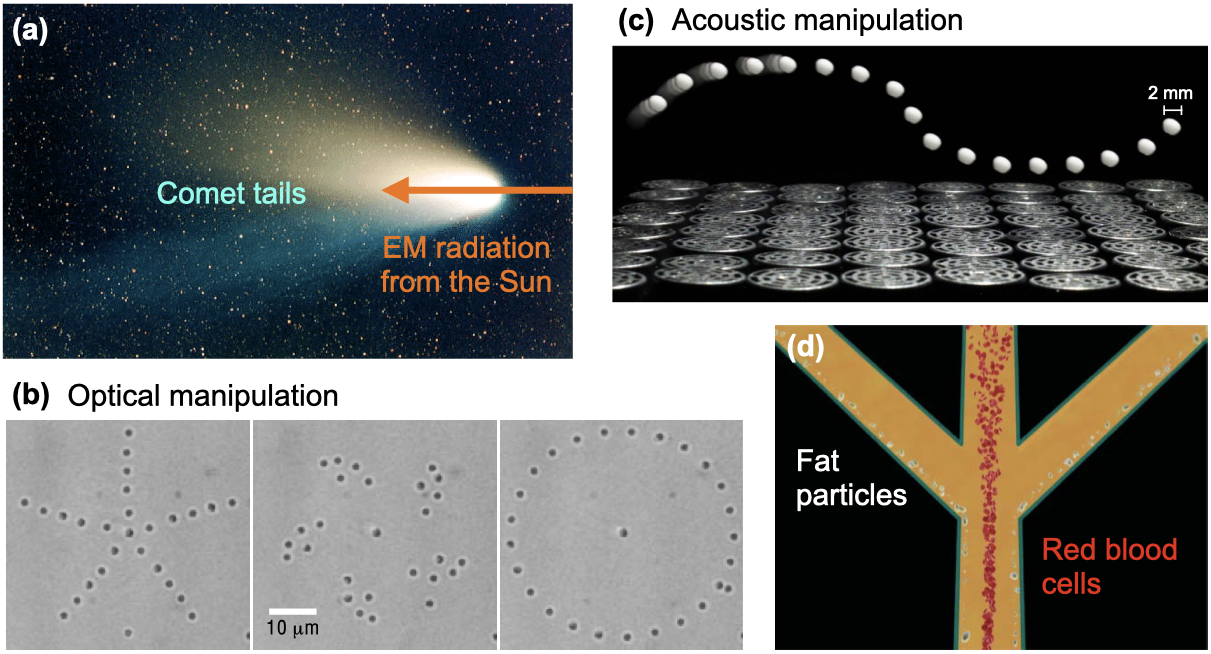}
\caption{Examples of manifestations of optical and acoustic forces. (a) The story began with Kepler's 17th-century suggestion that a comet's tail points away from the Sun due to radiation pressure from sunlight [image by Fred Espenak, NASA GSFC]. (b) Modern manipulation of dielectric microparticles using holographic optical tweezers. From \cite{Curtis2002OC}. (c) Acoustic holographic manipulation of millimeter-sized particles. From \cite{Marzo2015NC}. (d) Acoustofluidic (ultrasound) sorting of human lipid and erythrocytes cells. From \cite{Petersson2005LC}.}
\label{fig:overview}
\end{figure*}

The interaction between waves and matter plays a fundamental role in many areas of physics, from fluid mechanics to quantum field theory. The effect of matter on waves is obvious: for instance, a stone thrown into a water reservoir generates waves on the water surface. The reverse action of waves on matter is less obvious, yet it also occurs for all types of waves. For example, water-surface waves can push floating bodies in the direction of wave propagation. Similar effects arise for electromagnetic (light) and acoustic (sound) waves impinging on material objects.

Notably, over the past 150 years, research on optical and acoustic wave-induced mechanical effects has developed in parallel. It has progressed from 19th-century studies of radiation pressure and wave momentum, through advances in optical and acoustic trapping and forces on small particles, to impressive modern achievements in particle manipulation, such as optical and acoustic `tweezers', `spanners', levitation, sorting, and 3D holographic arrangements.

In this review, we present a unified description of theoretical fundamentals and major experimental advances in the mechanical action of electromagnetic and acoustic waves on various particles. We demonstrate that these optical and acoustic effects share a common physical origin, exhibit many parallel manifestations, and can be naturally considered within the same coherent framework.

\subsection{Historical overview}
\label{sec:historical}

The idea of mechanical pressure exerted by light dates back to 1619, when Kepler  observed that a comet's tail always points away from the Sun, as if blown by a `light wind' emanating from it \cite{Kepler1619}, Fig.~\ref{fig:overview}(a). In 1746, Euler described a similar effect for sound waves in air: ``{\it sound vigorously excites not only a vibratory motion in the air particles, but one also observes a real motion in small, very light dust particles which tumble in the air}'', and immediately extended this idea to light: ``{\it it cannot be doubted that the vibratory motion caused by the light produces a similar effect}'' \cite{Whittaker1989}. 

According to Newtonian mechanics, any force on a body originates from a {\it momentum} transfer, which leads us to think that light, or any other wave pushing the body, must carry momentum. The closely related concepts of light pressure and momentum were elaborated within Maxwell's electromagnetic theory. In 1873, Maxwell wrote: ``{\it in a medium in which waves are propagated, there is a pressure in the direction normal to the waves, and numerically equal to the energy in unit of volume}'' \cite{Maxwell1873}. In 1884, Poynting introduced the concept of electromagnetic {\it energy flow} (Poynting vector) \cite{Poynting1884}, and in 1905 he described a beam of light as ``{\it a stream of momentum, the direction of the momentum being along the line of propagation, and the amount of momentum passing per second through unit area cross-section of the beam being equal to the density of the energy in it}'' \cite{Poynting1905, Poynting1905_II}. Remarkably, a decade earlier, in 1874, Umov described an analogous energy flow for acoustic waves in fluids and solids \cite{Umov1874}. 

\begin{figure}
\centering
\includegraphics[width=\linewidth]{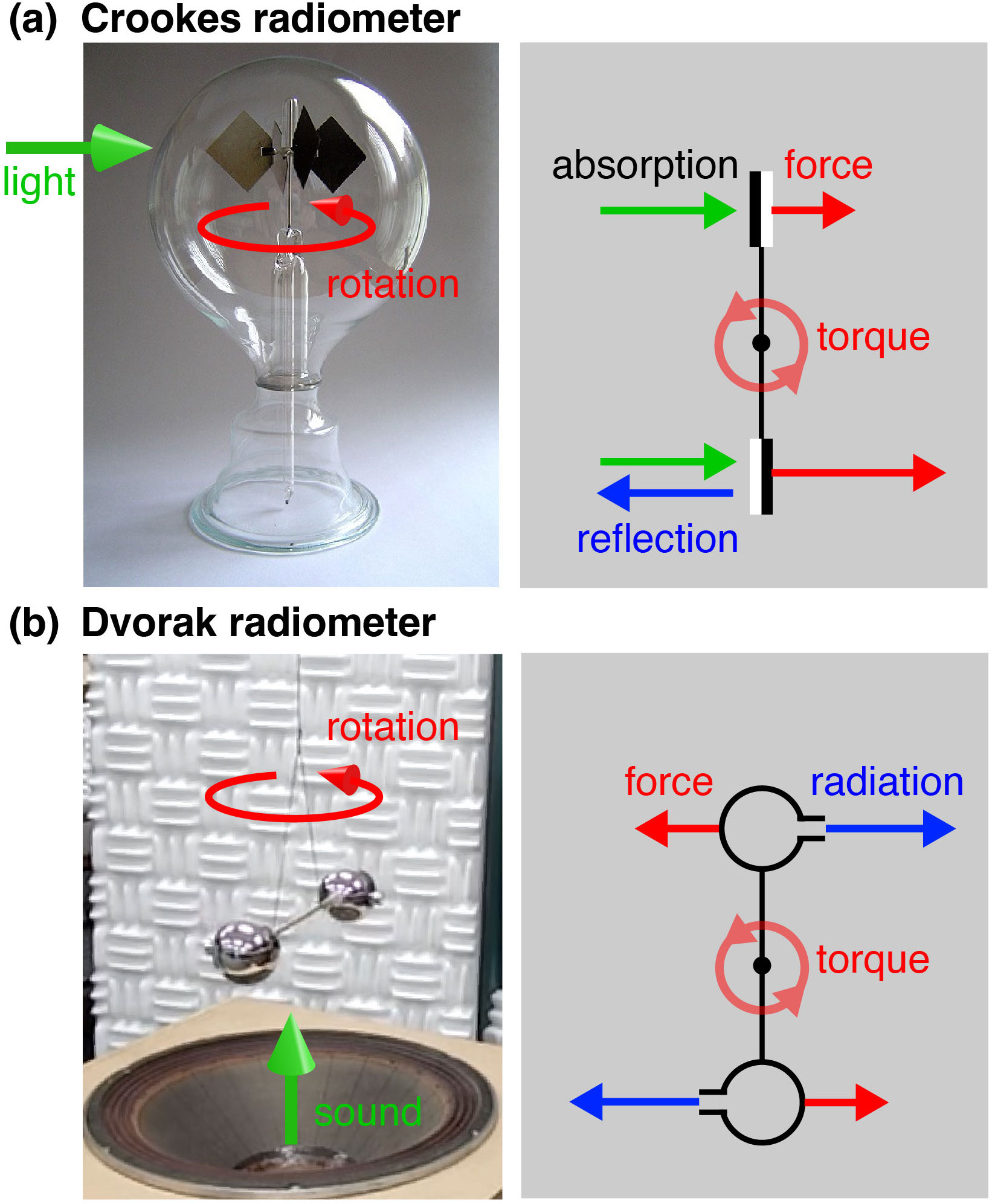}
\caption{`Executive toys' inspired by early 19th-century experiments on the mechanical action of electromagnetic and acoustic waves. (a) The Crookes radiometer featuring vanes with absorbing (black) and reflecting (white) surfaces. (b) The acoustic Dvorak radiometer using Helmholtz resonators that emit sound from their open sides. Schematics of the radiation-induced rotations (excluding photophoretic and nonlinear acoustic effects) are shown on the right. The photo images are adapted from \cite{crookes_wiki, Russell2011}.}
\label{fig:toys}
\end{figure}

Alongside these theoretical developments, there were numerous attempts to measure radiation pressure in a laboratory. While the presence of a radiation pressure force induced by sound was already obvious to Euler, detecting the pressure of light was far more challenging. In 1874, Crookes reported measurements of the ``{\it attraction and repulsion resulting from radiation}'' \cite{Crookes1874} using a so-called `light mill' (now known as the Crookes radiometer) \cite{Woodruff1966, Brush1969}, Fig.~\ref{fig:toys}(a). This, now available as an `executive toy', consists of a rotor inside a near-vacuum glass bulb, with its vanes painted white (reflective) on one side and black (absorptive) on the other. When this rotor is exposed to light coming from one side, the white (black) vane surfaces reflect (absorb) light, which corresponds to double (single) amount of the momentum transfer, as shown in Fig.~\ref{fig:toys}(a). This should produce a torque and rotation of the rotor proportional to the light momentum. Rotation is indeed observed, but detailed analysis reveals that it arises from a thermal effect (photophoretic force) due to heating of the black surfaces and the resulting motion of the residual gas in the bulb \cite{Woodruff1966, Brush1969}. Furthermore, the observed rotation often occurs in the opposite direction to that expected from light pressure. Thus, the Crookes radiometer failed to measure the radiation pressure. Nearly 30 years later, in 1901, Lebedew, as well as Nichols and Hull, succeeded in making the first conclusive laboratory measurements of electromagnetic radiation pressure consistent with Maxwell's theory \cite{Lebedew1901, Nichols1901, Nichols1903, Lebedew1910}. 

Remarkably, in 1877, around the same time as the Crookes radiometer, Dvo\v{r}\'{a}k performed somewhat similar experiments using acoustic open (Helmholtz) resonators mounted on a rotor and generating rotation when exposed to an external sound source \cite{Dvorak1878}, Fig.~\ref{fig:toys}(b). This effect is readily reproduced in room conditions \cite{Ingard2008, Russell2011, Russell2025}. It is usually explained by a flow of air induced by the sound and going out of the open side of the resonators (which is a nonlinear fluid-mechanical effect), but can also be described within linear acoustics and the asymmetric scattering of the sound. Namely, the sound, scattered preferentially in the direction of the open end of the resonator, generates an oppositely-directed {\it recoil} force, as shown in Fig.~\ref{fig:toys}(b). Note that a direct acoustic analogue of the Crookes radiometer can also be constructed \cite{Denardo2004, Russell2020}.

In 1902--1905, almost simultaneously with Poynting's seminal works, Rayleigh developed the theoretical foundations of acoustic radiation pressure \cite{Rayleigh1902, Rayleigh1905}. He also envisaged the universal character of wave-induced radiation forces: ``{\it it would be of interest to inquire whether other kinds of vibration exercise a pressure, and if possible to frame a general theory of the action}'' \cite{Rayleigh1902}.  

Next, alongside the electromagnetic wave momentum and the corresponding radiation force, it was noticed that a circularly (or, generally, elliptically) polarized electromagnetic wave carries an intrinsic {\it angular momentum} (now known as {\it spin}) which can exert a mechanical {\it torque} on bodies. The main features of this phenomenon for light passing through an anisotropic crystal plate were described in the pioneering 1899 work by Sadowsky \cite{Sadowsky1899}. Since this work was published in an antique Russian journal, it is worth translating and quoting its key conclusions: 
``{\it A crystal plate, ground parallel to the principal dielectric axes, which transforms the normally incident linearly-polarized electromagnetic waves into elliptically-polarized ones, tends to rotate in the direction opposite to the rotation of electromagnetic forces} (i.e., fields) {\it generated by the plate. ... These forces only tend to align the plate in a certain way but not to give it a continuous rotation.}'' and 
``{\it The forces developed during the transmission of circularly-polarized electromagnetic waves through a crystal plate, except for the $4/4\, \lambda$ plate, tend to lead it into continuous rotation in the direction of rotation of the electromagnetic forces in the incident waves. ... The largest torque is obtained for the $2/4\, \lambda$ plate}''. 
Sadowsky's work fully anticipated the optical torques on anisotropic plates that were first measured 37 years later \cite{Beth1935, Beth1936, Holbourn1936}, and---at a much finer scale---on small anisotropic particles a century later \cite{Friese1998Nat}. In 1909, using mechanical analogies, Poynting explicitly proposed that circularly-polarized light carries angular momentum \cite{Poynting1909}. 

However, despite these early milestones, systematic investigations of optical and acoustic torques emerged only recently \cite{Brasselet2023AP}. This is especially true for sound waves, which cannot exhibit circular polarizations in the paraxial, plane-wave-like regime. Although the presence of intrinsic angular momentum (spin) produced by the acoustic counterparts of electromagnetic elliptical polarizations in structured (non-plane-wave) sound fields was anticipated in 1973 \cite{Jones1973}, the first experimental demonstration of acoustic spin and torque was reported only in 2019 \cite{Shi2019}. 

Otherwise, one can see that the studies of electromagnetic and acoustic radiation pressure and forces have evolved in parallel ways. For reviews on the fascinating history of these investigations, see \cite{Jones1953, Schagrin1974, Beyer1978, Worrall1982, Loudon2012, Thomas2017, Sarvazyan2010UMB}.

\subsection{Modern development}

A revolution in the study of optical forces began with the invention of the laser in 1960 and the start of modern laser optics. In 1970, Ashkin reported the acceleration and trapping of micron-sized dielectric particles in focused laser beams \cite{Ashkin1970}. This discovery paved the way for the vast field of {\it optical manipulation}, including trapping, tweezers, levitation, and related techniques, which now has an enormous range of applications. In particular, optical manipulation of atoms \cite{Raab1987, Chu1998, Phillips1998} and biological objects \cite{Ashkin1987, Ashkin1987_II} ultimately led to Nobel Prizes in physics in 1997 and 2018. Notably, Ashkin's pioneering work already mentioned the possibility of optical trapping of atoms and molecules, as well as ``{\it angular acceleration} (i.e., torque) {\it of trapped particles based on optical absorption of circularly polarized light}''. Although Ashkin attributed optical trapping to ``radiation pressure'', the dominant mechanism is the {\it gradient force} proportional to the spatial gradient of the electromagnetic field intensity rather than the radiation-pressure force directed along the light's propagation (wave vector). In fact, these two forces are near-orthogonal in paraxial optical beams. 

The gradient force was first described in 1957 by Gaponov and Miller for the trapping of charged particles in a high-frequency electromagnetic field \cite{Gaponov1958}. For neutral dipole (Rayleigh) particles, optical forces, comprising both the gradient and radiation-pressure (scattering) parts, were examined in the 1970s by Gordon and Ashkin \cite{Gordon1973PRA, Gordon1980PRA, Ashkin1983OL}. 

Remarkably, small particles in a sound-wave field experience a very similar acoustic gradient force, first described by Gor'kov in 1961 \cite{Gorkov1962}. 
Two decades after Ashkin, Wu employed this force to achieve a stable acoustic trapping of latex particles and frog eggs in ultrasonic beams \cite{Wu1991JASA}. This initiated the rapidly developing field of {\it acoustic manipulation}, with numerous applications, particularly in life sciences and acoustofluidics \cite{Ozcelik2018NM, Meng2019JPD}. 

\begin{figure}
\includegraphics[width=0.75\linewidth]{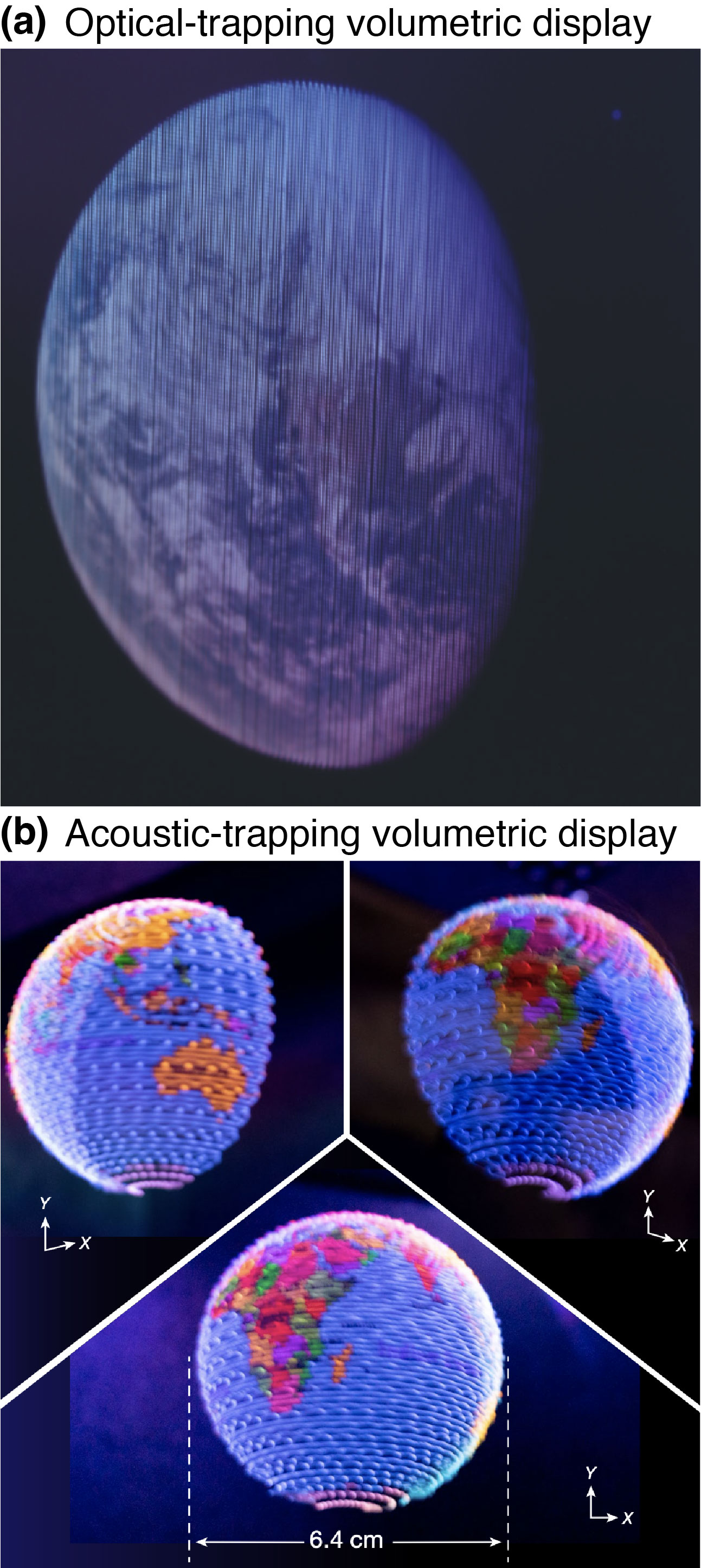}
\caption{Almost 150 years after the first experiments by Crookes and Dvorak: 3D images produced by volumetric displays based on optically (a) and acoustically (b) trapped small particles. From \cite{Smalley2018Nature} and \cite{Hirayama2019N}.}
\label{fig:displays}
\end{figure}

Another revolution in optical and acoustic manipulation came with the development of {\it structured} optical and acoustic fields, which can have arbitrarily space-varying phase and amplitude profiles and essentially differ from simple plane waves \cite{Grier2003N, Otte2020APR, Yang2021AP, Andrews2008, Rubinsztein-Dunlop2016JO, Bliokh2023JO}. A prominent example is provided by {\it vortex beams} generated in both optics and acoustics in the 1990s \cite{Bazhenov1990, Allen1992PRA, Hefner1999JASA}. Such beams feature a doughnut-shaped transverse intensity profile and an azimuthal phase gradient that produces {\it orbital angular momentum} (OAM). Vortex beams are perfectly suitable for radial trapping and azimuthal rotation of particles \cite{He1995PRL, Garces-Chavez2003PRL, Curtis2003PRL, Hong2015PRL, Baresch2018PRL, Baresch2016PRL}. 

Today, structured optical and acoustic fields enable impressive 2D and 3D {\it holographic} manipulation of multiple particles \cite{Curtis2002OC, Barredo2018N, Marzo2015NC, Melde2016N}, Fig.~\ref{fig:overview}(b,c), as well as efficient sorting of particles (e.g., biological molecules) according to size, shape, and other parameters \cite{Ozcelik2018NM, Wu2019MN, Laurell2007CSR}, Fig.~\ref{fig:overview}(d). 
Recently, almost 150 years after the first experiments of Crookes and Dvorak, optical and acoustic forces have been employed to trap 3D arrays of small particles for volumetric displays, Fig.~\ref{fig:displays} \cite{Smalley2018Nature, Hirayama2019N}. Furthermore, arrays of optically trapped atoms have provided a platform for quantum simulators \cite{Barredo2016S, Barredo2018N, Bernien2017N, Ebadi2021N}. 
In addition, optical forces and torques underpin numerous optomechanical systems, spanning both classical and quantum regimes \cite{Aspelmeyer2014RMP, Gonzalez-Ballestero2021S, Millen2020RPP, Chang2010PNAS, Tebbenjohanns2021N}. 

While optical and acoustic manipulations share many similarities, they also complement each other in their operational ranges: optical trapping is typically suited for particles from $100$~nm to 10~${\rm \mu m}$ scales, whereas acoustic trapping excels with larger objects from 1~${\rm \mu m}$ to 1~cm scales \cite{Dholakia2020NRP}. 
For example, the optical and acoustic volumetric displays in Fig.~\ref{fig:displays} are based on trapping particles of $10\,{\rm \mu m}$ and $2\,{\rm mm}$, respectively. 

It should be noted that most research on optical and acoustic manipulation has focused on radiation forces, while torques have received comparatively little attention (apart from studies in the context of OAM and vortex beams). Indeed, simple expressions for the dipole-approximation optical and acoustic torques on isotropic Rayleigh particles in arbitrary structured fields were derived only recently \cite{Chaumet2009OE, Canaguier-Durand2013PRA, Bliokh2014NC, Nieto-Vesperinas2015OL, Toftul2019PRL}. Although radiation forces are generally more important for trapping and tweezers, local torques have the same order of magnitude and can be crucial for inducing rotations and controlling orientation of the particles. Remarkably, recent experiments have achieved GHz-scale rotation speeds of optically trapped particles in vacuum \cite{Reimann2018PRL, Ahn2018PRL}, reaching the quantum-oscillator regime \cite{Tebbenjohanns2021N, Gonzalez-Ballestero2021S}. Likewise, recent acoustic experiments have produced kHz-scale rotations of trapped particles \cite{Zhang2022PRA}.

Let us recall that, from the outset, studies of optical and acoustic radiation forces and torques have been closely tied to the fundamental concepts of wave momentum and angular momentum. For structured fields and waves in media, these concepts have sparked long-standing debates: the famous Abraham-Minkowski dilemma for the electromagnetic plane-wave momentum in a medium \cite{Pfeifer2007, Barnett2010PTRS, Milonni2010AOP}, the problem of the sound-wave momentum \cite{Mcintyre1981, Peierls1979, Peierls1991}, and the separation of the spin and orbital angular momenta in optics \cite{Allen_book, Andrews_book, VanEnk1994JMO} and in field theory \cite{Leader2014PR}, among others. 

The past decade has brought significant progress in resolving these issues, particularly in the context of the mechanical action of structured wave fields. It has been shown that for monochromatic (single-frequency) and arbitrarily inhomogeneous optical and acoustic fields, one can define universal {\it canonical momentum and spin angular momentum densities} \cite{Bliokh2025CP}, such that the radiation pressure force and torque on isotropic Rayleigh particles are precisely determined by these densities \cite{Berry2009, Bliokh2014NC, Bliokh2015PR, Leader2016PLB, Shi2019, Toftul2019PRL}. This provides a unifying and intuitively clear framework for the local wave momentum/force and spin/torque properties. Notably, the universality of this approach makes it readily extendable to other types of waves, such as water-surface waves \cite{Bliokh2022SA, Wang2024}.

Importantly, in this framework, the local spin density and the associated torque are inherent to any kind of classical waves: electromagnetic, sound, water-surface, and others \cite{Bliokh2025CP}. This significantly extends the concept originally introduced by Sadowsky and Poynting, which for many decades was associated exclusively with electromagnetic waves. The explanation is as follows. A plane electromagnetic wave can have a transverse circular polarization and the corresponding longitudinal spin angular momentum, whereas a plane sound wave consists of purely longitudinal oscillations of the medium, with no circular motion and spin. However, {\it structured} sound fields (i.e., superpositions of multiple plane waves with oscillations in different directions) generically produce {\it elliptical} local oscillations of the medium, similar to elliptical polarizations in electromagnetism. This results in local spin angular momentum and its mechanical manifestations \cite{Jones1973, Shi2019, Bliokh2019b, Bliokh2022SA, Bliokh2025CP}. 

\subsection{About this review}
\label{sec:about}

Over the past decades, there have been numerous reviews dedicated to: optical forces and manipulation \cite{Dienerowitz2008, Molloy2002CP, Grier2003N, Bowman2013RPP, Ashkin2000IEEE, Ashkin_book, Woerdemann2013LPR, Dholakia2011NP, Sukhov2017RPP, Gao2017LSA, Yuan2020JPD, Volpe2023JPP, Gieseler2021AOP, Pesce2020EPJP, Polimeno2018JQSRT, Marago2013NN, Riccardi2023ChemRev, Zemanek2019AOP, Otte2020APR, Yang2021AP}, acoustic forces and manipulation \cite{Bruus2012a, Meng2019JPD, Ozcelik2018NM, Thomas2017, Baudoin2020AR, Lim2024RPP, Evander2012LC, Sarvazyan2010UMB}, acoustofluidic manipulation of bioparticles \cite{Friend2011RMP, Lenshof2012LC, Ding2013LC, Hossein2023BR, Wu2019MN, Laurell2007CSR}; and more specifically to: optical pulling forces \cite{Dogariu2013NP, Ding2019AP, Li2020AOP},  {optical lateral forces \cite{Shi2023AOP}, optical sorting in general \cite{Yang2025LSA}} and sorting of chiral particles \cite{Genet2022ACSPhot, Kakkanattu2021OE}, optical binding \cite{Forbes2020NP, Dholakia2010RMP, Bowman2013RPP}, optical levitation in vacuum \cite{Gonzalez-Ballestero2021S}, optical manipulation of liquid crystal droplets \cite{Brasselet2009JNOP} and single molecules \cite{Bustamante2021NR}, plasmonic tweezers \cite{Zhang2021LSA}, optical manipulation with metamaterial structures \cite{Shi2022APR}, optomechanics with levitated particles \cite{Millen2020RPP}, {and force-microscopy based on optical trapping \cite{Neuman2008NM, Stout1997}}.

Despite this extensive literature covering most of the specific aspects and applications of optical and acoustic forces, several general gaps are evident. 
First, only a few original papers and reviews have addressed optical and acoustic forces on equal footing \cite{Toftul2019PRL, Dholakia2020NRP, Abdelaziz2020PRR, Thomas2017}. Second, radiation torques have received comparatively little attention (mostly in optics) \cite{Padgett2011NP, Bustamante2021NR, Riccardi2023ChemRev}. Finally, theoretical introductions to optical and acoustic forces and torques usually lack clear connections to the local dynamical properties of wave fields: the momentum and spin angular momentum densities. 

In this review, we aim to fill these gaps by providing a clear yet comprehensive description of the main optical and acoustic forces and torques. We restrict ourselves to monochromatic but arbitrarily structured wave fields, considering fundamental radiation forces and torques in connection with the local momentum and spin properties of the waves. Additionally, we consider nontrivial forces and torques which appear in the cases of complex (anisotropic, chiral, resonant, etc.) particles. Finally, we overview key examples of optical and acoustic forces and torques in specific settings: evanescent waves, vortex beams, pulling forces, optical sorting of chiral particles, and acoustofluidic sorting of bioparticles, among others. 
We hope that this review will provide a unified and versatile perspective on both the fundamentals and diverse manifestations of the mechanical action of optical and acoustic fields.

It is worth noting that the nomenclature of radiation forces varies strongly across different studies and disciplines. For example, the term ``radiation pressure force'' can refer either to the gradient force or to the force associated with the momentum of the incident wave, while acoustic radiation forces are sometimes referred to as ``acoustoforetic''. Throughout this review, we adopt a unified nomenclature implying: (i) {\it gradient} forces associated with gradients of the energy density (intensity) of the incident wave field, (ii) {\it radiation-pressure} forces associated with the local momentum (wavevector) of the incident wave, and (iii) {\it recoil} forces associated with the momentum carried away by the scattered wave field.

\section{Dynamical Properties of Wave Fields}
\label{sec:momenta}

{In this Section, we outline the main dynamical properties of electromagnetic and sound waves, which determine the wave interaction with particles in Section~\ref{sec:forces}. A more detailed discussion of the momentum and angular momentum of various classical waves can be found in \cite{Bliokh2025CP}.}

\subsection{General features}
\label{sec:generalDynamicalWaveFeatures}

Throughout this review, we deal with monochromatic wave fields of the form $\mathbfcal{A}(\vb{r}, t) = \Re \left[\vb{A}(\vb{r}) e^{-i \omega t }\right]$, where $\vb{A}(\vb{r})$ is the complex space-variant field amplitude and $\omega$ is the frequency. Quadratic forms of the field, such as energy, momentum, force, etc., are evaluated assuming time-averaging over one oscillation period. 
The key quantities of interest are the wave {\it energy}, {\it momentum}, and {\it spin angular momentum (spin)} densities. 
The wave energy density is typically proportional to the field intensity: $U\propto |\vb{A}|^2$.
However, the definition of wave momentum has long been a subject of debate in both electromagnetism and acoustics \cite{Pfeifer2007, Brevik1979, Peierls1979, Peierls1991, Mcintyre1981, Barnett2010PTRS, Milonni2010AOP}, primarily because there are two distinct forms: kinetic and canonical momentum. 

\begin{figure}
\centering
\includegraphics[width=0.95\linewidth]{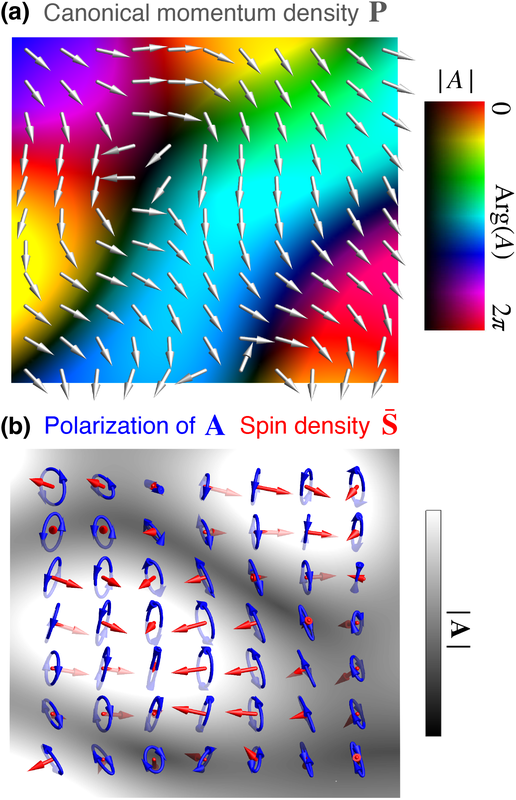}
\caption{(a) A complex scalar wave field $A(\vb{r})$ and its canonical momentum density $\vb{P} \propto \Im (A^* \grad A) = |A|^2 \grad {\rm Arg}(A)$. The phase ${\rm Arg}(A)$ and amplitude $|A|$ are presented by colors (online) and brightness, respectively \cite{Thaller_book}. (b) A vector wave field $\vb{A} (\vb{r})$ visualized via its amplitude $|\vb{A}|$ (grayscale) and polarization ellipses traced by $\mathbfcal{A}(\vb{r}, t) = \Re\! \left[\vb{A}(\vb{r}) e^{-i \omega t }\right]$ in each point $\vb{r}$. The normalized spin density $\bar{\vb{S}} = \vb{S}/ |\vb{A}|^2 \propto \Im (\vb{A}^* \times \vb{A}) / |\vb{A}|^2$ is directed along the normal to the polarization ellipse and is proportional to its ellipticity. The electromagnetic and acoustic canonical momentum and spin densities, Eqs.~(\ref{eq:PEcan})--(\ref{eq:SEM}) and (\ref{eq:PAcan})--(\ref{eq:notation}) are the natural counterparts of these quantities for the corresponding fields.}
\label{fig:momentum-spin}
\end{figure}

The density of the canonical momentum (sometimes called `pseudomomentum') is associated with the local {\it phase gradient} or {\it local wavevector} of the wave field, Fig.~\ref{fig:momentum-spin}(a). It can be written as $\vb{P} \propto \Im [\vb{A}^* \cdot (\grad) \vb{A}] \equiv \Sigma_i \Im ({A}_i^*  \grad {A}_i) = \Sigma_i |A_i|^2 \grad {\mathrm{Arg}} (A_i)$, $i=x,y,z$, which represents a weighted average of the phase gradients of the field components. This form is analogous to the probability current in quantum mechanics or to the local expectation value of the quantum-mechanical momentum operator $-i\grad$, averaged over the field components \cite{Berry2009,Berry2013,Bliokh2013NJP,Bliokh2014NC,Bliokh2025CP}. 

In turn, the kinetic momentum density is associated with the energy flux density (e.g., the Poynting vector for electromagnetic fields), which generally has a different form. Remarkably, the canonical momentum density $\vb{P}$ and kinetic momentum density $\vb{\Pi}$ are related through the universal Belinfante-Rosenfeld relation, originally derived in relativistic field theory \cite{Belinfante1940,Soper_book,Berry2009,Bliokh2013NJP,Bliokh2014NC,Bliokh2022SA,Bliokh2025CP}:
\begin{equation}
\vb{\Pi} = \vb{P} + \frac{1}{2}\grad\times \vb{S}\,.
\label{eq:BR}
\end{equation}
Here, $\vb{S}$ is the spin angular momentum (hereafter, spin) density of the field, which characterizes the local rotation (elliptical polarization) of the vector wave field: $\vb{S} \propto \Im (\vb{A}^* \times \vb{A})$. Namely, the time-dependent field $\mathbfcal{A}(\vb{r},t)$ traces an ellipse at each point of space, and the spin density is directed along the normal to this ellipse and has magnitude proportional to its ellipticity, Fig.~\ref{fig:momentum-spin}(b). 

Since the canonical and kinetic momentum densities differ by the curl of a vector field, both satisfy the stationary continuity equation in a homogeneous lossless medium: $\grad \cdot \vb{\Pi} = \grad \cdot \vb{P} =0$, and their integral values coincide for any localized wave state. However, their {\it local} values and directions can differ significantly in inhomogeneous wave fields, and this is crucial for local wave-particle interactions.

The total angular momentum density of the wave field comprises spin and orbital parts \cite{Allen_book, Andrews_book, Bliokh2015PR, Shi2019, Bliokh2019b, Burns2020}: $\vb{J} = \vb{S} + \vb{r}\times \vb{P}$, where the orbital part is determined by the canonical momentum density {and the position vector $\vb{r}$}. Since we are primarily concerned with wave-induced torques evaluated relative to the center of a small particle, $\vb{r} = \vb{0}$, such torque is naturally determined by the spin rather than orbital angular momentum density. The local Belinfante-Rosenfeld relation (\ref{eq:BR}) yields the integral identity for the total angular momentum of a localized wave state expressed via canonical and kinetic momenta: $\int (\vb{S} + \vb{r}\times \vb{P})\, dV = \int (\vb{r}\times \vb{\Pi})\, dV$. However, for the {\it local} characterization of the wave field, the canonical momentum density $\vb{P}$ and spin density $\vb{S}$ remain the two key independent quantities, describing the field's phase and polarization features, respectively. 

As the cross product ``$\times$'' produces pseudo-vectors, changing their signs under mirror reflections, one can notice that the energy density is a true scalar, the canonical and kinetic momentum densities are true vectors, while the spin and orbital angular momentum densities are pseudo-vectors. One can also construct a {\it pseudo-scalar} quadratic form: $\mathfrak{S} \propto \Re[\vb{A}^* \cdot (\grad \times \vb{A})]$, which characterizes the local {\it chirality} of the field \cite{Trueba1996,Tang2010,Bliokh2011PRA}. This quantity plays an important role in interactions with chiral particles that are not equivalent to their mirror reflections. 

\subsection{{Electromagnetic fields}}

Electromagnetic waves are described by two vector fields: electric $\vb{E}$ and magnetic $\vb{H}$. The electromagnetic energy density in an isotropic non-dispersive medium is \cite{jackson1998ClassicalElectrodynamics}
\begin{equation}
U_{{EM}} = \frac{1}{4} \left( {\varepsilon} |\vb{E}|^2 + {\mu} |\vb{H}|^2 \right) \equiv \frac{U^{(\vb{E})} + U^{(\vb{H})}}{2}\,,
\label{eq:UEM}
\end{equation}
where $\varepsilon$ and $\mu$ are the permittivity and permeability of the medium, and we write the total energy density as the arithmetic mean of the purely electric and purely magnetic contributions. 
The kinetic momentum density is described by the well-known Poynting vector \cite{jackson1998ClassicalElectrodynamics}:
\begin{equation}
\vb{\Pi}_{{EM}} = \frac{1}{2c_l^2}  \Re (\vb{E}^*\! \times \vb{H})\,.
\label{eq:PEMkin}
\end{equation}
Here we use the Minkowski (rather than Abraham) version of this quantity involving the speed of light {\it in the medium}, $c_l=1/\sqrt{\varepsilon\mu}$ \cite{Pfeifer2007, Milonni2010AOP, Barnett2010PTRS, Bliokh2017PRL}.

Standard electromagnetic field theory yields the canonical momentum and spin densities described by the general forms considered in Section~\ref{sec:generalDynamicalWaveFeatures} and involving the electric field \cite{Soper_book, Bliokh2013NJP, Bliokh2025CP}:
\begin{equation}
\vb{P}^{(\vb{E})} = \frac{1}{2\omega} \Im \!\left[ \varepsilon \vb{E}^*\! \cdot (\grad) \vb{E} \right],
\label{eq:PEcan}
\end{equation}
\begin{equation}
\vb{S}^{(\vb{E})} = \frac{1}{2\omega} \Im \!\left( \varepsilon\, \vb{E}^*\!\times \vb{E} \right).
\label{eq:SE}
\end{equation}
These quantities are convenient in problems where the matter interacts directly with the wave electric field (e.g., for electric-dipole particles), whereas magnetic coupling is inessential \cite{Gordon1980PRA, Ashkin1983OL, Berry2009, Bliokh2013NJP_II, Canaguier-Durand2013PRA, Antognozzi2016NP}. 

In more general cases involving both electric and magnetic interactions, it is advantageous to define the canonical momentum and spin densities incorporating both the electric and magnetic fields \cite{Berry2009, Cameron2012, Bliokh2013NJP, Bliokh2014NC, Bliokh2015PR, Aiello2015NP, Bliokh2017PRL}:
\begin{equation}
\vb{P}_{{EM}} = \frac{1}{4\omega} \Im \!\left[ \varepsilon \vb{E}^*\! \cdot (\grad) \vb{E} + \mu \vb{H}^*\! \cdot (\grad) \vb{H} \right] \equiv \frac{\vb{P}^{(\vb{E})} + \vb{P}^{(\vb{H})}}{2}.
\label{eq:PEMcan}
\end{equation}
\begin{equation}
\vb{S}_{{EM}} = \frac{1}{4\omega} \Im \!\left( \varepsilon\, \vb{E}^*\!\times \vb{E} + \mu\, \vb{H}^*\!\times \vb{H} \right) 
\equiv \frac{\vb{S}^{(\vb{E})} + \vb{S}^{(\vb{H})}}{2}.
\label{eq:SEM}
\end{equation}

\bgroup
\def\arraystretch{1.7}%
\begin{table*}[]
\caption{{The main local dynamical properties of monochromatic electromagnetic and acoustic wave fields.} \label{tab:momenta}}
    \begin{ruledtabular}
\begin{tabular}
{p{0.33\linewidth}>{\centering\arraybackslash}p{0.33\linewidth}>{\centering\arraybackslash}p{0.33\linewidth}}
\textbf{Quantity} & \textbf{Electromagnetism} & \textbf{Acoustics} \\  \hline
Wave fields
& Electric $\vb{E}$ and magnetic $\vb{H}$ 
& Velocity $\vb{v}$ and pressure $p$ \\
Energy density
& $U^{({\bf E})} = \dfrac{1}{2}{\varepsilon}   |\vb{E}|^2 $,~ 
$U^{({\bf H})} = \dfrac{1}{2} {\mu}   |\vb{H}|^2$
& $U^{({\bf v})} =\dfrac{1}{2} \rho |\vb{v}|^2$,~$U^{({p})} =\dfrac{1}{2} \beta |p|^2$ \\
Kinetic momentum density
& $\vb{\Pi}_{EM}=\dfrac{1}{2c_l^2} \Re (\vb{E}^* \times \vb{H})$
& $\vb{\Pi}_{A}=\dfrac{1}{2 c_s^2} \Re (\vb{v}^* p)$ \\
Canonical momentum density 
& $\vb{P}^{({\bf E})} = \dfrac{1}{2\omega} \Im\! \left[ {\varepsilon}\,  \vb{E}^{*}\!\cdot (\grad) \vb{E} \right]$, 
& $\vb{P}^{({\bf v})}=\dfrac{1}{2\omega} \Im\! \left[ \rho\, \vb{v}^* \!\cdot(\grad) \vb{v} \right]$, \\
& $\vb{P}^{({\bf H})} = \dfrac{1}{2\omega} \Im\! \left[{\mu}\, \vb{H}^{*}\!\cdot (\grad) \vb{H} \right]$ & $\vb{P}^{({p})}=\dfrac{1}{2\omega} \Re (\beta\,p^* \grad p) \equiv \vb{\Pi}_{A}$ \\
Spin density
&  $\vb{S}^{({\bf E})} = \dfrac{1}{2\omega} \Im\! \left(  {\varepsilon}\,  \vb{E}^*\! \times \vb{E} \right)$, 
&  $\vb{S}^{({\bf v})} = \dfrac{1}{2 \omega} \Im\! \left(\rho\, \vb{v}^*\! \times \vb{v} \right)$ \\
& $ \vb{S}^{({\bf H})} = \dfrac{1}{2\omega}\Im\! \left(   {\mu}\,  \vb{H}^*\! \times \vb{H} \right)$ & $\vb{S}^{(p)} \equiv {\bf 0}$ \\
Chirality (helicity)  & $\mathfrak{S}_{EM}=\dfrac{1}{2\omega c_l} \Im(\vb{H}^* \cdot \vb{E})$ & $\mathfrak{S}_{A}\equiv 0$ 
\end{tabular}
\end{ruledtabular}
\end{table*}
\egroup

Using Maxwell's equations $i\omega \mu \vb{H} = \grad \times \vb{E}$ and $i\omega \varepsilon \vb{E} = - \grad \times \vb{H}$, one can verify that the quantities (\ref{eq:PEcan}) and (\ref{eq:SE}), as well as (\ref{eq:PEMcan}) and (\ref{eq:SEM}), satisfy the Belinfante-Rosenfeld relation (\ref{eq:BR}) with the kinetic momentum (\ref{eq:PEMkin}). 
Note that the factors $\omega^{-1}$ in the canonical momentum and spin densities (\ref{eq:PEcan})--(\ref{eq:SEM}) indicate that these quantities are well defined only for monochromatic fields.

Finally, by using Maxwell's equations transforming the curls of the electric/magnetic field into the magnetic/electric field, the chirality density of an electromagnetic field can be expressed as \cite{Bliokh2011PRA,Bliokh2014PRL,Cameron2012,Alpeggiani2018}
\begin{equation}
\mathfrak{S}_{{EM}}=\dfrac{1}{2 \omega c_l} \Im(\vb{H}^* \cdot \vb{E}).
\label{eq:Helicity}
\end{equation}
This is the {\it helicity} density, characterizing the difference between the numbers of right-hand and left-hand circularly polarized photons \cite{Afanasiev1996,Trueba1996,Cameron2012,Bliokh2013NJP,Alpeggiani2018,Fernandez-Corbaton2013PRL}. For helicity eigenstates satisfying ${\sqrt{\varepsilon}}\,\vb{E} = i \sigma {\sqrt{\mu}}\, \vb{H}$, $\sigma =\pm 1$ (representing arbitrary superpositions of circularly-polarized plane waves of the same handedness $\sigma$), the helicity density (\ref{eq:Helicity}) equals $\mathfrak{S}_{EM} = \sigma\, U_{EM}/\omega$.

\subsection{{Acoustic fields}}
\label{subsec:acoustic_fields}

Consider now sound waves in a fluid or gas. These waves are described by the vector velocity field $\vb{v}$, characterizing the local motion of the medium molecules, and the scalar pressure field $p$. The acoustic energy density is given by~\cite{Landau_fluid}
\begin{equation}
U_{{A}} = \frac{1}{4} \left( {\rho} |\vb{v}|^2 + {\beta} |p|^2 \right) \equiv \frac{U^{(\vb{v})} + U^{(p)}}{2} \,,
\label{eq:UA}
\end{equation}
where $\rho$ and $\beta$ are the mass density and compressibility of the medium. The kinetic momentum density is described by the acoustic analogue of the Poynting vector \cite{Landau_fluid}: 
\begin{equation}
\vb{\Pi}_{{A}} = \frac{1}{2c_s^2}  \Re (\vb{v}^* p)\,,
\label{eq:PAkin}
\end{equation}
where $c_s = 1/\sqrt{\rho\beta}$ is the speed of sound.

Equations (\ref{eq:UA}) and (\ref{eq:PAkin}), involving the velocity and pressure fields, have forms similar to their electromagnetic counterparts (\ref{eq:UEM}) and (\ref{eq:PEMkin}). 
In turn, the canonical momentum and spin densities in sound waves are determined by the vector velocity field, analogously to Eqs.~(\ref{eq:PEcan}) and (\ref{eq:SE}) \cite{Shi2019, Bliokh2022SA, Bliokh2025CP}: 
\begin{equation}
\vb{P}^{({\bf v})} = \frac{1}{2\omega} \Im \!\left[ \rho\, \vb{v}^*\! \cdot (\grad) \vb{v} \right], 
\label{eq:PAcan}
\end{equation}
\begin{equation}
\vb{S}^{({\bf v})} = \frac{1}{2\omega} \Im \!\left( \rho\, \vb{v}^*\!\times \vb{v} \right). 
\label{eq:SA}
\end{equation}
Using the sound wave equations $i\omega\beta p = \grad\cdot \vb{v}$ and $i\omega \rho \vb{v} = \grad p$, one can verify that the quantities (\ref{eq:PAkin})--(\ref{eq:SA}) satisfy the Belinfante-Rosenfeld relation (\ref{eq:BR}). 

In addition to the general phase-gradient and polarization-ellipticity explanations, Eqs.~(\ref{eq:PAcan}) and (\ref{eq:SA}) have a clear {\it mechanical} interpretation in terms of the microscopic motion of the medium's molecules. Although in a plane sound wave the molecules oscillate along one longitudinal direction, in a generic inhomogeneous wave field they trace {\it elliptical trajectories}, analogous to the polarization ellipses of an optical field. This local motion generates a mechanical angular momentum density given by Eq.~(\ref{eq:SA}) \cite{Jones1973, Shi2019, Bliokh2022SA, Bliokh2025CP}. Moreover, in addition to the linear oscillatory motion, the molecules undergo a slow quadratic {\it Stokes drift} along the local wave-propagation (phase-gradient) direction \cite{Stokes1874, Bremer2018}. The mechanical momentum density corresponding to this drift is precisely the canonical momentum density (\ref{eq:PAcan}) \cite{Bliokh2022SA, Bliokh2022PRA, Bliokh2025CP}.   

Although canonical momentum and spin densities (\ref{eq:PAcan}) and (\ref{eq:SA}) do not explicitly depend on the scalar pressure field $p$, one can alternatively define these quantities with contributions from both the $\vb{v}$ and $p$ fields \cite{Burns2020, Toftul2019PRL}, analogously to the electric and magnetic contributions in Eqs.~(\ref{eq:PEMcan}) and (\ref{eq:SEM}). This is helpful in problems involving couplings between matter and both the velocity and pressure fields \cite{Toftul2019PRL, Bruus2012a, Meng2019JPD, Thomas2017}. Using the sound wave equations, the kinetic momentum density (\ref{eq:PAkin}) can be written in canonical form for the pressure field: 
\begin{equation}
\vb{P}^{(p)} = \frac{1}{2\omega} \Re (\beta\,p^* \grad p) \equiv \vb{\Pi}_{\text{A}}\,,
\label{eq:notation}
\end{equation}
whereas the spin density associated with the pressure field vanishes identically, ${\bf S}^{(p)}\equiv {\bf 0}$, due to the scalar nature of $p$.  
Finally, the chirality of the acoustic velocity field $\vb{v}$ vanishes as well, $\mathfrak{S}_{{A}} \equiv 0$, because this field is curl-less: $\grad \times \vb{v} = \vb{0}$ \cite{Bliokh2019b}.

\vspace{4mm}

For the sake of convenience, the above dynamical properties of monochromatic electromagnetic and acoustic wave fields are summarized in Table~\ref{tab:momenta}.

\section{Principal Wave-Induced Forces and Torques}
\label{sec:forces}

\subsection{General stress-tensor approach}
\label{sec:generalStressTensor}

\begin{figure}
\centering
\includegraphics[width=0.99\linewidth]{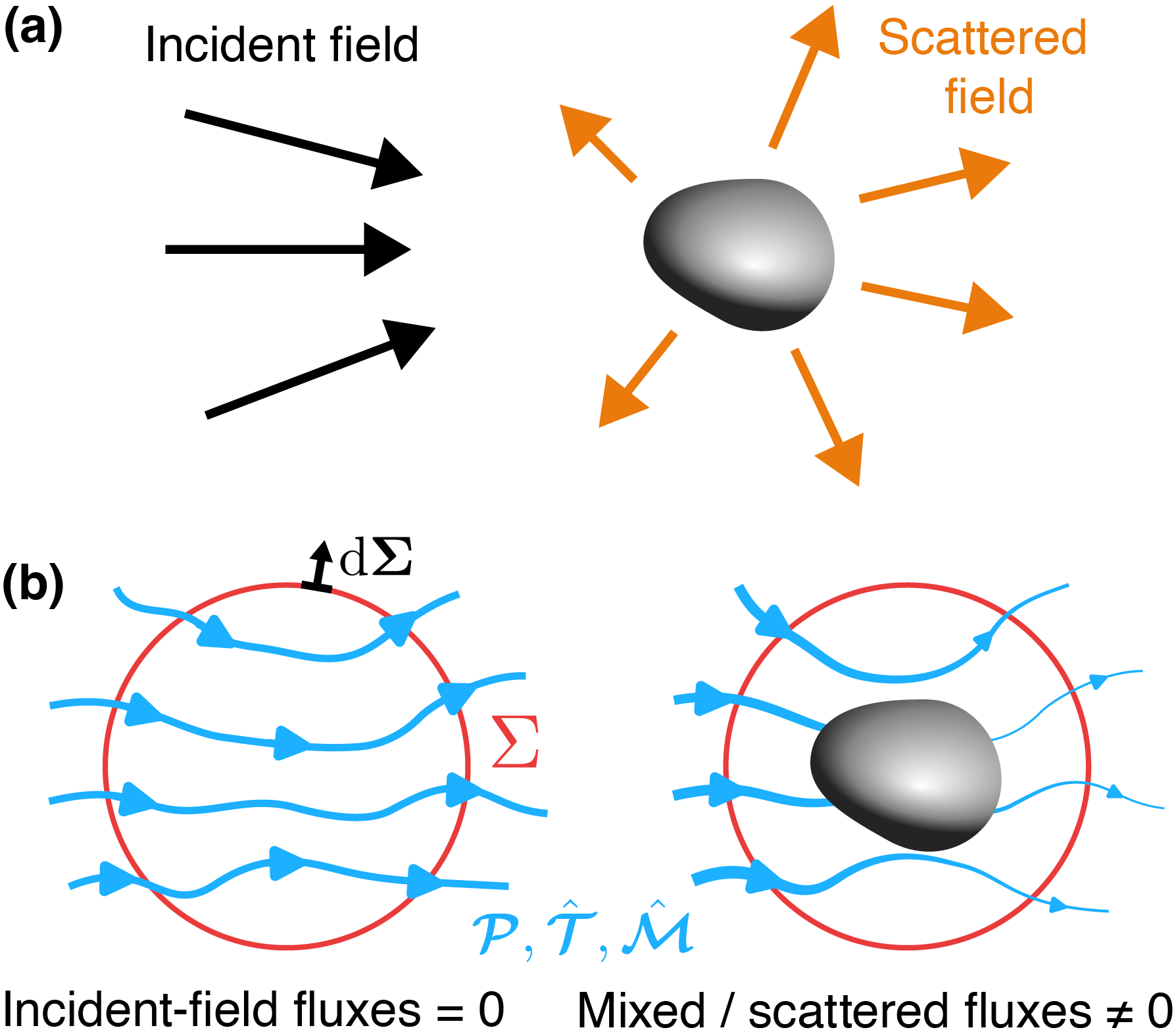}
\caption{(a) Schematic of the scattering problem, showing the incident wave field, the particle, and the resulting scattered field. The total field is a sum of the incident and scattered contributions, Eqs.~(\ref{eq:field-split}). (b) Schematic of the energy, momentum, and angular momentum fluxes, Eqs.~(\ref{eq:Eflux}) and (\ref{eq:stress}), which are quadratic in the wave fields. Separating the pure incident-field, pure scattered-field, and mixed contributions, Eq.~(\ref{eq:flux-split}), the integral `incident' fluxes through a closed surface $\Sigma$ vanish, while the integral `scattered' and `mixed' fluxes are generally nonzero. These latter fluxes determine the absorption rate, force, and torque on the particle, Eqs.~(\ref{eq:force_and_torque_average}).}
\label{fig:flux_simple}
\end{figure}

There are various methods to calculate the radiation forces and torques on matter, such as the Lorentz force approach in optics \cite{Dienerowitz2008} {or analysis based on optomechanical work effects \cite{Rakich2009OE}}. However, the most common method, universal to any wave fields, is the stress-tensor approach. When a wave field propagates in a lossless homogeneous isotropic medium, its total energy, momentum, and angular momentum are conserved. This implies that the {\it fluxes} of these quantities through any closed surface vanish. When the field interacts with a particle (e.g., via absorption or scattering), it can transfer energy, momentum, and angular momentum to the particle. The rates of these transfers per unit time correspond to the wave-induced absorption rate, force, and torque on the particle, respectively. They can be calculated as the corresponding fluxes through a closed surface $\Sigma$ enclosing the particle, Fig.~\ref{fig:flux_simple} \cite{Novotny2012, bohren1984AbsorbtionScatteringLight, Bruus2012a}:
\begin{equation}
A = - \oint \limits_\Sigma {\mathbfcal{P}} \cdot d \vb{\Sigma}, ~~
\vb{F} =  - \oint \limits_\Sigma \ten{\mathcal{T}} \cdot d \vb{\Sigma}, ~~
\vb{T} =  - \oint \limits_\Sigma \ten{\mathcal{M}} \cdot  d \vb{\Sigma}. 
\label{eq:force_and_torque_average}
\end{equation}
Here, ${\mathbfcal{P}}$ is the energy flux density related to the kinetic momentum density as:
\begin{equation}
{\mathbfcal{P}}_{{EM}} = c_l^2\, \vb{\Pi}_{{EM}}\,,\quad
{\mathbfcal{P}}_{{A}} = c_{{s}}^2\, \vb{\Pi}_{{A}}\,,
\label{eq:Eflux}
\end{equation}
$\ten{\mathcal{T}}$ is the momentum flux density or {\it stress tensor} (the flux of a vector quantity is a rank-2 tensor), 
and $\ten{\mathcal{M}} = \vb{r} \times \ten{\mathcal{T}}$ is the angular momentum flux density.
For monochromatic electromagnetic and acoustic wave fields, the {time-averaged} stress tensors take the form \cite{jackson1998ClassicalElectrodynamics, Novotny2012, Bruus2012a, Landau_fluid}:
\begin{align}
\ten{\mathcal{T}}_{{EM}} & = -\frac{1}{2} \Re\!
\left[  \varepsilon \vb{E}^*\!\otimes \vb{E} + \mu \vb{H}^*\!\otimes \vb{H} - \frac{\ten{I}}{2}\! \left( \varepsilon \abs{\vb{E}}^2 + \mu \abs{\vb{H}}^2\right)\right]\!, \nonumber \\
\ten{\mathcal{T}}_{{A}} & =  \frac{1}{2} \Re\! \left[ \rho \vb{v}^*\!\otimes \vb{v} - \frac{\ten{I}}{2}\! \left( \rho \abs{\vb{v}}^2 - \beta \abs{p}^2 \right) \right]\!,
\label{eq:stress}
\end{align}
where $\otimes$ is the dyadic product, $\ten{I}$ is the unit tensor, and we note that the electromagnetic momentum flux density $\ten{\mathcal{T}}_{{EM}}$ equals the Maxwell stress tensor with opposite sign.

Equations (\ref{eq:force_and_torque_average})--(\ref{eq:stress}) describe the most general case of an arbitrary particle in an arbitrary wave field. However, the fields in these equations are superpositions of the fields {\it incident} and {\it scattered} (or radiated) by the particle, Fig.~\ref{fig:flux_simple}: 
\begin{equation}
(\vb{E},\vb{H}) = (\vb{E},\vb{H})_{\rm inc} + (\vb{E},\vb{H})_{\rm sc}, ~~
(\vb{v},p) = (\vb{v},p)_{\rm inc} + (\vb{v},p)_{\rm sc}.
\label{eq:field-split}
\end{equation}
Since the energy, momentum, and angular momentum fluxes are quadratic functions of the fields, each of them can be decomposed into three contributions:
\begin{equation}
\resizebox{\hsize}{!}{$(\mathbfcal{P},\ten{\mathcal{T}},\ten{\mathcal{M}}) = (\mathbfcal{P},\ten{\mathcal{T}},\ten{\mathcal{M}})_{\text{inc}} + (\mathbfcal{P},\ten{\mathcal{T}},\ten{\mathcal{M}})_{\text{mix}} + (\mathbfcal{P},\ten{\mathcal{T}},\ten{\mathcal{M}})_{\text{sc}}$}
\label{eq:flux-split}
\end{equation}
where \qqoute{mix} denotes the cross terms between the incident and scattered fields.
The \qqoute{inc} terms in Eq.~(\ref{eq:flux-split}) make no contribution to the integrals (\ref{eq:force_and_torque_average}) because of the conservation laws in the unperturbed incident field. Therefore, the energy absorption, force, and torque on the particle are determined by the mixed and pure scattered contributions, Fig.~\ref{fig:flux_simple}.

To apply Eqs.~(\ref{eq:force_and_torque_average}), one must first solve the {\it scattering problem} to obtain the scattered fields. In general, this problem can be mathematically challenging, so various approximations are often employed. 
In particular, the following scattering regimes can be distinguished depending on the particle's characteristic size $a$ with respect to the wavelength $\lambda = 2\pi/k$ ($k$ is the wavenumber): 
{\begin{align}   
\label{eq:ka_regimes}
ka \ll 2\pi & \quad\text{---}\quad \text{Rayleigh approximation} \nonumber \\
ka\, \sim\, 2\pi & \quad\text{---}\quad \text{Mie resonant regime}  \\
ka \gg 2\pi & \quad\text{---}\quad \text{ray approximation} \nonumber
\end{align}}
Below we examine various simplified cases in which the scattering problem can be solved analytically. In these situations, the forces and torques can be expressed directly in terms of the {\it incident} fields. 

\subsection{General Rayleigh particles}

\begin{figure}
\centering
\includegraphics[width=1.0\linewidth]{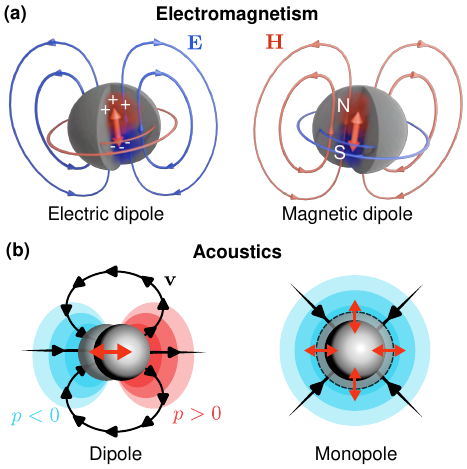}
\caption{The lowest-order multipoles in optics and acoustics. (a) The electric and magnetic dipoles are generated by the separation of electric charges and magnetic poles, respectively. The instantaneous electric and magnetic field distributions are shown. (b) The acoustic dipole and monopole are produced by a linear displacement of the particle and its uniform compression, respectively. The corresponding instantaneous velocity and pressure field distributions are shown. These electromagnetic and acoustic multipoles are characterized by the corresponding dipole and monopole moments $\vb{e}$, $\vb{m}$, $\vb{D}$, and $M$.}
\label{fig:dipoles}
\end{figure}

We start with the case of small (Rayleigh) particles with $ka\ll 2\pi$. In this regime, the scattering is relatively weak, and the scattered field can be approximated by the lowest-order point multipoles. For electromagnetic waves, these are {\it electric and magnetic dipoles} (oscillating electric/magnetic monopoles do not radiate) \cite{Nieto-Vesperinas2010OE,jackson1998ClassicalElectrodynamics,bohren1984AbsorbtionScatteringLight}, whereas for sound waves these are {\it monopole and dipole}  \cite{williams1999FourierAcousticsSound, blackstock2000FundamentalsPhysicalAcoustics, Russell1998}, Fig.~\ref{fig:dipoles}. These scattered fields can be characterized by the corresponding dipole and monopole moments: $(\vb{e},\vb{m})$ in electromagnetism and $(M,\vb{D})$ in acoustics. Here we denote the electric dipole moment by $\vb{e}$, rather than the more conventional $\vb{p}$, for consistency with a unified notation.

By adding the dipole and monopole scattered fields to the incident fields, expanding the incident fields in the first-order Taylor series about the particle's position, and substituting the resulting total fields into Eqs.~(\ref{eq:force_and_torque_average})--(\ref{eq:stress}), one can calculate (see Appendix~\ref{app:tensor_force}) the absorption rate, force, and torque on the particle. The resulting expressions are summarized in Table~\ref{tab:forces}~\cite{Nieto-Vesperinas2010OE,Toftul2019PRL,Smagin2024PRAppl}. From now on, we omit the \qqoute{inc} subscript, and $\{\vb{E},\vb{H},p,\vb{v}\}$ refer to the {\it incident} wave fields evaluated at the position of the particle.

The equations in Table~\ref{tab:forces} are rather general: they only require the particle to be small, with the scattering dominated by the dipole and monopole terms. All other properties of the particle are encoded in its dipole and monopole moments. 
The `mixed' contributions in Table~\ref{tab:forces} correspond to the minimal-coupling interaction between the point dipoles/monopole and the incident field. In turn, the `scattered' contributions can be interpreted as cross- or self-interaction of the multipoles. These `scattered' terms are of a higher order in $ka$ and will be discussed in detail in Section~\ref{subsec:recoil}.

\bgroup
\def\arraystretch{1.7}%
\begin{table*}[]
\caption{Optical and acoustic absorption rates, forces, and torques for the interaction of a generic monochromatic wave field with a generic Rayleigh particle, limited to the lowest-order dipole and monopole contributions in the scattered field. The ``mixed  (extinction)'' terms originate from the `mixed' energy, momentum, and angular momentum fluxes (\ref{eq:flux-split}) involving both the incident and scattered fields. These terms can be interpreted as the interaction between the induced dipole and monopole moments of the particle and the incident wave field. The ``scattered (recoil)'' terms arise from the pure scattered field contributions to the fluxes (\ref{eq:flux-split}); these are quadratic in the induced dipole and monopole moments of the particle. Note that here $\vb{E}$, $\vb{H}$, $\vb{v}$ and $p$ denote the \emph{incident} wave fields, and the factors $g$, $g_{M}$, and  $g_{D}$ are given in Eqs.~\eqref{eq:g}.\label{tab:forces}}
\begin{ruledtabular}
\begin{tabular}{p{0.1\linewidth}p{0.1\linewidth}p{0.65\linewidth}}
\multirow{2}{*}{\textbf{Absorption}} 
& Electromagnetic & \( A_{{EM}} = \underbrace{- {\dfrac{\omega}{2}\Im ( \vb{e}^{*} \cdot\vb{E} )} -  {\dfrac{\omega}{2} \Im ( \vb{m}^* \cdot \vb{H} )}}_{\text{mixed (extinction)}}   \underbrace{-\dfrac{\omega g}{2} \!\left(  \dfrac{1}{\varepsilon}\abs{\vb{e}}^2 +  \dfrac{1}{\mu}\abs{\vb{m}}^2 \right)}_{\text{scattered (recoil)}} \) \\[3em]
& Acoustic  & \( A_A = \underbrace{-  \dfrac{\omega}{2} \Im  ( \vb{D}^* \cdot \vb{v} ) -\dfrac{\omega}{2} \Im (  M^* p ) }_{\text{mixed (extinction)}}   \underbrace{- \dfrac{\omega}{2} \!\left( \dfrac{g_D}{\rho}\abs{\vb{D}}^2  + \dfrac{g_M}{\beta}\abs{M}^2 \right)}_{\text{scattered (recoil)}} \) \\ \hline
\multirow{2}{*}{\textbf{Force}}  
& Electromagnetic
&  \(\vb{F}_{EM} = \underbrace{{\dfrac{1}{2}\Re\!\left[ \vb{e}^{*} \berrydot \vb{E} \right]} + {\dfrac{1}{2} \Re\! \left[ \vb{m}^* \berrydot \vb{H}\right]}}_{\text{mixed}} \underbrace{-\frac{\omega g}{2} \Re( \vb{e}^* \!\times \vb{m})}_{{\text{scattered (recoil)}}} \white{\frac{\dfrac{1}{1}}{1}} \) \\[3em]
& Acoustic  
& \( \vb{F}_A = \underbrace{{ \dfrac{1}{2} \Re\! \left[ \vb{D}^* \cdot (\grad) \vb{v} \right]} + { \dfrac{1}{2} \Re\! \left[  M^* \grad p \right]} }_{\text{mixed}}   \underbrace{-\frac{\omega g}{2} \Re ( M^* \vb{D} )}_{{\text{scattered (recoil)}}}  \)  \\ \hline
\multirow{2}{*}{\textbf{Torque}} 
& Electromagnetic     
& \( \vb{T}_{EM} =  \underbrace{{\dfrac{1}{2} \Re ( \vb{e}^* \!\times \vb{E} ) + \dfrac{1}{2} \Re ( \vb{m}^* \!\times \vb{H} )}}_{\text{mixed}} \underbrace{- \frac{g}{2}\left[ \dfrac{1}{\varepsilon }\Im (\vb{e}^* \!\times \vb{e})  + \dfrac{1}{\mu} \Im (\vb{m}^* \!\times \vb{m})\right]}_{{\text{scattered (recoil)}}} \white{\frac{\dfrac{1}{1}}{1}} \)  \\[3em]
& Acoustic & \(\vb{T}_A = \underbrace{{\dfrac{1}{2} \Re ( \vb{D}^* \!\times \vb{v} )}}_{\text{mixed}} \underbrace{- \frac{g_D}{2 \rho} \Im ( \vb{D}^*\! \times \vb{D} )}_{{\text{scattered (recoil)}}}\)  
\end{tabular}
\end{ruledtabular}
\end{table*}
\egroup

\subsection{Isotropic Rayleigh particles}
\label{subsec:Isotropic_Rayleigh}

Consider now the simplest case of {\it isotropic} Rayleigh particles, whose dipole and monopole moments, induced by the incident wave, are directly proportional to the corresponding incident fields \cite{jackson1998ClassicalElectrodynamics, Nieto-Vesperinas2010OE, Toftul2019PRL, Jordaan2018}:
\begin{align}
\label{eq:polarizabilities}
\begin{pmatrix}
{\vb{e}}/{\sqrt{\varepsilon}} \\  
\vb{m}/\sqrt{\mu}
\end{pmatrix} &= 
\begin{pmatrix}
{\alpha}_{{e}} & 0 \\ 
0  & {\alpha}_{{m}} 
\end{pmatrix}  
\begin{pmatrix}
\sqrt{\varepsilon}\,\vb{E} \\  
\sqrt{\mu}\,\vb{H}
\end{pmatrix}\!, \nonumber \\[5pt] 
\begin{pmatrix}
\vb{D}/{\sqrt{\rho}} \\  
M/{\sqrt{\beta}}
\end{pmatrix} &= 
\begin{pmatrix}
{\alpha}_D & 0 \\ 
0  & {\alpha}_M 
\end{pmatrix}  
\begin{pmatrix}
\sqrt{\rho}\, \vb{v} \\  
\sqrt{\beta}\, p
\end{pmatrix}\!.
\end{align}
Here, the complex-valued quantities $\alpha_{\{e,m,M,D\}}$ are the corresponding {\it polarizabilities} of the particle, which are defined such that all have the dimension of volume: $[\alpha_{\{e,m,M,D\}}]=\text{m}^3$.
We write relations (\ref{eq:polarizabilities}) in the general form using {\it polarizability matrices}, which are diagonal in this case. 
Since there are different conventions for defining electromagnetic and acoustic dipole and monopole moments, we present the fields produced by the moments adopted in this review in Appendix~\ref{app:multipoles}.

The polarizabilities can be connected with the \textit{scattering}, \textit{absorption}, and \textit{extinction cross-sections}, related as $\sigma^{\text{ext}} = \sigma^{\text{sc}} + \sigma^{\text{abs}}$. In particular, for all polarizabilities in Eqs.~\eqref{eq:polarizabilities}, we have:
\begin{align}
\label{eq:cross-sections}
\sigma^{\text{ext}}_{e,m} &= k \Im (\alpha_{e,m})\,, \quad
\sigma^{\text{sc}}_{e,m} = k\, g\, 
\abs{\alpha_{e,m}}^2\,, \nonumber \\
\sigma^{\text{ext}}_{D,M} &= k \Im (\alpha_{D,M}),~~
\sigma^{\text{sc}}_{D,M} = k\,g_{D,M}\,
\abs{\alpha_{D,M}}^2,
\end{align}
where
\begin{equation}
g =\frac{k^3}{6\pi}\,,\qquad
g_D =\frac{k^3}{12\pi}\,, \qquad
g_M =\frac{k^3}{4\pi}\,.
\label{eq:g}
\end{equation}
These relations follow from the direct calculation of the `mixed' and `scattered' energy fluxes through a closed surface around the particle, normalized to the absolute value of the `incident' energy flux density, assuming an incident plane-wave field \cite{Tretyakov2014Aug, LeRu2013, Toftul2019PRL, bohren1984AbsorbtionScatteringLight}.
For particles without gain, $\sigma^{\text{abs}} \ge 0$ and $\sigma^{\text{ext}} \ge \sigma^{\text{sc}}$, where equality holds in the lossless case. 
These relations impose constraints on the polarizabilities, sometimes referred to as the {\it optical theorem} \cite{Novotny2012, Belov2003TPL, Quan2018PRL}. Note also that the absorption cross-section equals the normalized absorption rate: $\sigma^{\text{abs}} = A/|\mathbfcal{P}|$, in the case of an incident plane wave.

Explicit formulas for the isotropic Rayleigh-particle polarizabilities in terms of the material parameters can be derived from the electromagnetic Mie theory and its acoustic analogue. The results are summarized in Appendix~\ref{app:polarizabilities} and Table~\ref{tab:polarizabilities}. 

\begin{figure}
\centering
\includegraphics[width=1.0\linewidth]{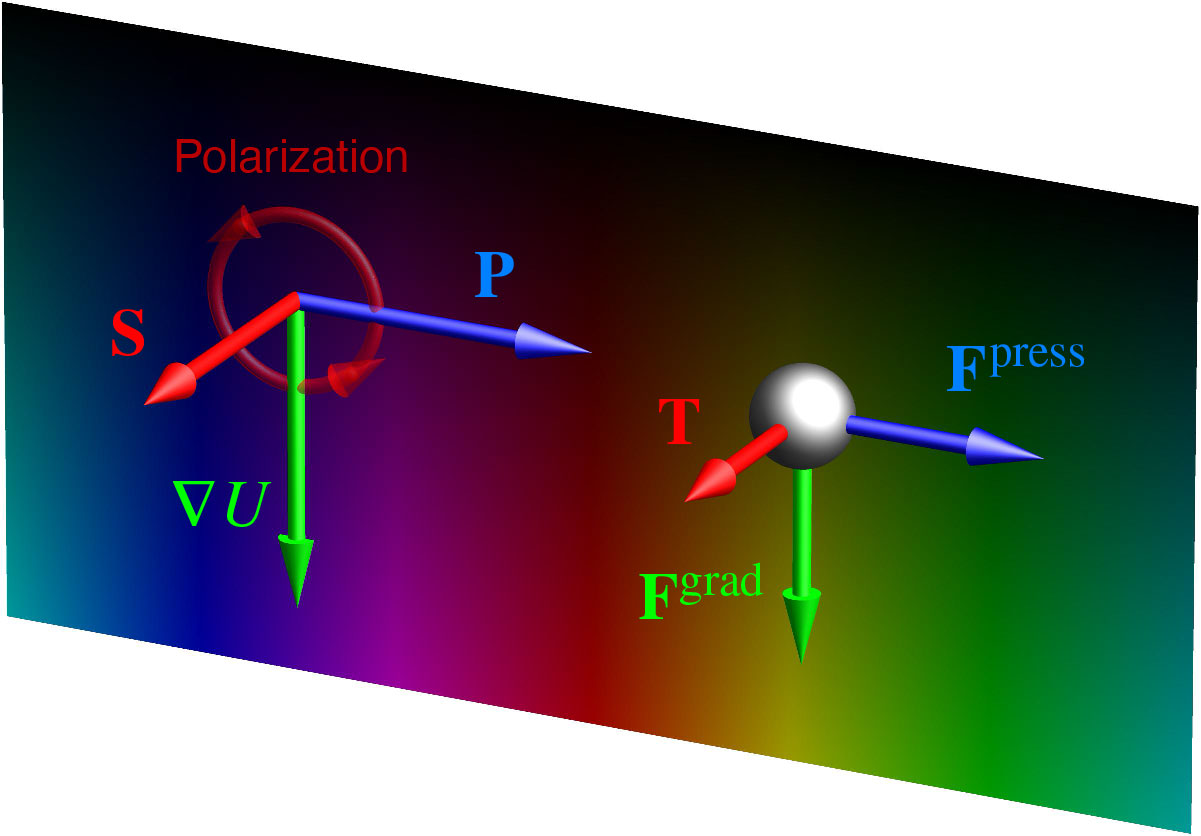}
\caption{{Schematics of the main radiation-pressure force ${\bf F}^{\rm press}$, gradient force ${\bf F}^{\rm grad}$, and torque ${\bf T}$ on an isotropic Rayleigh particle, Eqs.~(\ref{eq:F}) and (\ref{eq:T}). These are produced by the canonical momentum (phase gradient) ${\bf P}$, the energy density (intensity) gradient $\bm{\nabla}U$, and the spin density (elliptical polarization) ${\bf S}$ of the incident field, respectively.} The background colors and their brightness represent the phase and intensity of the wave field. In a generic structured wave field, the vectors in these triads, which are shown here as orthogonal, can have arbitrary directions.}
\label{fig:concept_complex}
\end{figure}

Substituting Eqs.~(\ref{eq:polarizabilities}) into the `mixed' terms (linear in the dipole and monopole moments) in Table~\ref{tab:forces}, we obtain the absorption rates, forces, and torques on the particle conveniently expressed via the energy, canonical momentum, and spin densities of the incident wave field, Eqs.~(\ref{eq:UEM}), (\ref{eq:PEMcan}), (\ref{eq:SEM}), (\ref{eq:UA}), (\ref{eq:PAcan}), (\ref{eq:SA}), and (\ref{eq:notation}) {(summarized in Table~\ref{tab:momenta})}:
\begin{align}
A_{EM}^{\rm mix} &= \omega \!\left[ \Im({\alpha_e})U^{(\vb{E})} + \Im({\alpha_m})U^{(\vb{H})} \right], \nonumber \\
A_{A}^{\rm mix} &= \omega\! \left[ \Im({\alpha_D})U^{(\vb{v})} + \Im({\alpha_M})U^{(p)} \right], 
\label{eq:A} 
\end{align}
\begin{align}
\vb{F}_{EM}^{\rm mix} &= \frac{1}{2}\left[\Re(\alpha_e) \grad U^{(\vb{E})} + \Re(\alpha_m) \grad U^{(\vb{H})} \right] \nonumber \\
&+ \omega\! \left[ \Im({\alpha_e})\vb{P}^{(\vb{E})} + \Im({\alpha_m})\vb{P}^{(\vb{H})} \right] 
\equiv \vb{F}_{EM}^{\text{grad}} + \vb{F}_{EM}^{\text{press}}, \nonumber \\
\vb{F}_{A}^{\rm mix} &= \frac{1}{2}\left[\Re(\alpha_D) \grad U^{(\vb{v})} + \Re(\alpha_M) \grad U^{(p)} \right] \nonumber \\
&+ \omega\! \left[ \Im({\alpha_D})\vb{P}^{(\vb{v})} + \Im({\alpha_M})\vb{P}^{(p)} \right] 
\equiv \vb{F}_{A}^{\text{grad}} + \vb{F}_{A}^{\text{press}},
\label{eq:F} 
\end{align}
\begin{align}
\vb{T}_{EM}^{\rm mix} &= \omega\! \left[ \Im({\alpha_e})\vb{S}^{(\vb{E})} + \Im({\alpha_m})\vb{S}^{(\vb{H})} \right], \nonumber \\
\vb{T}_{A}^{\rm mix} &= \omega  \Im({\alpha_D})\vb{S}^{(\vb{v})} .
\label{eq:T}
\end{align}
These are the most important optical and acoustic forces acting on small particles, schematically illustrated in Fig.~\ref{fig:concept_complex}. The remarkable symmetry between optical and acoustic quantities justifies the unified framework employed here. 
In Eqs.~(\ref{eq:F}), one can distinguish between the {\it gradient forces} ${\bf F}^{\rm grad}$, determined by gradients of the corresponding energy-density terms, and the {\it radiation-pressure forces} ${\bf F}^{\rm press}$, governed by the canonical momentum density terms. Notably, in paraxial optical fields, the energy density gradients are nearly orthogonal to the canonical momentum (propagation) direction, causing the gradient and radiation-pressure forces to act in nearly perpendicular directions. The torques (\ref{eq:T}) are naturally determined by the corresponding spin densities. Importantly, Eqs.~(\ref{eq:A})--(\ref{eq:T}) are valid for an arbitrarily inhomogeneous (structured) incident field, including near-fields and evanescent waves. 

The forces and torques (\ref{eq:F}) and (\ref{eq:T}) underpin the optical and acoustic manipulation of small particles, enabling trapping, tweezers, spanners, levitation, and related techniques. 
\cite{Dienerowitz2008, Molloy2002CP, Grier2003N, Bowman2013RPP, Ashkin2000IEEE, Woerdemann2013LPR, Dholakia2011NP, Sukhov2017RPP, Gao2017LSA, Yuan2020JPD, Volpe2023JPP, Gieseler2021AOP, Pesce2020EPJP, Polimeno2018JQSRT, Marago2013NN, Zemanek2019AOP, Yang2021AP, Bruus2012a, Meng2019JPD, Ozcelik2018NM, Thomas2017, Baudoin2020AR, Lim2024RPP, Friend2011RMP, Lenshof2012LC, Ding2013LC, Hossein2023BR, Wu2019MN, Dholakia2020NRP, Laurell2007CSR, Gonzalez-Ballestero2021S, Bustamante2021NR, Zhang2021LSA, Shi2022APR}
Pioneering studies of the dipole radiation forces were performed by Gordon and Ashkin in 1980 \cite{Gordon1980PRA,Ashkin1983OL} (considering interaction of an electric-dipole particle with an incident electromagnetic wave) and by Gor'kov in 1961 for the acoustic gradient force \cite{Gorkov1962}. Since then, numerous works have extended these results \cite{Chaumet2000OL, Nieto-Vesperinas2010OE, Ruffner2012PRL, Dienerowitz2008, Sukhov2017RPP, Bekshaev2013JO, Bliokh2014NC, Bliokh2015PR, Bruus2012a, Toftul2019PRL,  Silva2014JASA, Shi2019}, until they acquired the modern form (\ref{eq:F}). Surprisingly, Eqs.~(\ref{eq:T}) for optical and acoustic torques were derived only relatively recently \cite{Chaumet2009OE, Canaguier-Durand2013PRA, Bliokh2014NC, Bliokh2015PR, Nieto-Vesperinas2015OL, Silva2014JASA, Toftul2019PRL}.

At lowest order, the imaginary parts of the polarizabilities in Eqs.~(\ref{eq:A})--(\ref{eq:T}) are commonly associated with absorption in the particle. However, according to Eqs.~(\ref{eq:cross-sections}), they are in fact determined by the {\it extinction} cross-sections: $\Im(\alpha) = k^{-1} \sigma^{\text{ext}} = k^{-1}( \sigma^{\text{abs}} + \sigma^{\text{sc}})$ which include both absorption \emph{and scattering}. This results in nonzero imaginary parts of the polarizabilities even for lossless particles with $\sigma^{\rm abs} =0$: ${\rm Im}(\alpha) = k^{-1} \sigma^{\text{sc}} \propto |\alpha|^2$, Eqs.~(\ref{eq:cross-sections}) (see Appendix~\ref{app:polarizabilities}). 
In particular, this means that the radiation-pressure forces in Eqs.~(\ref{eq:F}) do not vanish for such lossless particles \cite{Harada1996OC}. This phenomenon is sometimes termed ``radiation friction'' \cite{Sipe1974PRA, Nieto-Vesperinas2010OE, Simpson2010JOSA, Albaladejo2010OE, LeRu2013, Toftul2019PRL}. 
In contrast, the absorption rates and torques on lossless isotropic particles do vanish, but demonstrating this rigorously requires taking into account the contributions from the `scattered' fluxes, which are analyzed in the next Section~\ref{subsec:recoil}.

It is worth noting that some studies of optical forces \cite{Albaladejo2009PRL, Marago2013NN, Jones_book, Gao2017LSA, Wang2014NC, Nieto-Vesperinas2010OE, Sukhov2017RPP} use an alternative yet equivalent form of the electromagnetic radiation-pressure force ${\bf F}_{EM}^{\rm press}$ in Eq.~(\ref{eq:F}). This form is obtained by expressing the canonical momentum densities via the difference of the kinetic Poynting momentum (\ref{eq:PEMkin}) and curls of the corresponding spin densities, according to the Belinfante-Rosenfeld relation (\ref{eq:BR}): ${\bf P}^{({\bf E},{\bf H})} = {\vb{\Pi}}_{EM} - (1/2) \grad \times {\bf S}^{({\bf E},{\bf H})}$. However, this  decomposition, meant to introduce a term proportional to the Poynting vector and an additional ``spin-curl'' force, can be delusive. Indeed, the canonical momentum density has a clear physical interpretation as a natural measure of the mean local phase gradient (weighted by the field intensity) \cite{Berry2009, Berry2013, Bliokh2014NC, Bliokh2015PR}, universal for any wave field \cite{Bliokh2022SA} and independent of polarization in paraxial uniformly-polarized fields. 
In contrast, decomposing the force into Poynting-vector and spin-curl contributions introduces two polarization-dependent force terms even in simple paraxial fields, which partially cancel each other, resulting ultimately in the same radiation-pressure force. 
 
\subsection{Recoil force and torque}
\label{subsec:recoil}

\begin{figure}
\centering
\includegraphics[width=0.95\linewidth]{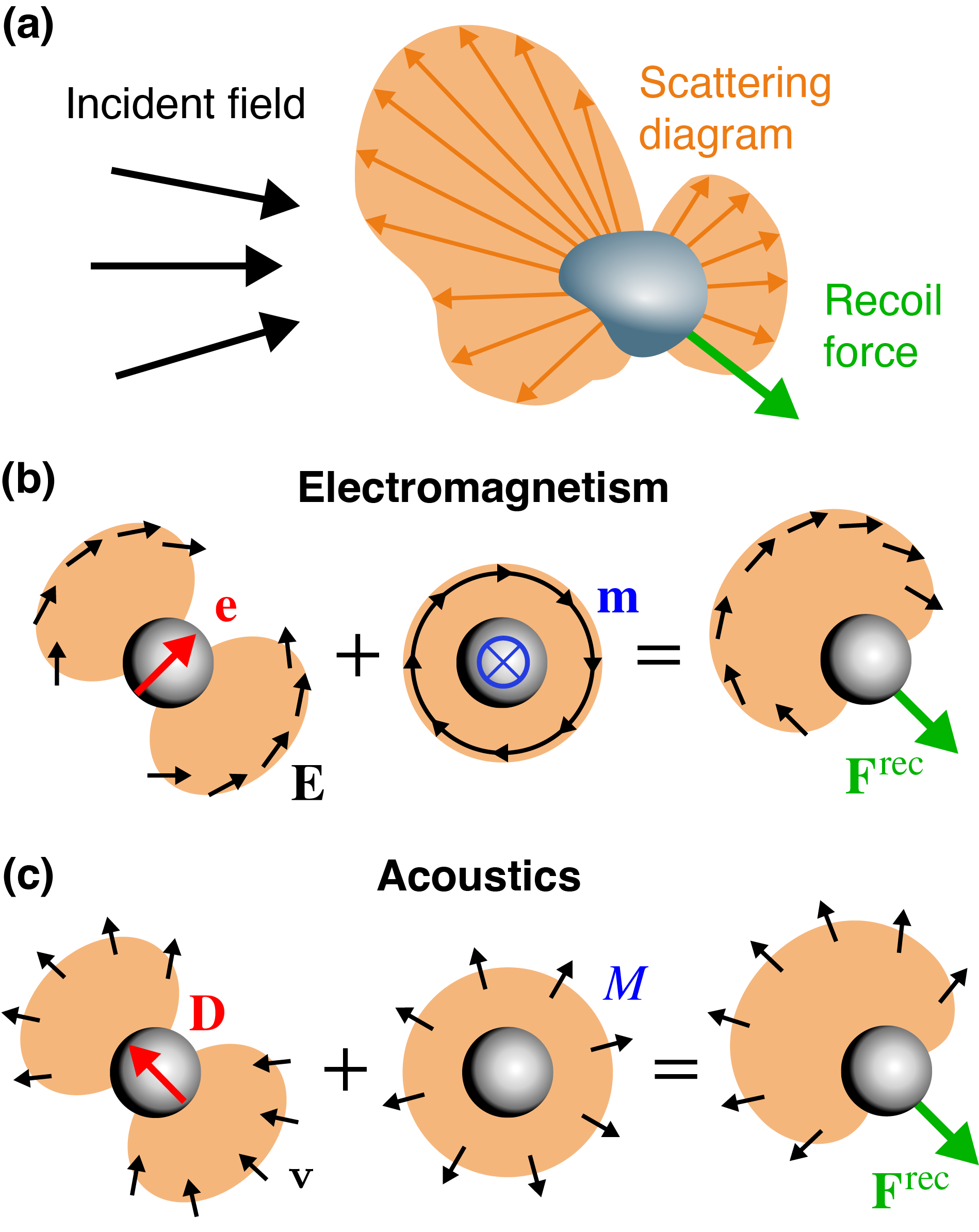}
\caption{(a) Schematics of the recoil force produced by the pure scattered-field contribution to the momentum flux. An asymmetric scattering diagram generates a recoil force directed opposite to the mean scattering direction. Recoil contributions to the torque and absorption rate are defined similarly using the `scattered' angular-momentum and energy fluxes. (b) Electromagnetic recoil force arising from the interference of the electric and magnetic dipole contributions in the scattered field, Table~\ref{tab:forces} and Eqs.~(\ref{eq:F_recoil}). Here the corresponding instantaneous electric fields (black arrows) and the scattering diagram are shown. (c) Acoustic recoil force arising from the interference of the dipole and monopole contributions in the scattered field, Table~\ref{tab:forces} and Eqs.~(\ref{eq:F_recoil}). The instantaneous vector velocity field and the scattering diagram are shown.}
\label{fig:recoil}
\end{figure}

We now consider the second-order forces and torques arising from the pure scattering-field fluxes, which are quadratic in the dipole and monopole moments. These contributions are labeled in Table~\ref{tab:forces} as ``scattered (recoil)''. Figure~\ref{fig:recoil} schematically illustrates the nature of the recoil force. It originates from the asymmetry of the scattered momentum flux, which in turn stems from the interference of the symmetric electric and magnetic dipole or acoustic dipole and monopole contributions \cite{Nieto-Vesperinas2010OE, Smagin2024PRAppl, Wei2020NJP} (see also Appendix~\ref{app:recoil_and_radiation_diagram}).   

Substituting the dipole and monopole moments (\ref{eq:polarizabilities}) into the `scattered' terms in Table~\ref{tab:forces}, we obtain the recoil forces:
\begin{align}
&\vb{F}^{\text{rec}}_{EM} \!=\! - {\omega g}\! \left[ \Re(\alpha_e \alpha^{*}_m) \Re\tilde{\vb{\Pi}}_{EM} + \Im(\alpha_e \alpha^{*}_m) \Im\tilde{\vb{\Pi}}_{EM} \right]\!, \nonumber \\
&\vb{F}^{\text{rec}}_{A} \! = \! - {\omega g}\! \left[ \Re(\alpha_D \alpha^{*}_M) \Re\tilde{\vb{\Pi}}_{A} + \Im(\alpha_D \alpha^{*}_M) \Im\tilde{\vb{\Pi}}_{A} \right]\!,
\label{eq:F_recoil}
\end{align}
where $\tilde{\vb{\Pi}}_{EM} = (\vb{E}^*\! \times \vb{H})/2c_l^2$ and $\tilde{\vb{\Pi}}_{A} = (\vb{v}^* p )/2c_s^2$ are the natural complex extensions of the electromagnetic and acoustic kinetic momentum densities (\ref{eq:PEMkin}) and (\ref{eq:PAkin}).

The recoil force (\ref{eq:F_recoil}) has been studied in optics since 2009 
\cite{Chaumet2009OE, Nieto-Vesperinas2010OE, Bekshaev2013JO,Bliokh2014NC,Bekshaev2015PRX, Jones_book, Antognozzi2016NP, Gao2017LSA}, and only recently it was derived in acoustics \cite{Smagin2024PRAppl}.
Remarkably, a special case of {\it negative} recoil force, antiparallel to the radiation-pressure force, provides one of the main mechanisms of {\it pulling} forces and tractor beams in optics \cite{Brzobohaty2013NP} and acoustics \cite{Demore2014PRL, Marston2006JASA} (see Section \ref{sec:pulling}).

In terms of the particle's dipole moments, the optical recoil force (\ref{eq:F_recoil}) is maximized when the electric $\mathbf{e}$ and magnetic $\mathbf{m}$ dipole moments are orthogonal, in phase, and have amplitudes satisfying $|\mathbf{m}| = |\mathbf{e}|\sqrt{\mu/\varepsilon}$. These relations are known as the {\it Kerker conditions} \cite{Liu2018OE, Kerker1982JOSA}, and such an electromagnetic source is termed a {\it Huygens source} \cite{Milligan2005}. Under these conditions, the far-field radiation patterns of the two dipoles interfere constructively in the direction of $\mathbf{e} \times \mathbf{m}$ and destructively in the opposite direction, Fig.~\ref{fig:recoil}(b).
Most commonly, the Rayleigh-particle polarizabilities can be designed such that, in an incident plane-wave field with a wave-vector $\mathbf{k}$, the Kerker condition is achieved with maximal forward scattering, $(\mathbf{e} \times \mathbf{m}) \parallel {\bf k}$ \cite{Kerker1982JOSA}, thereby producing a backward pulling force \cite{Chen2011NP}. Conversely, the anti-Kerker condition, $(\mathbf{e} \times \mathbf{m}) \parallel -{\bf k}$, yields maximal back-scattering \cite{Alu2010JN, Geffrin2012NC}. 

An acoustic analogue of the Kerker condition occurs when the acoustic monopole and dipole moments are in phase and have amplitudes satisfying $|M|=|\mathbf{D}|\sqrt{\beta/\rho}$ \cite{Wei2020NJP, Long2020NSR, Wu2021ApplPhysExpress}. Their radiation diagrams interfere constructively in the direction of $\mathbf{D}$, and destructively in the opposite direction, Fig.~\ref{fig:recoil}(c), enhancing the acoustic recoil force. Similar interference effects can also happen for higher-order multipoles, often referred to as the {generalized Kerker conditions} (see Section~\ref{subsec:multipoles}).

From the perspective of the incident optical field, the real and imaginary parts of the complex Poynting momentum density $\tilde{\vb{\Pi}}_{EM}$ in Eq.~(\ref{eq:F_recoil}) are generally independent of the canonical momentum densities and the energy density gradients in structured fields. Therefore, the recoil force can have a nonzero component in a `{\it third}' direction, orthogonal to both the radiation and gradient forces (\ref{eq:F}) \cite{Bliokh2014NC, Bekshaev2015PRX, Antognozzi2016NP, Liu2018PRL, Shi2022SA, Xu2019PRL, Zhou2022PNAS, Chen2020PRL, Shi2023AOP}. In particular, using the Belinfante-Rosenfeld relation (\ref{eq:BR}) for $\Re \tilde{\vb{\Pi}}_{EM} = {\vb{\Pi}}_{EM}$, one can see that this quantity differs from the canonical momentum density by a curl of the spin density. Thus, in contrast to the artificially-introduced `spin-curl' term in the radiation-pressure force \cite{Albaladejo2009PRL}, the recoil force does contain a polarization-dependent spin-curl component \cite{Bliokh2014NC, Bekshaev2015PRX, Antognozzi2016NP, Liu2018PRL, Angelsky2012OE, Angelsky2012OE_II, Shi2022SA}. {Notably, a transverse spin-dependent recoil force can also arise in a simple plane-wave field interacting with a specially designed anisotropic object side-scattering different spin components in opposite directions \cite{Magallanes2018NP}.}

In acoustics, using the relation $i\omega\rho \vb{v} = \grad p$, the recoil force (\ref{eq:F_recoil}) is determined by the quantities $\Re \tilde{\vb{\Pi}}_A = \vb{P}^{(p)}$ and $\Im \tilde{\vb{\Pi}}_A = - (1/2\omega) \grad U^{(p)}$. Therefore, these recoil-force terms are always aligned with the corresponding gradient and radiation-pressure terms in Eqs.~(\ref{eq:F}), and cannot produce a force in the `third' direction.

Calculating now the recoil torque terms from Eqs.~(\ref{eq:polarizabilities}) and Table~\ref{tab:forces}, we obtain:
\begin{align}
\begin{split}
\vb{T}^{\text{rec}}_{EM} &= - {\omega\, g} \left( |\alpha_e|^2 {\vb{S}}^{(\vb{E})} + |\alpha_m|^2 {\vb{S}}^{(\vb{H})} \right)\!,  \\
\vb{T}^{\text{rec}}_{A} &= - {\omega\, g_D\,} |\alpha_D|^2 {\vb{S}}^{(\vb{v})}\,,
\end{split}
\label{eq:T_recoil}
\end{align}
Thus, the recoil torques are also associated with the corresponding spin densities of the incident wave field. Furthermore, from Eqs.~(\ref{eq:cross-sections}), one can see that the proportionality coefficients are determined by the corresponding scattering cross-sections: $-(\omega/k)\, \sigma^{\rm sc}$. Adding the recoil torques (\ref{eq:T_recoil}) to Eqs.~(\ref{eq:T}), where the proportionality coefficients are $(\omega/k)\, \sigma^{\rm ext}$, and using $\sigma^{\rm ext} = \sigma^{\rm abs} + \sigma^{\rm sc}$, we find the total torque:
\begin{align}
\begin{split}
\vb{T}^{\text{abs}}_{EM} &= c_l \left( \sigma^{\text{abs}}_e\, \vb{S}^{(\vb{E})} + \sigma^{\text{abs}}_m\, \vb{S}^{(\vb{H})}\right) ,  \\
\vb{T}^{\text{abs}}_{A} &= c_s\, \sigma^{\text{abs}}_D\, \vb{S}^{(\vb{v})} .
\end{split}
\label{eq:T_tot}
\end{align}
These important results show that the optical and acoustic torques on isotropic particles are determined by the corresponding spin densities and the {\it absorption} cross-sections \cite{Marston1984PRA, Nieto-Vesperinas2015OL, Nieto-Vesperinas2015PRA, Toftul2019PRL, Smagin2024PRAppl}. Consequently, the torque vanishes identically for isotropic lossless particles, such as dielectric particles in an optical field, and this remains true for particles of {\it arbitrary size} \cite{Marston1984PRA, Zhang2011PRE_II}. 
This contrasts with the radiation-pressure forces in Eqs.~(\ref{eq:F}), which are proportional to the extinction cross-sections and do not vanish even for lossless isotropic particles.

Finally, adding the recoil terms in the absorption rates from Table~\ref{tab:forces} to the `mixed terms' \eqref{eq:A}, we find that the total absorption rates are also determined by the absorption cross-sections: 
\begin{align}
\begin{split}
A^{\text{abs}}_{EM} &= c_l \left( \sigma^{\text{abs}}_e\, U^{(\vb{E})} + \sigma^{\text{abs}}_m\, U^{(\vb{H})} \right) ,  \\
A^{\text{abs}}_{A} &= c_s \left( \sigma^{\text{abs}}_D\, U^{(\vb{v})} + 
\sigma^{\text{abs}}_M\, U^{(p)} \right) .
\end{split}
\label{eq:A_tot}
\end{align}
A seeming paradox arises from these expressions. According to the general Eqs.~(\ref{eq:force_and_torque_average}), the absorption rate quantifies the energy transfer from the field to the particle. For a lossless isotropic particle, the absorption rates (\ref{eq:A_tot}) vanish. However, the radiation-pressure forces in Eq.~(\ref{eq:F}) (proportional to the extinction cross-section $\sigma^{\rm ext}$), as well as the gradient and recoil forces in Eqs.~(\ref{eq:F}) and (\ref{eq:F_recoil}) do not vanish in this case. These forces can accelerate the particle and increase its kinetic energy. 
The resolution of this paradox is that the above derivations assume perfectly {\it monochromatic} incident and scattered fields, neglecting Doppler effects due to the finite velocity of the particle. Under this assumption, Eqs.~(\ref{eq:A_tot}) describe only the thermal energy absorbed by the particle, whereas changes in its kinetic energy require accounting for non-monochromaticity of the scattered field.

\section{Complex Forces and Torques}
\label{sec:complex_forces_and_torques}

We are now in a position to consider more sophisticated optical and acoustic interactions with particles, beyond the isotropic Rayleigh case. 

\begin{figure*}
\centering
\includegraphics[width=0.8\linewidth]{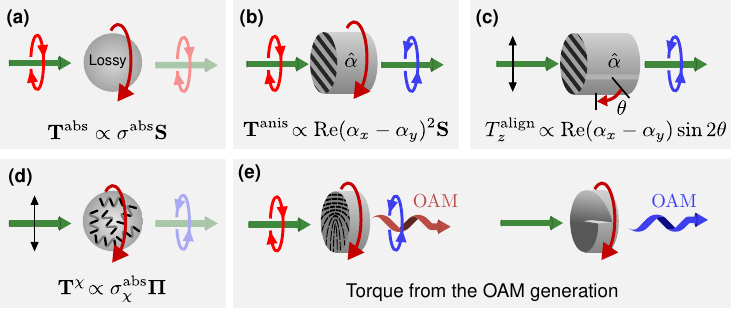}
\caption{Schematics of different types of optical and acoustic torques. (a) Torques (\ref{eq:T}) and (\ref{eq:T_tot}) induced by absorption of the spin angular momentum of the incident wave by an isotropic lossy particle. (b,c) Torques (\ref{eq:T_aniz}) and (\ref{eq:T_align}) resulting from changes in the spin angular momentum of a wave interacting with an anisotropic particle (e.g., a wave plate). (d) Torque (\ref{eq:T_chiral}) produced by helicity-dependent absorption (chiral dichroism) of an isotropic chiral particle. (e) Torques on structured wave elements generating orbital angular momentum (OAM) in the scattered (e.g., transmitted or reflected) wave, such as q-plates (left) or spiral-phase plates (right) \cite{Hakobyan2014NP, Wunenburger2015NJP, Magallanes2018NP}.
}
\label{fig:types_of_torque}
\end{figure*}

\subsection{Anisotropic particles}
\label{subsec:anisotropic_particles}

 We first examine {\it anisotropic} Rayleigh particles. Anisotropy can originate either from a non-spherical shape of the particle or from the intrinsic anisotropy of its material (described by tensor parameters $\hat{\varepsilon}$, $\hat{\mu}$, $\hat{\rho}$, and $\hat{\beta}$). When the particle is small, {$ka \ll 2\pi$}, its interaction with the wave field can still be described in terms of the dipole and monopole moments, but the dipole polarizabilities (\ref{eq:polarizabilities}) now become $3\times 3$ {\it tensors}, which capture the anisotropic response of the particle: 
\begin{equation}
\alpha_{e,m,D} \to \ten{\alpha}_{e,m,D}\,.
\label{eq:tensors}
\end{equation}
As a consequence, the induced dipole moments are generally not aligned with the local incident fields. Importantly, all equations in Table~\ref{tab:forces} remain valid, but with the tensor polarizabilities they no longer reduce to Eqs.~(\ref{eq:A})--(\ref{eq:A_tot}). Hence, the resulting forces and torques cannot be expressed via the basic dynamical properties of the incident field. Thus, for anisotropic particles, the gradient forces, radiation forces, and torques are generally not aligned with the energy-density gradients, canonical momenta, and spin densities.  

Since generic dipole-polarizability tensors carry many degrees of freedom, one has to analyze each case of an anisotropic particle in a given incident field individually. Even the simplest uniaxial anisotropy (e.g., spheroidal particles) in a uniform plane wave can produce qualitatively new types of forces and torques, enabling new functionalities \cite{Smagin2024PRAppl,Nan2023NC}.

Some general remarks can also be made regarding torques on anisotropic particles. Recall that the transmission of a polarized electromagnetic wave through an anisotropic plate underpinned the pioneering studies of optical spin angular momentum and torque \cite{Sadowsky1899,Poynting1909, Beth1935, Beth1936, Holbourn1936, Brasselet2023AP} (see Section~\ref{sec:historical}). The main effect of an anisotropic material is that it modifies the polarization of the scattered (e.g., transmitted) light. Since the spin angular momentum of the field (\ref{eq:SEM}) or (\ref{eq:SA}) is determined by the degree of circular polarization (Fig.~\ref{fig:momentum-spin}), changes in the polarization of the scattered field affect the angular momentum balance and thereby generate torques. 
In contrast to the case of isotropic particles, where torques (\ref{eq:T}), (\ref{eq:T_recoil}), and (\ref{eq:T_tot}) arise only from {\it absorption} of the spin angular momentum, torques on anisotropic particle can exist without absorption, due to {\it changes} in the scattered-field spin. Within this mechanism, two distinct situations can be identified, already described in \cite{Sadowsky1899} and experimentally observed in \cite{Friese1998Nat}. 

First, when a circularly or elliptically polarized wave interacts with an anisotropic particle, a constant torque can be generated, similar to the torque on a half-wave plate in experiments \cite{Beth1935, Holbourn1936, Beth1936, Brasselet2023AP} {or on an electric-dipole antenna in \cite{Allen1966AJP}}. For a non-absorbing particle with a diagonal dipole-polarizability tensor $\ten{\alpha}= \operatorname{diag}\left\{ \alpha_x, \alpha_y, \alpha_z \right\}$ and incident-field spin $\vb{S}$ aligned with the $z$-axis, this torque has the form \cite{Toftul2023PRL}
\begin{equation}
{T}^{\text{anis}}_z \propto \Re (\alpha_x - \alpha_y)^2 {S}_z\,.
\label{eq:T_aniz}
\end{equation}
This equation follows from the torque formulas in Table~\ref{tab:forces} supplied with Eqs.~(\ref{eq:cross-sections}) for a non-absorbing particle (see Appendix~\ref{app:torque_anisotropic}).

Second, for a linearly or elliptically polarized incident wave, there is a torque that vanishes at certain orientations of the particle. Such torque acts to {\it align} the particle with the polarization of the incident field \cite{Friese1998Nat, Smagin2024PRAppl}. 
This torque appears even when the incident spin vanishes, because the scattered/transmitted field acquires a non-zero spin when the incident polarization is not aligned with the particle’s optical axes. For a diagonal polarizability tensor $\ten{\alpha}$ and a linear polarization of the incident field, tilted by an angle $\theta$ with respect to the $x$-axis in the $(x,y)$-plane, the torque is given by \cite{Smagin2024PRAppl,Riccardi2023ChemRev}
\begin{equation}
{T}^{\text{align}}_z \propto  \Re(\alpha_x - \alpha_y) \sin(2\theta)\,.
\label{eq:T_align}
\end{equation}
This result follows directly from the main torque contributions in Table~\ref{tab:forces}.

Remarkably, although these anisotropy-induced torques have been considered exclusively in optics (apart from \cite{Smagin2024PRAppl}), the above considerations are universal for any type of vector field and associated dipoles: electric, magnetic, or acoustic. Equations~(\ref{eq:T_aniz}) and (\ref{eq:T_align}) involve the corresponding dipole polarizability tensors and the incident-field polarization.  

Similarly to torques originating from the change in the spin angular momentum of light transmitted through an anisotropic particle/plate, there are also torques on more complex particles/plates that generate {\it orbital} angular momentum (OAM) in the transmitted or reflected wave. 
Such wave elements have either space-variant anisotropy axes (q-plates) or anisotropic chiral geometries (spiral-phase plates). The corresponding optical and acoustic torques on such structures have been observed in \cite{Hakobyan2014NP, Wunenburger2015NJP, Magallanes2018NP}. 

Figure~\ref{fig:types_of_torque} schematically illustrates the main mechanisms of wave-induced torques. A fundamental condition of the existence of a nonzero torque is that the combined system ``incident field + particle'' must be mirror-asymmetric, i.e., effectively ``chiral'' in this sense \cite{Liu2010NN,Achouri2023ACS}.           

\subsection{Bi-isotropic particles in optics}
\label{sec:chiral}

\subsubsection{Chiral particles}

Another important case, specific to electromagnetism, is the cross-coupling between the induced electric and magnetic dipoles and the corresponding incident fields. This coupling occurs for isotropic {\it chiral} particles that lack mirror and space-inversion (P) symmetries. Such particles typically consist of chiral micro-elements (e.g., molecules) with a prevailing handedness, randomly oriented to ensure macroscopic isotropy. In this case, the induced electric and magnetic dipoles are given by \cite{Barron_book} 
\begin{equation}
{\begin{pmatrix}
\vb{e}/\sqrt{\varepsilon} \\
\vb{m}/\sqrt{\mu} \\
\end{pmatrix}
= 
\begin{pmatrix}
\alpha_{e} & i \alpha_{\chi} \\
-i \alpha_{\chi} & \alpha_{m} \\
\end{pmatrix}
\begin{pmatrix}
\sqrt{\varepsilon}\, \vb{E} \\
\sqrt{\mu}\, \vb{H} 
\end{pmatrix}},
\label{eq:polarizabilities_chiral}
\end{equation}
where $\alpha_{\chi}$ quantifies the chirality of the particle~\cite{Lindell1994,Wang2014NC}. 

Substituting Eq.~(\ref{eq:polarizabilities_chiral}) into the general equations of Table~\ref{tab:forces}, and using definitions (\ref{eq:PEMkin})--(\ref{eq:Helicity}), we obtain additional absorption rates, forces, and torques produced by the chirality and linear in $\alpha_{\chi}$ \cite{Chen2017OE,Genet2022ACSPhot,Golat2024PRR}: 
\begin{align}
\label{eq:A_chiral} 
    A^{\chi} = &\:  2 \omega^2 \Im(\alpha_{\chi}) \mathfrak{S}_{\text{EM}}  
\nonumber \\ 
    &-  2{\omega^2 g}  \Re\!\left[ \left(\alpha_{\chi} \alpha^*_e - \alpha^*_{\chi} \alpha_m \right)\tilde{\mathfrak{S}}_{{EM}} \right],
     \\
\label{eq:F_chiral}
\vb{F}^{\chi} = & \: {\omega} \Re(\alpha_{\chi}) \grad \mathfrak{S}_{EM} 
+  2\omega \Im(\alpha_{\chi})\, \boldsymbol{\mathfrak{P}}_{EM} \nonumber \\
&- {\omega k g} \left[ \Re(\alpha^*_{\chi}\alpha_e) \vb{S}_e
+ \Re(\alpha^*_{\chi}\alpha_m) \vb{S}_m \right], \\
\vb{T}^{\chi} = & \frac{c_l}{k} \sigma_{\chi\,1}^{\text{abs}} \Re \tilde{\vb{\Pi}}_{EM} + \frac{c_l}{k} \sigma_{\chi\,2}^{\text{abs}} \Im \tilde{\vb{\Pi}}_{EM}.
\label{eq:T_chiral}
\end{align}
These equations include both the main (`mixed') and recoil terms, 
and require further explanation. 

First, the chirality-dependent absorption rate (\ref{eq:A_chiral}) describes {\it chiral dichroism}, which is central to optical sensing of chiral molecules \cite{Barron_book,Tang2010,Tang2011,Hendry2010,Choi2012PRA,Bliokh2014PRL}. The leading term is proportional to the electromagnetic helicity density (\ref{eq:Helicity}), which can also be associated with ``electromagnetic chirality'' \cite{Tang2010,Bliokh2011PRA,Cameron2012}. The additional recoil term involves a natural complex extension of the helicity (\ref{eq:Helicity}): 
$\tilde{\mathfrak{S}}_{{EM}} = i\vb{H} \cdot \vb{E}^* /(2 \omega c_l)$, discussed in \cite{Bliokh2014PRL, Kamenetskii2015APB, Nieto-Vesperinas2021PRR}. 

Second, the chiral force (\ref{eq:F_chiral}) contains three terms that parallel the gradient, radiation-pressure, and recoil forces in Eqs.~(\ref{eq:F}) and (\ref{eq:F_recoil}). Specifically, the chiral gradient force is determined by the gradient of the helicity density in the incident field, whereas the chiral radiation-pressure force is proportional to the quantity $\boldsymbol{\mathfrak{P}}_{EM}$, which can be called ``chiral momentum'' density \cite{Canaguier-Durand2013NJP,Bliokh2014PRL,Vernon2024LSA}: 
\begin{align}
\boldsymbol{\mathfrak{P}}_{EM} &=  \frac{1}{4\omega c} \Re\! \left[ \vb{E}^* \cdot (\grad) \vb{H} - \vb{H}^* \cdot (\grad) \vb{E} \right] \nonumber \\
&= k \vb{S}_{EM} - \frac{1}{2k} \curl \vb{\Pi}_{EM}\,,
\label{eq:P_chiral}
\end{align}
where Maxwell's equations were used. 
The chiral momentum density characterizes the difference between the momenta of the right-hand and left-hand circularly-polarized photons. 
For the helicity eigenstates ${\sqrt{\varepsilon}}\,\vb{E} = i \sigma {\sqrt{\mu}}\, \vb{H}$, $\sigma =\pm 1$, it equals $\boldsymbol{\mathfrak{P}}_{EM} = \sigma \vb{P}_{EM}$. 
Importantly, the chiral gradient and radiation-pressure forces in Eq.~(\ref{eq:F_chiral}) have been successfully exploited for {\it optical sorting} (separation) of chiral particles \cite{Canaguier-Durand2013NJP, Bliokh2014PRL, Cameron2014NJP, Tkachenko2013PRL, Tkachenko2014NC, Wang2014NC, Kravets2019PRL} (see Section~\ref{sec:sorting}).

Third, we expressed the chiral optical torque (\ref{eq:T_chiral}) using the chiral absorption cross-sections \cite{Golat2024PRR}
\begin{align}
\sigma_{\chi\,1}^{\text{abs}} &= 2 k \!\left[ \Im (\alpha_{\chi}) -  g \Re(\alpha_{{e}} \alpha_{\chi}^*) -  g \Re(\alpha_{{m}} \alpha_{\chi}^*) \right], \nonumber \\
\sigma_{\chi\,2}^{\text{abs}} &= 2 k g \!\left[  \Im (\alpha_{{m}} \alpha_{\chi}^*)  -  \Im (\alpha_{{e}} \alpha_{\chi}^*)  \right] .
\end{align}
As in the isotropic-particle case, Eqs.~(\ref{eq:T_tot}), the chiral torque is determined by absorption and vanishes for lossless particles. This holds for isotropic chiral spheres of any size \cite{Chen2017OE}, consistent with the optical theorem for chiral particles~\cite{Belov2003TPL}. The torque (\ref{eq:T_chiral}) provides an additional mechanism of optical torque, Fig.~\ref{fig:types_of_torque}(d). Even a linearly-polarized incident plane wave can induce such a torque, because an absorptive chiral particle scatters it into a field carrying non-zero angular momentum. To our knowledge, this type of torque has not yet been observed experimentally, but it should be accessible in systems with chiral metallic nano-structures \cite{Liu2010NN, Gautier2009Chemphyschem, Yamanishi2022SA}.

In addition to the chiral effects linear in $\alpha_{\chi}$, there are small corrections $\propto |\alpha_{\chi}|^2$ to the isotropic-particle force and torque, Eqs.~(\ref{eq:F_recoil}) and (\ref{eq:T_tot}): $\Re({\alpha_e \alpha_m^*}) \to \Re({\alpha_e \alpha_m^*}) + |\alpha_{\chi}|^2$ and 
$\sigma_{e,m}^{\text{abs}} \to \sigma_{e,m}^{\text{abs}} - k g |\alpha_{\chi}|^2$. However, these corrections are insensitive to the sign of chirality $\alpha_{\chi}$ and therefore cannot distinguish enantiomers; they only provide small, usually negligible corrections to the magnitudes of the isotropic recoil force and torque. 

Note that some works group the chiral and non-chiral force terms differently, introducing additional `vortex' and `spin-curl' contributions \cite{Mun2020LSA, Wang2014NC, Li2019PRA, Chen2016PRA}. This difference arises from not explicitly using the canonical momentum and chiral momentum (\ref{eq:P_chiral}). The framework adopted here, based on these quantities, provides a clear and physically intuitive description.  A detailed comparison of equivalent formulations can be found in \cite{Golat2024PRR}.

\subsubsection{Magnetoelectric coupling}

Notably, the cross-coupling between electric and magnetic quantities in isotropic particles is not limited to chirality. In the chiral case (\ref{eq:polarizabilities_chiral}), the off-diagonal components of the polarizability tensor are anti-symmetric. However, they may also contain a symmetric part, corresponding to the {\it magnetoelectric} coupling \cite{Barron_book, Bliokh2014PRL}:
\begin{equation}
\begin{pmatrix}
\vb{e}/\sqrt{\varepsilon} \\ 
\vb{m}/\sqrt{\mu}
\end{pmatrix} = 
\begin{pmatrix}
{\alpha}_{{e}} & {\alpha}_{\text{me}} \\ 
 {\alpha}_{\text{me}}  & {\alpha}_{{m}} 
\end{pmatrix}  
\begin{pmatrix}
\sqrt{\varepsilon}\,\vb{E} \\ 
\sqrt{\mu}\,\vb{H}
\end{pmatrix},
\label{eq:polarizabilities_me}
\end{equation}
where ${\alpha}_{\text{me}}$ quantifies the magnetoelectric response of the particle. In contrast to chirality, which breaks space-inversion P-symmetry but preserves time-inversion T-symmetry, the magnetoelectric effect violates both. Consequently, it produces a {\it non-reciprocal} response and is sometimes referred to as ``false chirality'' \cite{Barron2001ACR, Barron1986CPL, Barron1986CSR}. Isotropic continuous media with magnetoelectric coupling are known as ``Tellegen media'' \cite{Tellegen, Raab1997, Tretyakov1998, Jazi2024NC, Asadchy2020, Ghosh2008}. Remarkably, the isotropic magnetoelectric effect was first predicted for molecules by Curie and Debye more than a century ago \cite{Curie1894, Debye}. However, it has remained elusive in direct experimental observations.

\begin{table*}[]
\caption{Summary of various types of particle according to their geometric, optical and acoustic properties.}
\begin{tabular}{@{}c@{}}
\hline \hline
\includegraphics[width=1.0\linewidth]{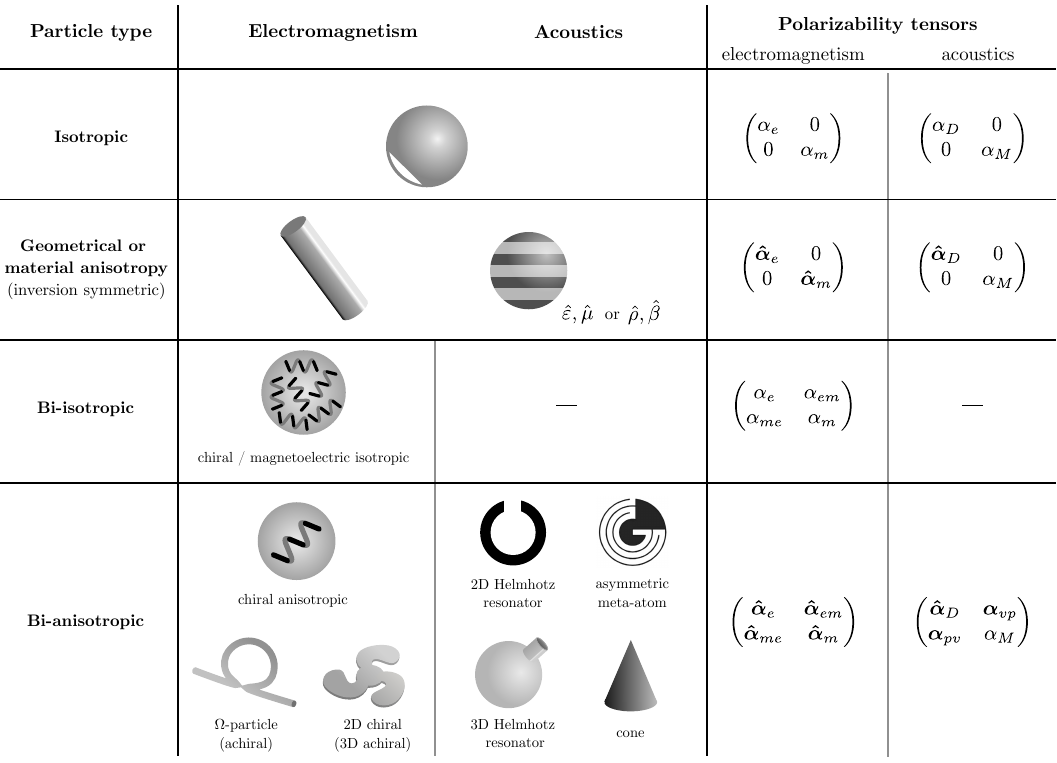}
\label{tab:particles_overview} \\
\hline \hline
\end{tabular}
\end{table*}

Absorption rates, forces, and torques produced by magnetoelectric coupling can be obtained by substituting Eq.~(\ref{eq:polarizabilities_me}) into the general equations of Table~\ref{tab:forces}. The main expressions have been described in \cite{Bliokh2014PRL, Golat2024PRR}. Importantly, while material chirality couples naturally to the electromagnetic helicity $\mathfrak{S}_{EM} = \Re \tilde{\mathfrak{S}}_{EM}$ (nonzero in a circularly-polarized plane wave and giving rise to optical chiral dichroism), the magnetoelectric effect couples to the quantity $\Im \tilde{\mathfrak{S}}_{EM} \propto \Re (\vb{H}^*\cdot \vb{E})$, which vanishes in {\it any} plane wave. Nonetheless, this quantity (called ``magnetoelectric density'' or ``reactive helicity density'') is generally nonzero in {\it structured} electromagnetic waves with locally non-orthogonal electric and magnetic fields \cite{Bliokh2014PRL, Kamenetskii2015APB, Nieto-Vesperinas2021PRR, Ghosh2024}. Therefore, the local magnetoelectric response of small particles can be revealed in suitably structured optical fields.

\subsection{Acoustic Willis coupling}
\label{sec:Willis}

As we mentioned in Section~\ref{subsec:acoustic_fields}, there is no acoustic counterpart of electromagnetic helicity: the chirality of a sound wave field vanishes identically due to its curl-less nature. Moreover, an isotropic cross-coupling, described by a single scalar parameter, as in Eqs.~(\ref{eq:polarizabilities_chiral}) and (\ref{eq:polarizabilities_me}), cannot be introduced. Indeed, cross-coupling between the scalar pressure and vector velocity fields requires a {\it vector} quantity, which implies {\it anisotropy}. (Alternatively, effective vector properties may arise for isotropic particles in the presence of {\it spatial dispersion}, i.e., a nonlocal response proportional to the wavevector $\vb{k}$ of the incident field.) As a consequence, isotropic chiral particles of opposite handedness cannot be distinguished in linear interactions with a monochromatic sound wave field. (This is reminiscent of Lord Kelvin's ``isotropic helicoid'': a hypothetical isotropic particle that couples translation and rotation when interacting with a fluid \cite{Collins2021Phys.Rev.Fluids}.)  

The acoustic cross-coupling between the pressure and velocity degrees of freedom is known as {\it Willis coupling} \cite{Willis1981, Willis1985, Milton2007PRSA}, which has recently attracted intense interest \cite{Muhlestein2016PRSA, Sieck2017PRB, Muhlestein2017NC, Quan2018PRL, Melnikov2019NC, Liu2019PRX, Quan2021NC, Sepehrirahnama2021PRE, Sepehrirahnama2022PRL}. The Willis coupling is described by the constitutive relations
\begin{equation}
\begin{pmatrix} 
{\vb{D}}/\sqrt{\rho} \\
M/\sqrt{\beta}
\end{pmatrix} = 
\begin{pmatrix}
\ten{\alpha}_{D} & \boldsymbol{\alpha}_{vp}  \\
{\boldsymbol{\alpha}}_{pv} & \alpha_{M} 
\end{pmatrix}
\begin{pmatrix}
\sqrt{\rho}\, \vb{v} \\
\sqrt{\beta}\, p 
\end{pmatrix},
\label{eq:polarizabilities_Willis1}
\end{equation}
where ${\boldsymbol{\alpha}}_{pv}$ and $\boldsymbol{\alpha}_{vp}$ are vector quantities, while the dipole polarizability $\ten{\alpha}_{D}$ generally becomes a $3 \times 3$ tensor due to particle anisotropy. The reciprocal (T-symmetric) case is characterized by the following relations
\cite{Quan2018PRL,Sepehrirahnama2021PRE}:
\begin{eqnarray}
\boldsymbol{\alpha}_{vp} = - \boldsymbol{\alpha}_{pv}^T \equiv \boldsymbol{\alpha}_W, \qquad 
\ten{\alpha}_{D} = \ten{\alpha}_{D}^{T}\,.
\label{eq:polarizabilities_Willis}
\end{eqnarray}
This can be regarded as the acoustic counterpart of the chiral electromagnetic coupling (\ref{eq:polarizabilities_chiral}). A non-reciprocal T-odd Willis coupling with $\boldsymbol{\alpha}_{pv} = \boldsymbol{\alpha}_{vp}^T$ is also possible \cite{Quan2021NC}, analogous to the magnetoelectric coupling (\ref{eq:polarizabilities_me}).

Substituting Eqs.~(\ref{eq:polarizabilities_Willis1}) and  (\ref{eq:polarizabilities_Willis}) into the general equations of Table~\ref{tab:forces}, and assuming for simplicity a scalar dipole polarizability $\ten{\alpha}_{D} = {\alpha}_{D}$, we obtain the leading corrections to the acoustic forces and torques induced by the Willis coupling (excluding recoil terms):
\begin{align}
\vb{F}_{W} &= \frac{1}{2c_s} \Re\!\left[ - (\boldsymbol{\alpha}_W^* \cdot \vb{v}^*) \grad p + \boldsymbol{\alpha}_W^* p^* \cdot (\grad) \vb{v}\right], 
 \label{eq:F_Willis} \\
\vb{T}_{W} &= \frac{1}{2c_s} \Re\!\left[ (\boldsymbol{\alpha}_W^* p^*) \times \vb{v}\right].
\label{eq:T_Willis}
\end{align}
Studies of Willis-coupling-induced forces and torques are just emerging \cite{Sepehrirahnama2021PRE, Sepehrirahnama2022PRL}. 
These effects depend sensitively on the direction of the vector polarizability $\boldsymbol{\alpha}_W$, and can thus be attributed to anisotropy-induced forces and torques \cite{Smagin2024PRAppl}.   
{Notably, since the acoustic recoil force terms in Eqs.~(\ref{eq:F_recoil}) are aligned with the main gradient and radiation-pressure forces, the Willis-coupling force (\ref{eq:F_Willis}) can serve as the main mechanism for generating a nonzero force component in the `third' direction}.

\subsection{Generic bi-anisotropic particles}

The most generic small particles with anisotropy and cross-coupling between different degrees of freedom (electric and magnetic in optics or pressure and velocity in acoustics) are known as bi-anisotropic particles \cite{Semchenko2001}. Such particles are still described by the induced dipole and monopole moments, and the general force and torque expressions in Table~\ref{tab:forces} remain applicable. However, in this case 
the polarizability tensors take their most general form: a $6\times 6$ complex-valued tensor for the electric–magnetic response in optics, and a $4\times 4$ tensor for the monopole–dipole response in acoustics, see Table~\ref{tab:particles_overview}. 
Spatial and temporal symmetries of the particle impose constraints on the tensor components, but even very simple geometries (such as spirals, cones, or 2D anisotropic and chiral shapes) exhibit a bi-anisotropic response.  Explicit values of optical and acoustic forces and torques should be analyzed in each specific case using the corresponding polarizability tensor and  expressions in Table~\ref{tab:forces}. Table~\ref{tab:particles_overview} summarizes representative examples and corresponding polarizability tensors of isotropic, anisotropic, bi-isotropic, and bi-anisotropic particles in optics and acoustics. 

\subsection{Role of resonances}
\label{subsec:resonances}

We now return to the case of isotropic particles to examine the effects of frequency dependence (dispersion) in their polarizabilities, as well as in the resulting forces and torques. These effects are most pronounced in the vicinity of various {\it resonances}, such as the first (dipole and monopole) Mie resonances, plasmonic resonances, atomic resonances, etc. A typical resonance is characterized by a resonant frequency $\omega_{\text{res}}$ and a quality factor $Q$, which is related to the decay rate parameter $\gamma$ by $Q = \omega_{\text{res}} / (2\gamma)$. In the vicinity of a typical resonance, the corresponding polarizability can be written in the form~\cite{Sipe1974PRA,Rahimzadegan2020OE,Poshakinskiy2019PRX} (see also Appendix~\ref{app:alpha_res})
\begin{align}
\alpha(\omega) = g^{-1} \frac{\gamma_{\text{rad}}}{\omega_{\text{res}} - \omega - \iu (\gamma_{\text{rad}} + \gamma_{\text{abs}})}\,.
\label{eq:alpha_res}
\end{align}
Here, for the acoustic dipole and monopole polarizabilities one should use the corresponding factors $g_D$ and $g_M$, Eq.~(\ref{eq:g}), whereas $\gamma_{\text{rad}}$ and $\gamma_{\text{abs}}$ are the decay rates associated with radiation and absorption losses, respectively \cite{Rahimzadegan2020OE, Lalanne2018LPR, Xu2000PRE, Bliokh2008RMP, Aspelmeyer2014RMP}.  
The total decay rate is $\gamma = \gamma_{\text{rad}} + \gamma_{\text{abs}}$. The polarizability (\ref{eq:alpha_res}), like any response function in a linear physical system, must satisfy the Kramers–Kronig relations \cite{Toll1956PhysRev, LeRu2013}. 

\begin{figure}
\centering
\includegraphics[width=\linewidth]{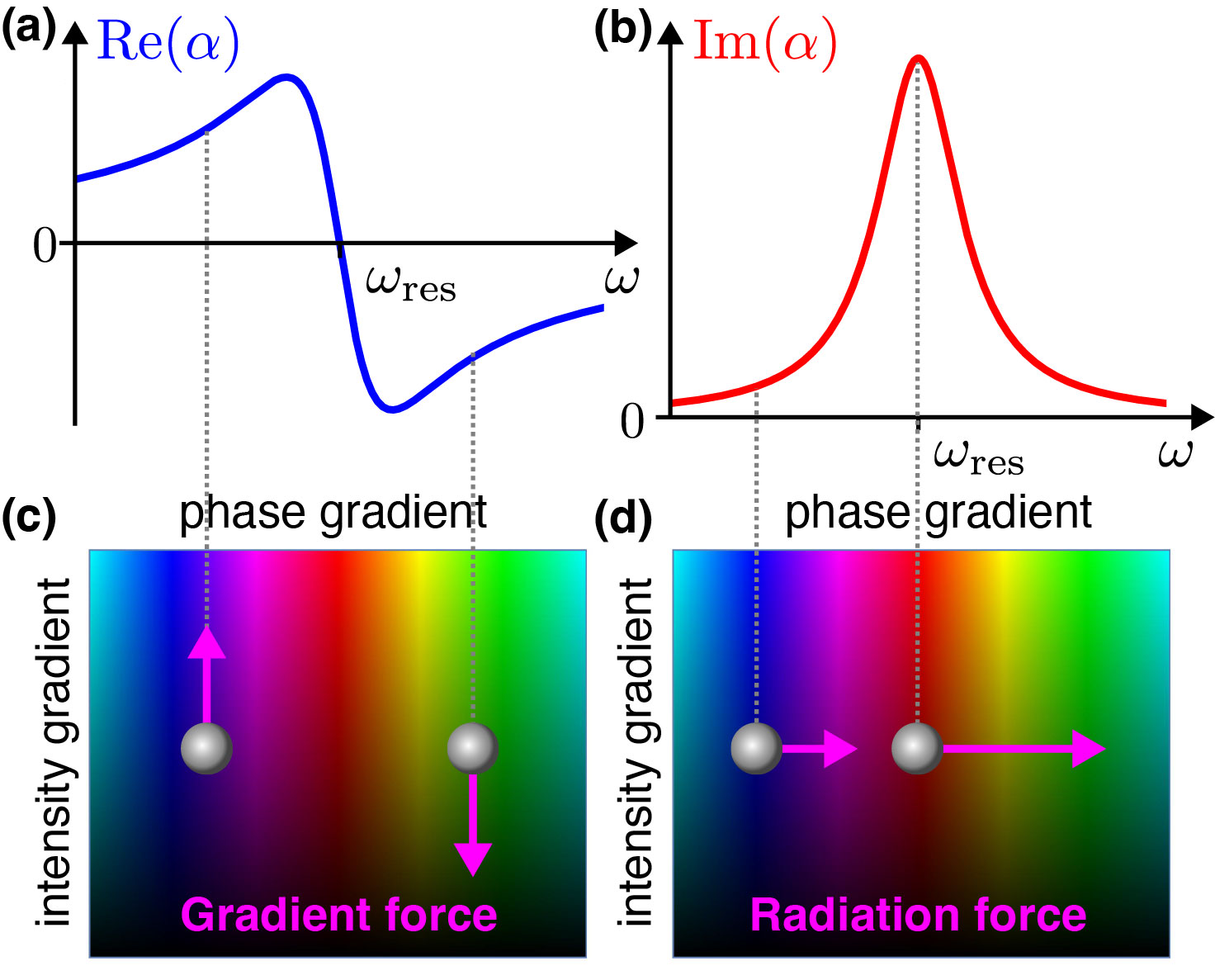}
\caption{{(a,b) Typical frequency dependencies of the real and imaginary part of a particle's polarizability $\alpha$ (electromagnetic or acoustic, dipole or monopole) near the resonance $\omega = \omega_{\rm res}$, Eq.~\eqref{eq:alpha_res}.} (c) The opposite signs of $\Re (\alpha)$ for $\omega > \omega_{\rm res}$ and $\omega < \omega_{\rm res}$ allows one to control the direction of the corresponding gradient force, Eqs.~(\ref{eq:F}). (d) The behaviour of $\Im (\alpha)$ offers resonant enhancement of the corresponding radiation-pressure force at $\omega = \omega_{\rm res}$.}
\label{fig:resonant-simple}
\end{figure}

The frequency dependencies of the real and imaginary parts of the polarizability (\ref{eq:alpha_res}) are shown in Fig.~\ref{fig:resonant-simple}(a,b). Both parts can be significantly enhanced in the vicinity of the resonant frequency. Consequently, the optical or acoustic absorption rates, forces, and torques exhibit resonant behavior. Importantly, $\Re (\alpha)$, and hence the gradient forces in Eqs.~(\ref{eq:F}), change sign when crossing the resonance. As a result, the particle can either be attracted to or repelled from high-intensity field regions, Fig.~\ref{fig:resonant-simple}(c) \cite{Rockstuhl2004OL, Mao2024LPR}. This effect can be employed for various trapping applications, such as atom trapping near a nanofiber with evanescent optical fields of two frequencies with $\omega < \omega_{\text{res}}$ and $\omega > \omega_{\text{res}}$ \cite{LeKien2004PRA, Dowling1996, Vetsch2010PRL, Goban2012PRL}. 

In turn, $\Im (\alpha)$ does not change sign and reaches its peak value at $\omega = \omega_{\text{res}}$. This provides conditions for the resonant enhancement of absorption rates (\ref{eq:A}), radiation-pressure forces in Eqs.~(\ref{eq:F}) \cite{Ashkin1977PRL, Li2015LPR, Li2013LSA}, and torques (\ref{eq:T}) \cite{Shi2019}, Fig.~\ref{fig:resonant-simple}(d). The maximum peak value of $k \Im (\alpha) = \sigma^{\rm ext} \simeq k g^{-1}$ is achieved when absorption is negligible, $\gamma_{\text{abs}} \ll \gamma_{\text{rad}}$. This corresponds to the maximum radiation-pressure force \cite{Rahimzadegan2017PRB}. 

In turn, the radiation torques (\ref{eq:T_tot}) is determined by the absorption cross-section $\sigma^{\text{abs}} = k \Im(\alpha) - kg \abs{\alpha}^2$, Eqs.~\eqref{eq:cross-sections}. Using Eq.~(\ref{eq:alpha_res}), we obtain
\begin{align}
\sigma^{\text{abs}}(\omega) = kg^{-1} \frac{\gamma_{\text{rad}} \gamma_{\text{abs}}}{(\omega_{\text{res}} - \omega)^2 + (\gamma_{\text{rad}} + \gamma_{\text{abs}})^2},
\label{eq:sigma_abs_res}
\end{align}
Remarkably, the maximum peak value $\sigma^{\text{abs}} = kg^{-1}/4$ is achieved under the so-called {\it critical coupling} conditions, when $\gamma_{\text{abs}} = \gamma_{\text{rad}}$ and $\omega = \omega_{\text{res}}$ \cite{Rahimzadegan2017PRB, Cheng2021ACSPhot}. At critical coupling, absorption is maximized while the scattered field amplitude vanishes \cite{Suh2004IEEE,Ruan2010PRL}. In some systems, this can lead to the total absorption of the incident field \cite{Xu2000PRE, Bliokh2008RMP, Bliokh2006PRL, Asano2016NC, Aspelmeyer2014RMP}, also known as ``coherent perfect absorption'' \cite{Chong2010PRL, Wan2011Science, Cao2022NP}. 
In the context of our work, the critical coupling maximizes the optical and acoustic radiation torques \cite{Rahimzadegan2017PRB}.
Note that this single-resonance limit can be surpassed when multiple resonances overlap, a phenomenon known as \qqoute{super-absorption} \cite{Ladutenko2015Nanoscale, Ruan2010PRL}.

Since the Mie resonance frequencies $\omega_{\text{res}}$ depend on the particle size $a$, the resonant behavior of radiation forces and torques can be used for size-selective sorting of particles \cite{Shilkin2017ACSPhot}. 
It has also been proposed that Helmholtz-resonator particles can provide resonant enhancement of Willis-coupling-induced acoustic forces and torques (\ref{eq:F_Willis}) and (\ref{eq:T_Willis}) \cite{Sepehrirahnama2022PRL}.

\subsection{Higher multipoles}
\label{subsec:multipoles}

So far, we have mostly considered the Rayleigh-particle regime, $ka \ll 2\pi$, where the scattered fields are well described by the lowest-order monopole and dipole modes. For larger Mie particles with $ka \sim 2\pi$, higher-order multipoles contribute significantly to the scattered fields and, hence, to the wave-induced forces and torques. In this case, the expressions in Table~\ref{tab:forces} are no longer applicable.

\begin{figure}
\centering
\includegraphics[width=\linewidth]{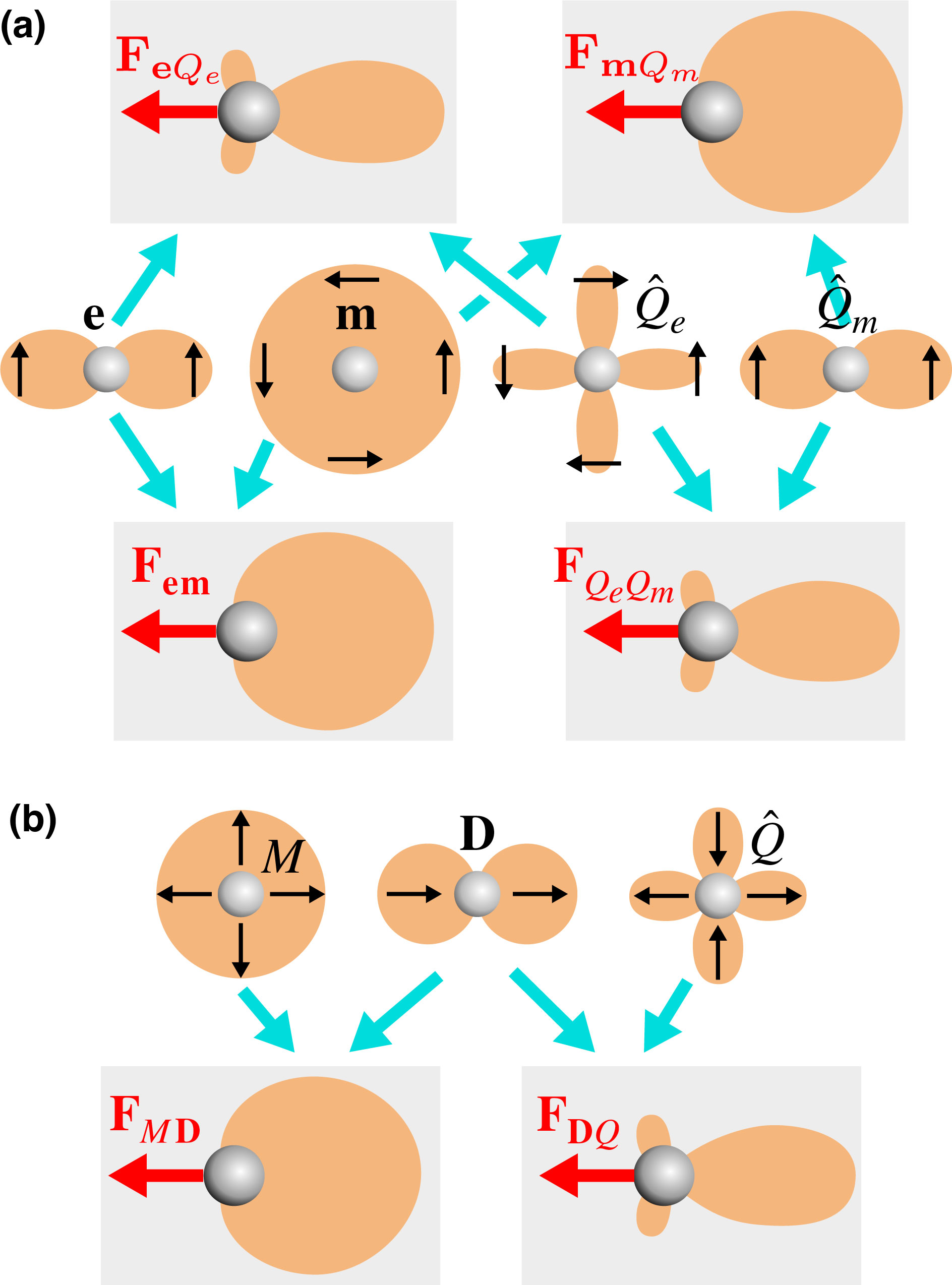}
\caption{Schematics of the generalized Kerker effect and recoil forces for higher-order multipoles (here, up to quadrupoles). (a) In optics, interference between electric and magnetic dipoles ($\vb{e}$ and $\vb{m}$) and quadrupoles ($\hat{Q}_{e}$ and $\hat{Q}_{m}$) in the scattered field results in four recoil-force terms associated with the interference of: $\vb{e}$ and $\vb{m}$ [Fig.~\ref{fig:recoil} and Eq.~(\ref{eq:F_recoil})],  $\vb{e}$ and $\hat{Q}_{e}$, $\vb{m}$ and $\hat{Q}_{m}$, $\hat{Q}_{e}$ and $\hat{Q}_{m}$ \cite{Liu2018OE}. (b) In acoustics, interference between the monopole $M$, dipole $\vb{D}$, and quadrupole $\hat{Q}$ in the scattered field produces two recoil-force terms corresponding to the interference of: $M$ and $\vb{D}$ [Fig.~\ref{fig:recoil} and Eq.~(\ref{eq:F_recoil})], $\vb{D}$ and $\hat{Q}$ \cite{Wu2021ApplPhysExpress, Wei2020NJP}.}
\label{fig:higher_multipoles_recoil}
\end{figure}

In the general case, both in optics and acoustics, the scattered field can be expanded into an infinite series of multipoles (spherical harmonics) \cite{bohren1984AbsorbtionScatteringLight, williams1999FourierAcousticsSound}. By substituting this multipole expansion into the general flux equations (\ref{eq:force_and_torque_average})--(\ref{eq:flux-split}) and evaluating the integrals in a manner similar to Appendix~\ref{app:tensor_force}, one can calculate the contributions of different multipoles to the absorption rate, force, and torque. 
Such approach has been implemented for optical forces \cite{Yu2019PRA, Zhou2023LPR, Jiang2015Dec, Chen2011NP} and torques \cite{Wei2022PRB,Xu2024NC}, whereas its application in acoustics remains an open challenge. A related framework, which involves the multipole decompositions of both incident and scattered fields, is known as the T-matrix formalism \cite{Nieminen2011JMO, gouesbet2011generalized, Barton1988,Barton1989, Gong2019PRE, Silva2011,Silva2012a,Silva2012EPL}.

Here we do not present cumbersome expressions for higher-order multipole contributions to the radiation forces and torques, but instead highlight their main features. 
First, the force and torque expressions are still split into `mixed' and `scattered' (recoil) terms.
The `mixed' terms contain independent contributions from the coupling of the $n$-th multipole with the incident field, whereas the recoil force terms arise from the interference between the neighboring $n$-th and $(n\pm 1)$-th multipoles, Fig.~\ref{fig:higher_multipoles_recoil}. In electromagnetism, interference between electric and magnetic multipoles of the same order also contributes to the recoil force terms. 
{By contrast, the higher-order contributions to the recoil optical torque include only interference between multipoles of the same type (either electric or magnetic), and only within the same order \cite[see SM]{Xu2024NC}.}
Note that higher-order multipole contributions to forces and torques involve higher-order derivatives of the incident field. 

One can also define multipole moments and the corresponding multipole polarizabilities of the particle, similarly to the dipole and monopole quantities considered earlier. For spherical Mie particles, the $n$-th multipole polarizability is proportional to the $n$-th Mie scattering coefficient \cite{LeRu2013}. For complex-shaped particles, the polarizabilities can be computed numerically \cite{BernalArango2014ACSPhot}. 

The role of higher multipoles in radiation forces and torques has been explored both theoretically \cite{Kislov2021AdvPhotRes, Stilgoe2008OE, Rezaei2022JOSA, Shilkin2017ACSPhot,Toftul2025arXiv_acusort, Shi2022NanoLett} and experimentally \cite{Ashkin1977PRL, Mao2024LPR, Toftul2025arXiv_hopping, Qi2013ApplPhysLett, Zhou2022PNAS, Zhou2023LPR}.
In particular, the multipolar decomposition  has proven useful in describing 
{chiral forces \cite{Guzatov2011QE, Shang2013OE, Canaguier-Durand2015PRA, Zheng2020PRA, Patti2019SR},}
pulling and lateral forces \cite{Chen2011NP, Chen2015ACSNano, Nan2023NC}, 
trapping effects \cite{Lepeshov2023PRL, Mao2024LPR, Toftul2025arXiv_hopping}, 
{as well as torque engineering in the linear \cite{Chen2014SR, Toftul2025ACSPhot} and nonlinear \cite{Toftul2023PRL} regimes}.

\subsection{Forces near interfaces}
\label{sec:interfaces}

\begin{figure}
\centering
\includegraphics[width=\linewidth]{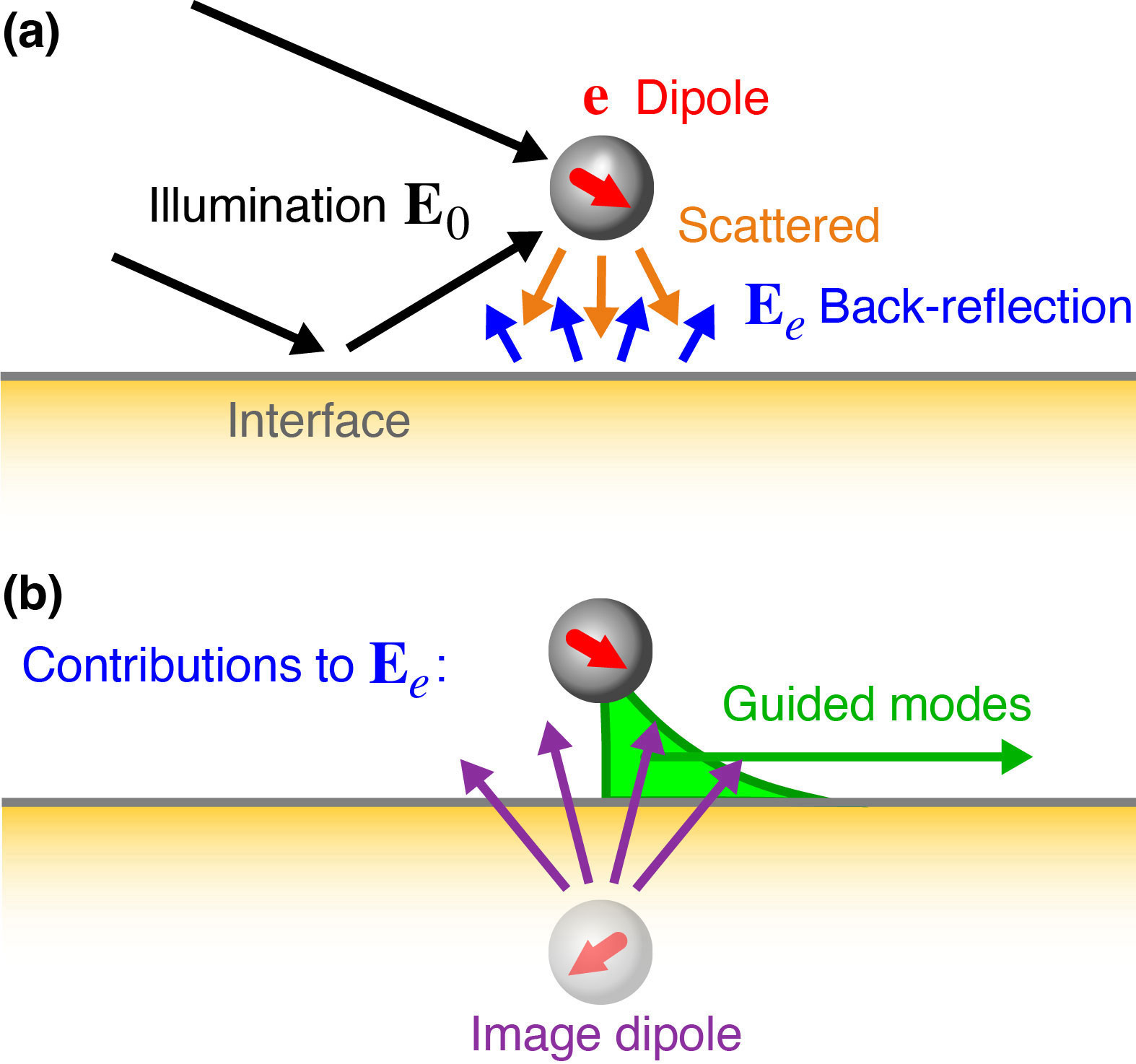}
\caption{(a) Schematics of the fields involved in the problem with a particle near an interface. $\vb{E}_0$ is the external incident field, including its reflection from the surface. The scattered dipole field and its reflection from the surface (the back-reflection field $\vb{E}_e$) are also shown.
The induced dipole $\vb{e}$ is determined by the superposition $\vb{E}_0 + \vb{E}_e$, or, equivalently, by the effective polarizability in Eq.~(\ref{eq:polarizability_eff}). (b) The back-reflected field $\vb{E}_e$ can include contributions from the `image dipole' located behind the interface, as well as near-fields of guided modes excited at the interface. Both contributions are encompassed in the environment's Green function $\ten{G}_{\text{sc}}$.}
\label{fig:near_surface}
\end{figure}

{When a particle is placed near an interface or another object, the force it experiences is no longer limited to the
principal forces described in Section~\ref{sec:forces}. The presence of the surface introduces additional contributions coming from the back-reflected scattered fields, Fig.~\ref{fig:near_surface}.} 
Manipulating particles near a surface is a common scenario in plasmonic and metamaterial-based tweezers \cite{Zhang2021LSA, Shi2022APR}.
For simplicity, we focus on the interaction of an electric dipole with an electromagnetic wave field; this approach extends to magnetic or acoustic dipoles and monopoles. 

In the presence of an interface, the total wave field at the dipole's location can be written as $\vb{E}_0 + \vb{E}_{e}$, where $\vb{E}_0$ is the total incident field due to the illumination and its scattering from the surface (i.e. the field that would exist without the particle), and $\vb{E}_{e}$ is the dipole's own field reflected back from the surface. The latter can be expressed using the Green's function: $\vb{E}_{e} (\vb{r}) = \varepsilon^{-1} k^2 \ten{G}_{\text{sc}}(\vb{r}, \vb{r}_0)\, \vb{e}$ with $\vb{r}_0$ denoting the dipole’s position. The Green's function $\ten{G}_{\text{sc}}$ encodes all information about the surrounding environment \cite{Sukhov2015NP, Petrov2016LPR, Kostina2020ACSPhot, Paul2019PRB,Ivinskaya2017LSA, Giron-Sedas2019PRB, Toftul2020ACSPhot, Kleshchenko2024APL}. Note that $\vb{E}_{e}$ does \emph{not} include the primary fields radiated/scattered by the dipole, Eq.~(\ref{eq:e-dipole}), which diverge at the dipole’s location in the point-dipole model; the self-action of such fields is taken into account in the recoil force terms, which remain unchanged. 

Two effects must be taken into account near a surface. First, the reflected dipole field induces self-polarization, leading to a self-consistent dipole moment $\vb{e} =\varepsilon \ten{\alpha}_e (\vb{E}_0 + \vb{E}_{e})$, which includes all orders of multiple scattering. Substituting $\vb{E}_{e}$, yields an effective polarizability \cite{Petrov2016LPR,Toftul2020ACSPhot}: 
\begin{equation}
\vb{e} = \varepsilon\, \ten{\alpha}_{e}^{\text{eff}}\, \vb{E}_0\,, \quad  
\ten{\alpha}_{e}^{\text{eff}} = \ten{\alpha}_e \!\left[ 1 - \varepsilon^{-1} k^2 \ten{G}_{\text{sc}}(\vb{r}_0, \vb{r}_0) \right]^{-1}. 
\label{eq:polarizability_eff}
\end{equation} 
Second, The reflected field also acts directly as a local field that modifies the force on the dipole in the general equations of Table~\ref{tab:forces}: $\vb{E} \to \vb{E}_0 + \vb{E}_{e}$. As a result, the main force on an electric dipole near a surface (without recoil terms) is given by \cite{Petrov2016LPR,Paul2019PRB,Ivinskaya2017LSA,Rodriguez-Fortuno2018May,Kostina2020ACSPhot}:
\begin{equation}
\label{eq:forcenearsurface}
\vb{F} = {\frac{1}{2} \Re \!\left[ \vb{e}^* \cdot(\grad) \vb{E}_0 \right]}
+ {\frac{1}{2} \Re \!\left[ \vb{e}^* \cdot(\grad) \vb{E}_{e} \right]} \equiv {\vb{F}_{0}} + {\vb{F}_{e}}.
\end{equation}

The additional force term $\vb{F}_{e}$ includes both the reflected far-field and the near-field contained in $\vb{E}_{e}$, giving rise to a variety of phenomena, some of which are illustrated in Fig.~\ref{fig:near_surface}(b). For instance, the interaction with the reflected field can be interpreted in terms of an `image dipole' behind the surface, which may either attract or repel the real dipole
\cite{Rodriguez-Fortuno2014Jan}. Furthermore, the near-field reflections account for the excitation of surface or guided modes at the interface. By momentum conservation, these modes lead to new `recoil-type' forces \cite{Wang2014NC,Rodriguez-Fortuno2015Nov,Petrov2016LPR, Li2013LSA, Maslov2014PRL} (which originate from the `mixed' rather than recoil term in Table \ref{tab:forces}). 

Other remarkable consequences of the force $\vb{F}_{e}$ include particle levitation, \cite{Rodriguez-Fortuno2014Jan,Krasikov2014RRL,Giron-Sedas2016Dec,Kingsley-Smith2020Oct,Rodriguez-Fortuno2018May}, optical binding, trapping or anti-trapping near metallic surfaces {and nanostructures} \cite{Juan2009NP, Ivinskaya2017LSA, Kostina2019PRB, Maslov2020AP, Zhang2021LSA}, optical manipulation using {metamaterials \cite{Shi2022APR},} hyperbolic metasurfaces \cite{Shalin2015Jun, Ivinskaya2018ACSPhot, Rodriguez-Fortuno2016Jan, Paul2019PRB}, magneto-optical surfaces \cite{Giron-Sedas2019PRB}, and novel Casimir forces \cite{Manjavacas2017PRL}. 
{For instance, particle levitation is produced by the phase-dependent interaction between the source dipole and its image; by appropriately tuning the reflection coefficients, one can induce a repulsive regime in which the dipole and its image effectively repel each other. 
In turn, Casimir forces for an electric dipole near a planar surface arise from stochastic fluctuations of both the dipole moment ${\bf e}$ and the field $\vb{E}_{e}$ generated by currents in the surface, both of which result in a stochastic contribution to the force ${\bf F}_e$.} 
The framework described here can also be extended to electric-magnetic dipoles \cite{Kingsley-Smith2019PRB} and to dipole-monopole acoustic interactions.

\subsection{Wavefront shaping of radiation forces and torques}
\label{sec:wavefron_shaping}

In the general scattering approach to wave-induced forces and torques, Section~\ref{sec:generalStressTensor} and Fig.~\ref{fig:flux_simple}, we assume that the incident wave field is given, while the scattered field is to be determined. Moreover, we have assumed that the particle is embedded in a homogeneous lossless medium, so that the incident-field fluxes through a closed surface vanish. In practice, however, many setups involve a non-ideal surrounding medium. In such cases, it becomes important to adjust the incident field to optimize the desired effect on the particle. 

One strategy is aberration correction, which compensates distortions introduced by the medium in order to reproduce a target incident field (e.g., a tightly focused beam) at the particle’s location \cite{Cizmar2010NP, Mazilu2011OE}. Another approach, which was introduced recently \cite{Ambichl2017PRL, Horodynski2020NP, Orazbayev2024NP, Hupfl2023PRL}, employs smart wavefront shaping outside the scattering medium and particle \cite{Cao2022NP} to directly optimize the {\it desired action on the particle} (e.g., the radiation force in a given direction) rather than a specific incident field. 
This method requires measuring the system’s scattering properties and then adapting the illumination accordingly.

\begin{figure}
\centering
\includegraphics[width=0.75\linewidth]{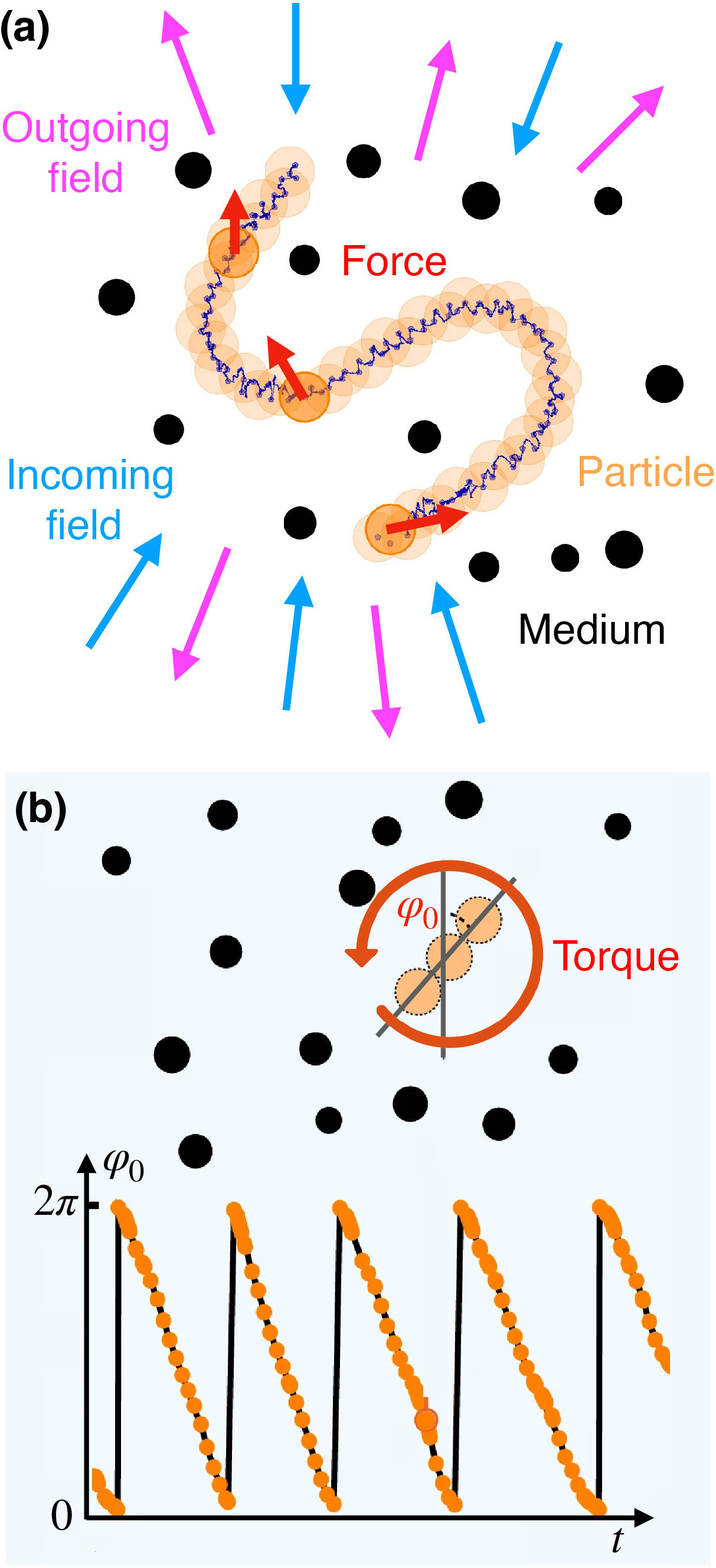}
\caption{(a) Smart wavefront shaping of the radiation force on a particle in a complex scattering medium, based on scattering-matrix measurements and adjustment of the incident field after each step \cite{Ambichl2017PRL, Horodynski2020NP, Orazbayev2024NP}. The central part shows experimental measurements of the particle (ping-pong ball in this case), which was forced to follow a prescribed S-shaped trajectory in an acoustic wave field. (b) Analogous wavefront shaping of the radiation torque on a three-ball cluster in a complex scattering medium. The experimental plot at the bottom demonstrates uniform rotation produced by the properly tailored incident acoustic field. Adapted from \cite{Orazbayev2024NP}.}
\label{fig:Rotter}
\end{figure}

Since this method is universal for optical, acoustic, or other waves, it is convenient to explain it within a universal quantum-mechanical-like formalism. Assuming that the system is lossless and supports $N$ wave modes, the scattering problem is described by a unitary $N \times N$ {\it scattering matrix} $\hat{S}$ that connects incoming and outgoing waves: $| \psi_{\rm out} \rangle = \hat{S}\, | \psi_{\rm in} \rangle$, where the wavefunction $\psi$ includes all necessary field components and the state vector $| \psi \rangle$ represents the field decomposition into the $N$ modes. (Here `incoming' and `outgoing' refer to modal propagation directions and should not be confused with the `incident' and `scattered' fields, since part of the incident field also contributes to outgoing channels.)
The interaction of the wave with the scatterers (medium and particle) can be characterized using the Hermitian {\it generalized Wigner-Smith operator} $\hat{Q}_\alpha = - i \hat{S}^{\dagger} d\hat{S}/d\alpha$, defined with respect to any suitable parameter of the system, $\alpha$ \cite{Ambichl2017PRL, Horodynski2020NP, Orazbayev2024NP, Bliokh2025arxiv}. The original Wigner-Smith operator was introduced in the context of the Wigner time delay, where $\alpha = \omega$ is the frequency, and $\hat{Q}_\omega$ describes the temporal delay of the scattered field \cite{Wigner1955PR, Smith1960PR}. More generally, the operator $\hat{Q}_\alpha$ quantifies the shift in the variable canonically conjugate to $\alpha$.

For example, choosing $\alpha = x_0$ (the $x$-coordinate of the particle), the operator $\hat{Q}_{x_0}$ characterizes the shift in the {\it momentum} component $p_x$ of the particle, i.e., the $F_x$ component of the wave-induced force. The maximum force $F_x$ is achieved when the incoming field $| \psi_{\rm in} \rangle$ is the eigenvector of $\hat{Q}_{x_0}$ corresponding to the largest eigenvalue. To find this optimal incoming field, one has to determine the scattering matrix $\hat{S}$ for two neighboring $x$-positions of the particle (to calculate its $x$-derivative in the operator $\hat{Q}_{x_0}$). Once the operator $\hat{Q}_{x_0}$ is known, 
the incident field can be tailored to realize the desired force on the particle, Fig.~\ref{fig:Rotter}(a) \cite{Ambichl2017PRL, Horodynski2020NP, Orazbayev2024NP}. 

In a similar manner, choosing $\alpha = \varphi_0$ (the azimuthal orientation of the particle with respect to the $z$-axis) allows one to control the torque $T_z$ using the operator $\hat{Q}_{\varphi_0}$ and its eigenvectors, Fig.~\ref{fig:Rotter}(b) \cite{Horodynski2020NP, Orazbayev2024NP}. Furthermore, choosing $\alpha = t$, the operator $\hat{Q}_t$ quantifies energy shifts in the system. Choosing the incident field as the eigenvector of $\hat{Q}_t$ associated with the most negative eigenvalue enables optimal {\it cooling} in a multi-particle system \cite{Hupfl2023PRL}. 

This smart wavefront shaping of radiation forces and torques has been demonstrated in both electromagnetic \cite{Horodynski2020NP} and acoustic \cite{Orazbayev2024NP} systems. It has also been applied to realize optimal pulling forces \cite{Horodynski2023PRA} and to enhance the stiffness of optical tweezers \cite{Butaite2024SA}. The advantages of this method are its universality and ability to optimize action for particles of any shape in arbitrary scattering environments. 
Its limitations are the assumption of a lossless system (needed to ensure that $\hat{Q}_\alpha$ is Hermitian) and time-consuming measurements of the full scattering matrix $\hat{S}$ at each step of system evolution.

\subsection{Photophoretic and acoustophoretic forces}
\label{subsec:photo_and_acoustophoretic}

The medium surrounding a particle can also influence its dynamics via thermal effects. Non-uniform absorption of radiation produces temperature gradients within the particle and surrounding medium, which in turn generate radiation-induced flows and {\it photophoretic} forces \cite{Ehrenhaft1917, Yalamov1976, Rohatschek1985, Tehranian2001, Horvath2014, Davis_book}.    

Research on photophoretic forces has traditionally focused on atmospheric aerosol particles \cite{Tong1975, Reed1977, Kerker1982JOSA, Williams1986, Chernyak1993, Rohatschek1995, Desyatnikov2009OE, Cheremisin2011, Keith2010PNAS}. More recently, however, the scope has broadened to include applications such as optical manipulation of nanoclusters \cite{Shvedov2009OE} and `giant' particles of order $100\,{\rm{\mu m}}$  \cite{Shvedov2010PRL}, chiral nanotube sorting \cite{Smith2014PCCP}, controllable pulling forces \cite{Lu2017PRL}, volumetric displays \cite{Smalley2018Nature}, and levitation and near-space flights \cite{Cortes2020AM, Azadi2021SA}.

In acoustics, analogous {\it acoustophoretic} forces arise from the interplay between sound waves and the properties of the surrounding medium, and they play a crucial role in acoustofluidics \cite{Lenshof2012LC, Lin2012LC, Bruus2012, Bruus2012a, Karlsen2015PRE}. However, 
in the acoustofluidics literature, the term ``acoustophoretic'' is often used broadly to describe {\it all} sound-wave-induced forces, including those covered in previous sections.
Within the framework of this review, it is useful to distinguish mechanisms: we reserve the term acoustophoretic forces for those not captured by linear acoustic-scattering theory, such as forces arising from nonlinear hydrodynamic, viscous, or thermal effects in the medium.

\subsection{Ponderomotive forces on the medium particles}
\label{sec:ponderomotive}

So far, we have considered the interaction of an incident wave field with a particle immersed in an optical or acoustic medium. However, a propagating wave also acts on the microscopic constituents of the medium itself (e.g., molecules or fluid elements). To describe such effects, one has to analyze the microscopic equations of motion governing the medium's particles. 

Let us consider a sound wave propagating in a fluid or gas. To describe the most relevant force effects, we consider a quasi-monochromatic wave with central frequency $\omega$, whose amplitude may vary both in space and in time (e.g., a propagating wavepacket). The corresponding cycle-averaged force density acting on the fluid particles can be expressed as
\cite{Bliokh2022PRA, Bliokh2025CP}:
\begin{align}
\vb{F}_{\text{pond}} &= -\dfrac{\rho}{4}\grad |\vb{v}|^2 + \dfrac{\rho}{2\omega} \dfrac{\partial}{\partial t} \Im\!\left[ (\vb{v}^* \cdot \grad) \vb{v}\right] \nonumber \\
&= - \dfrac{1}{2}\grad U^{(\vb{v})} + \dfrac{\partial \vb{P}^{({\bf v})}}{\partial t}\,.
\label{eq:F_pond}
\end{align}
Here the first term represents a ponderomotive gradient force, originally derived for a charged particle in an oscillating electromagnetic field \cite{Gaponov1958}, while the second term corresponds to the temporal variation of the local canonical momentum density in the wave field. Note the similarity between Eq.~(\ref{eq:F_pond}) and the dipole acoustic force on a particle, Eq.~(\ref{eq:F}). The gradient force tends to expel the medium's particles from regions of high kinetic wave energy density. 

\begin{figure}
\centering
\includegraphics[width=0.89\linewidth]{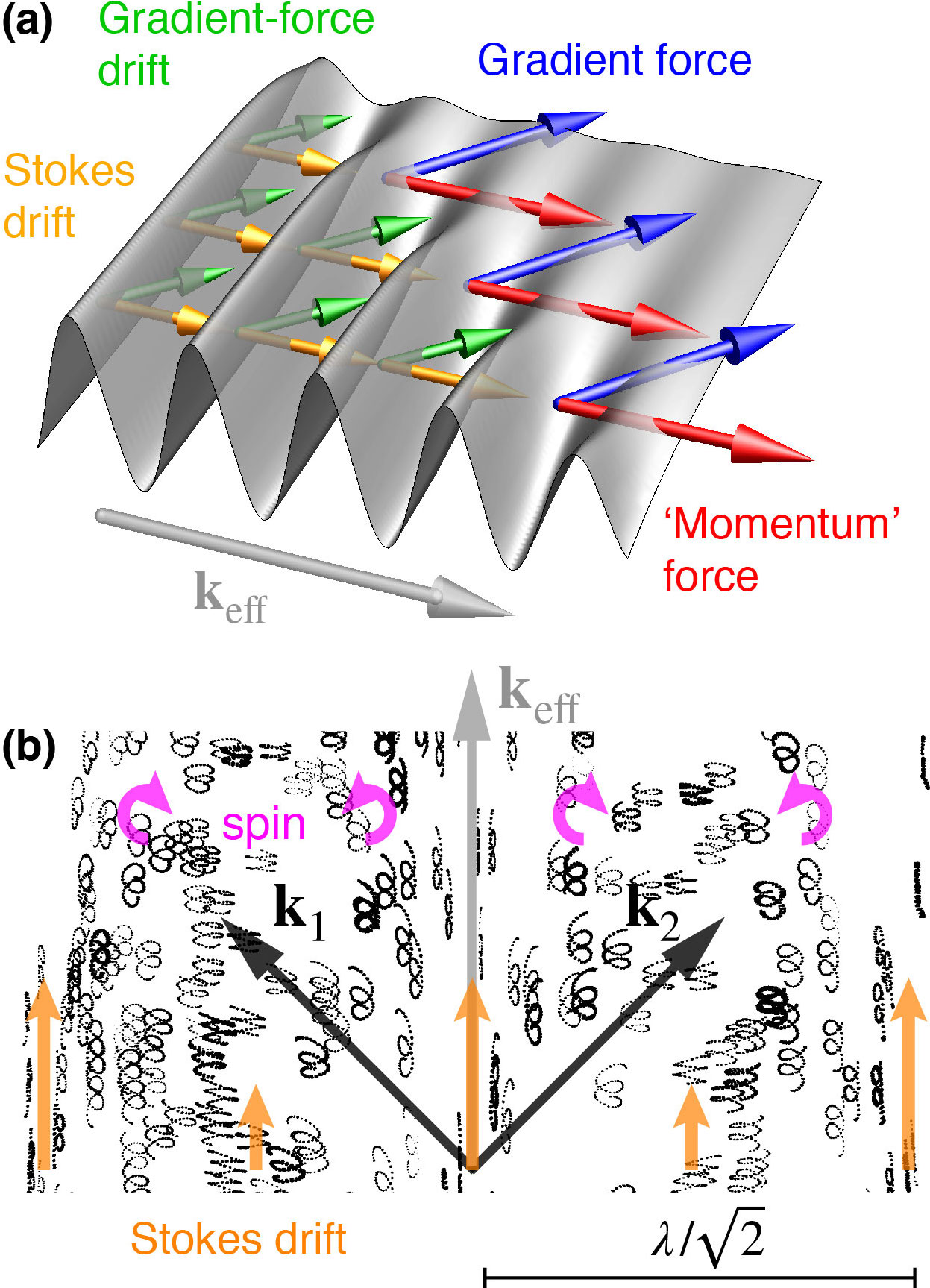}
\caption{Ponderomotive forces (\ref{eq:F_pond}) acting on microscopic constituents of the medium and producing the Stokes drift (\ref{eq:V_Stokes}). (a) Schematics of a wave front propagating through an initially quiescent medium. Spatial intensity gradients and temporal variations of the wave momentum density accelerate the particles, producing slow drifts in the wave zone. From \cite{Bliokh2022PRA}. (b) Experimentally observed Stokes drift of water-surface particles in the interference of two gravity water waves [see also Fig.~\ref{fig:vortex_beams}(c)]. Note that the local elliptical motions of the particles correspond to elliptical `polarizations' and the associated spin density (normal to the polarization plane) \cite{Jones1973, Shi2019, Bliokh2022SA, Bliokh2025CP} [see also Figs.~\ref{fig:momentum-spin}(b) and Fig.~\ref{fig:torque_exp}(a)]. Adapted from \cite{Bliokh2022SA}}
\label{fig:ponderomotive}
\end{figure}

Consider a wavefront arriving at a given point, starting from zero amplitude and saturating at a constant value. In this case, the second term in Eq.~(\ref{eq:F_pond}) accelerates the medium's particles to a steady velocity
\begin{align}
\vb{V}_{S} = \dfrac{1}{2\omega} \Im\!\left[ (\vb{v}^* \cdot \grad) \vb{v}\right] 
= \dfrac{\vb{P}^{({\bf v})}}{\rho}\,.
\label{eq:V_Stokes}
\end{align}
This is the Stokes drift known in fluid mechanics \cite{Bremer2018, Bliokh2022SA, Stokes1874, Bliokh2025CP}, which provides the microscopic mechanical origin of the acoustic canonical momentum density $\vb{P}^{({\bf v})}$, Fig.~\ref{fig:ponderomotive}. Thus, the ponderomotive force (\ref{eq:F_pond}) is intimately linked to the origin of canonical wave momentum.

Now consider the propagation of an electromagnetic wave in a plasma-like medium (e.g., a metal) with freely moving electrons. Remarkably, the ponderomotive force acting on the electrons is given by the first line in Eq.~(\ref{eq:F_pond}), with $\vb{v}$ and $\rho$ representing the electron velocity and density \cite{Bliokh2022PRA}. The gradient and time-derivative parts of this force correspond to the electron contributions to the electromagnetic wave energy and canonical momentum in a dispersive medium \cite{Bliokh2022PRA, Bliokh2017NJP, Bliokh2022PRE}. The medium dispersion, expressed through $\varepsilon = \varepsilon (\omega)$, arises from the microscopic electron-field interaction. In this case, the electromagnetic energy and canonical momentum equations (\ref{eq:UEM}) and (\ref{eq:PEMcan}) are modified by the substitution $\varepsilon \to \tilde{\varepsilon} = \varepsilon + \omega\, d \varepsilon / d \omega$ \cite{Landau_ECM, Bliokh2017NJP, Bliokh2022PRE}. In dielectric media, analogous results can be derived by considering the interaction of an electromagnetic wave field with polarizable atoms behaving as electric dipoles \cite{Milonni2010AOP, Gordon1973PRA}.

The ponderomotive forces acting on medium constituents are a fundamental ingredient of the theory of wave momentum in matter \cite{Gordon1973PRA, Brevik1979, Nelson1991PRA, Obukhov2003PLA, Milonni2010AOP, Mansuripur2012PRL, Bliokh2017NJP, Partanen2022OE, Astrath2022LSA}, including the Abraham-Minkowski dilemma in electromagnetism and the sound-wave momentum problem \cite{Pfeifer2007, Mcintyre1981, Peierls1979, Peierls1991, Milonni2010AOP, Barnett2010PTRS}.

\section{Examples of Forces and Torques in Structured Fields}
\label{sec:examples}

In this section, we describe key examples of optical and acoustic forces and torques that arise in specific systems and applications.

\subsection{Evanescent waves}

We begin with a simple yet illustrative example that exhibits nontrivial forces and torques. This case is important both as a model system and because of its relevance to numerous real systems. Namely, we consider a particle immersed in a single {\it evanescent wave} \cite{Fornel_book}, i.e., a wave propagating along the $z$-axis and decaying in the $x$-direction: $\propto \exp (ik_z z - \kappa x) \equiv \psi (z,x)$, where $k_z^2 - \kappa^2 = k^2$. Such a wave can be viewed as a plane wave with a complex wavevector $\vb{k} = k_z \bar{\vb{z}} + i \kappa \bar{\vb{x}}$. Here and hereafter, the overbars denote unit vectors along the corresponding axes. 

\begin{figure*}
\centering
\includegraphics[width=\linewidth]{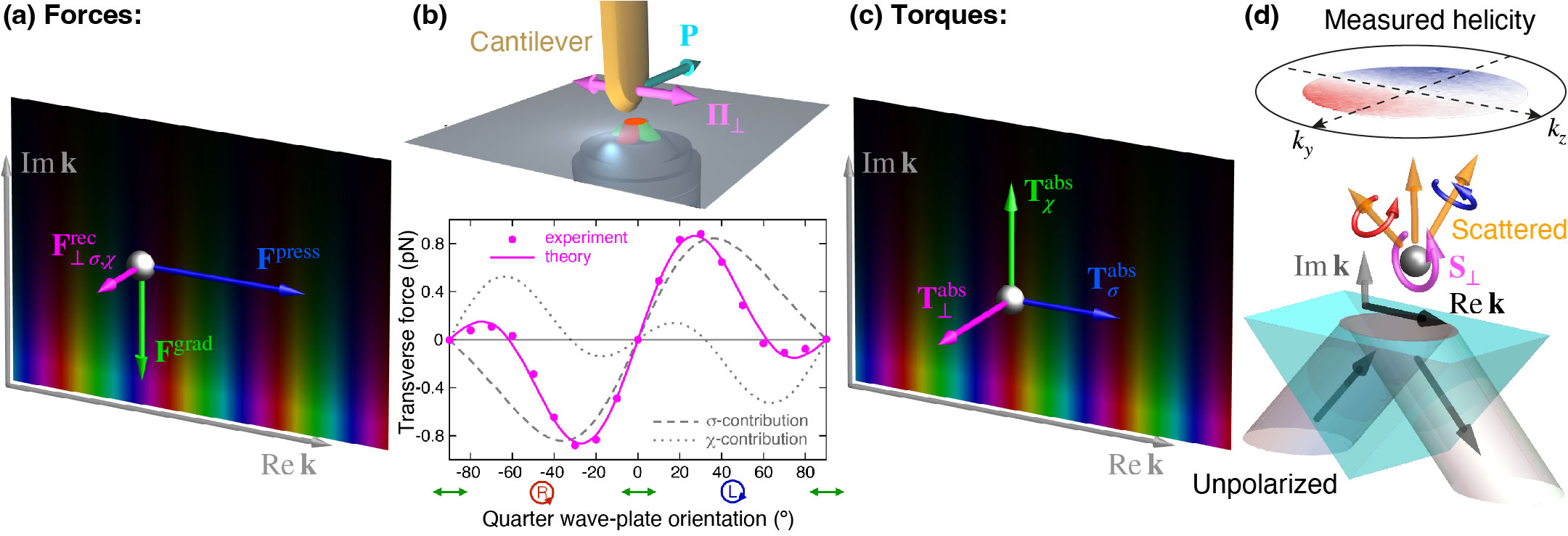}
\caption{Optical forces and torques in evanescent wave fields. (a) Gradient, radiation-pressure, and transverse recoil forces acting on a small isotropic particle in an evanescent wave. (b) Experimental measurements of the polarization-dependent transverse recoil force, Eqs.~\eqref{eq:F_recoil} and \eqref{eq:evanescent_EM_Pkin}. Adapted from \cite{Antognozzi2016NP}. (c) Three components of the absorption-related torque on an isotropic particle in an evanescent wave, Eqs.~\eqref{eq:T_tot} and \eqref{eq:evanescent_EM_UPS}. (d) Experimentally measured helicity of the side-scattered field from an isotropic particle in an evanescent wave generated by unpolarized light. This can be interpreted as an indirect measurement of the transverse torque on the particle, which follows from the angular momentum conservation. Adapted from \cite{Eismann2021NP}.}
\label{fig:evanescent}
\end{figure*}

\subsubsection{Optics}

An electromagnetic evanescent wave field can be written as \cite{Bliokh2014NC, Bliokh2015PR}
\begin{align}
\sqrt{\varepsilon}\vb{E} &\propto \dfrac{1}{\sqrt{1+|m|^2}}\!
\left( \bar{\vb{x}} + m \dfrac{k}{k_z} \bar{\vb{y}} - i \dfrac{\kappa}{k_z} \bar{\vb{z}} \right)\! \psi \,, \nonumber \\
\sqrt{\mu}\vb{H} &\propto \dfrac{1}{\sqrt{1+|m|^2}}\!
\left( -m \bar{\vb{x}} + \dfrac{k}{k_z} \bar{\vb{y}} + i m \dfrac{\kappa}{k_z} \bar{\vb{z}} \right)\! \psi \,.
\label{eq:evanescent_EM}
\end{align}
Here $m$ is a complex number that determines the polarization state of the wave. The corresponding normalized Stokes parameters are: $\tau = (1-|m|^2)/(1+|m|^2)$, $\chi = 2 \Re (m)/(1+|m|^2)$, and $\sigma = 2 \Im (m)/(1+|m|^2)$, with $\tau^2 + \chi^2 + \sigma^2 =1$. In the limiting plane-wave case ($\kappa=0$), $\tau$, $\chi$, and $\sigma$ describe the degrees of vertical/horizontal, diagonal ($\pm 45^{\circ}$), and right-hand/left-hand circular polarizations, respectively. 

Substituting the fields (\ref{eq:evanescent_EM}) into Eqs.~(\ref{eq:UEM})--(\ref{eq:Helicity}), we obtain the the energy, momentum, and spin densities:
\begin{align}
U^{(\vb{E},\vb{H})} &\propto \frac{1}{2}\!\left( 1 \pm \tau \frac{\kappa^2}{k_z^2} \right)\! e^{-2\kappa x},~~ 
\vb{P}^{(\vb{E},\vb{H})} = \frac{k_z \bar{\vb{z}}}{\omega} \, U^{(\vb{E},\vb{H})},
\nonumber \\ 
\vb{S}^{(\vb{E},\vb{H})} &\propto \frac{1}{2\omega} \!\left[ \pm \chi \frac{\kappa k}{k_z^2} \bar{\vb{x}} + (1\pm \tau) \frac{\kappa}{k_z} \bar{\vb{y}} + \sigma \frac{k}{k_z} \bar{\vb{z}} \right]\! e^{-2\kappa x}.
\label{eq:evanescent_EM_UPS}
\end{align}
The helicity density (\ref{eq:Helicity}) and the ``chiral momentum'' density (\ref{eq:P_chiral}) are determined by the parameter $\sigma$: $\mathfrak{S} \propto (\sigma/2\omega) \exp (-2\kappa x)$ and $\boldsymbol{\mathfrak{P}} = \mathfrak{S}\,k_z  \bar{\vb{z}}$. 

Substituting Eqs.~(\ref{eq:evanescent_EM_UPS}) into Eqs.~(\ref{eq:F}) and (\ref{eq:T}) or (\ref{eq:T_tot}), we find that an isotropic Rayleigh particle in an evanescent electromagnetic field experiences: (i) an $x$-directed gradient force due to the energy-density (intensity) gradient, (ii) a $z$-directed radiation-pressure force due to the canonical momentum (phase gradient), and (iii) a radiation torque due to the spin density, which generally has all three components, see Fig.~\ref{fig:evanescent}. 

In the simplest case of a vertically polarized wave ($m=\chi=\sigma=0$, $\tau=1$), an absorbing electric-dipole particle experiences a transverse $y$-directed torque caused by the {\it transverse spin} $S^{(\vb{E})}_y$, Fig.~\ref{fig:evanescent}(c,d). Notably, this helicity-independent spin component was recognized only recently \cite{Bliokh2012PRA, Bliokh2014NC, Bliokh2015PR, Aiello2015NP, Neugebauer2015PRL, Eismann2021NP}. It has since found applications for efficient spin-momentum coupling using evanescent and guided waves, both in optics \cite{Rodriguez-Fortuno2013Science, Petersen2014Science, leFeber2015NC, Bliokh2015Science, Bliokh2015NP, Lodahl2017Nature} and acoustics \cite{Shi2019, Bliokh2019a, Xu2020SA, Yuan2021NC, Yang2023PRL, Wei2020NJP}. Note that the geometry of the main optical forces and torque in a vertically-polarized evanescent wave is entirely similar to that shown schematically in Fig.~\ref{fig:concept_complex}.

In addition to the main forces and torque, the evanescent wave produces a nontrivial recoil force (\ref{eq:F_recoil}). The complex kinetic momentum density of the evanescent field (\ref{eq:evanescent_EM}) is
\begin{align}
\tilde{\vb{\Pi}} \propto \frac{1}{2c_l} \!\left[ - i \tau \frac{\kappa k}{k_z^2} \bar{\vb{x}} + (\sigma - i \chi) \frac{\kappa}{k_z} \bar{\vb{y}} + \frac{k}{k_z} \bar{\vb{z}} \right]\! e^{-2\kappa x}.
\label{eq:evanescent_EM_Pkin}
\end{align}
Here, the $\sigma$-dependent $y$-directed term corresponds to the transverse, circular-polarization-dependent component of the Poynting vector, first identified by Fedorov and Imbert \cite{Fedorov2013JO, Fedorov1955, Imbert1972PRD}. The $\tilde{\Pi}_y$ component of the complex kinetic momentum (\ref{eq:evanescent_EM_Pkin}) can produce a transverse recoil force (\ref{eq:F_recoil}) orthogonal to both the gradient and radiation-pressure forces. This lateral recoil force was described theoretically in \cite{Bliokh2014NC} and measured experimentally in \cite{Antognozzi2016NP, Liu2018PRL}, Fig.~\ref{fig:evanescent}(a,b). 

Furthermore, the presence of transverse spin $S_y$ suggests that an evanescent wave can exert a transverse chiral force on chiral particles. The leading terms in the chiral force (\ref{eq:F_chiral}) yield an $x$-directed helicity-gradient force and a $z$-directed force associated with the ``chiral momentum''. In turn, the recoil terms in Eq.~(\ref{eq:F_chiral}) have a nonzero $y$-component due to the transverse spin. This chiral recoil force, suitable for chiral particle sorting, has been analyzed in \cite{Wang2014NC, Hayat2015PNAS,Martinez-Romeu2024PR,Golat2024PRR}. 

Note that since evanescent waves appear near interfaces (e.g., under total internal reflection or in surface waves), particles in such fields are located close to an interface. Therefore, the interface-induced forces, described in Section~\ref{sec:interfaces}, can also play an important role in these systems. 

\subsubsection{Acoustics}

The main features of acoustic evanescent waves, forces, and torques are quite similar to the electromagnetic ones. Using the same waveform $\psi(z,x) = \exp (ik_z z - \kappa x)$, the pressure and velocity wave fields can be written as \cite{Bliokh2019a}
\begin{align}
\sqrt{\beta}p \propto \psi \,, ~~
\sqrt{\rho}\vb{v} = \frac{\vb{k}}{k} \sqrt{\beta}p \propto 
\left( \frac{i\kappa}{k} \bar{\vb{x}} + \dfrac{k_z}{k} \bar{\vb{z}} \right)\! \psi \,.
\label{eq:evanescent_A}
\end{align}
Substituting these fields into Eqs.~(\ref{eq:UA})--(\ref{eq:notation}), we obtain
\begin{align}
U^{(\vb{v},p)} &\propto \frac{1}{2}\!\left( \frac{k_z^2}{k^2} \pm \frac{\kappa^2}{k^2} \right)\! e^{-2\kappa x}, ~~ 
\vb{P}^{(\vb{v},p)} = \frac{k_z \bar{\vb{z}}}{\omega} \, U^{(\vb{v},p)}, \nonumber \\ 
\vb{S}^{(\vb{v})} &\propto \frac{1}{2\omega} \, \frac{\kappa k_z}{k^2}\, \bar{\vb{y}}\, e^{-2\kappa x}.
\label{eq:evanescent_A_UPS}
\end{align}

These equations are similar to electromagnetic expressions (\ref{eq:evanescent_EM_UPS}) with $\tau=1$ and $\sigma=\chi=0$, i.e., without polarization degrees of freedom. Consequently, they result in analogous basic forces (\ref{eq:F}) and torque (\ref{eq:T}) or (\ref{eq:T_tot}) acting on an isotropic Rayleigh particle: (i) the $x$-directed gradient force, (ii) the $z$-directed radiation-pressure force, and (iii) the $y$-directed torque due to the transverse spin $S_y$. 
Such forces and torque have been analyzed in \cite{Toftul2019PRL, Alhaitz2023PRL, Farhat2021PRB, Rondon2021JPhysCommun}. 

The recoil force (\ref{eq:F_recoil}) is produced by the complex kinetic momentum density $\tilde{\vb{\Pi}} \propto \dfrac{1}{2c_s} \! \left( \dfrac{i\kappa}{k} \bar{\vb{x}} + \dfrac{k_z}{k} \bar{\vb{z}} \right)$. Unlike the optical case (\ref{eq:evanescent_EM_Pkin}), this momentum has only $x$ and $z$ components, and, hence, the recoil force produces only small corrections to the dominant gradient and radiation-pressure forces. In acoustics, obtaining a recoil force with a nontrivial transverse direction requires either anisotropic particles \cite{Smagin2024PRAppl} or Willis coupling \cite{Sepehrirahnama2021PRE}. 

The evanescent-wave example is valuable not only {\it per se} but also as a generic model for fields where the intensity and phase gradients are orthogonal. For instance, the interference of two plane waves with equal amplitudes and frequencies but different propagation directions produces a field with its mean wavevector (phase gradient) directed along, say, the $z$-axis, and with periodic intensity variations along the orthogonal $x$-axis. Such fields produce analogous local forces and torques, both in optics \cite{Bekshaev2015PRX, Chen2016PRA, Achouri2020PRB, Sukhov2014O} and acoustics \cite{Shi2019, Bliokh2022SA}.

\subsection{Vortex beams}
\label{subsec:vortex_beams}

Let us consider another important class of structured wave fields: paraxial {\it vortex beams}. Such beams carrying orbital angular momentum (OAM) were introduced in optics in the early 1990s \cite{Bazhenov1990, Allen1992PRA}. Since then, they have found widespread applications in optics \cite{Allen_book, Torres_book, Bekshaev_book, Andrews_book, Molina-Terriza2007NP, Franke-Arnold2008LPR, Yao2011AOP, Shen2019LSA}, acoustics \cite{Hefner1999JASA, Lekner2006JASA, Guo2022JAP, Chaplain2022CP}, and quantum matter waves \cite{Bliokh2007PRL, Verbeeck2010Nature, Bliokh2017PR, Lloyd2017RMP, Larocque2018CP, Clark2015Nature, Luski2021Science}. 
Independently of their physical nature, paraxial vortex beams are characterized by the scalar vortex factor $\Psi(r,\varphi, z) = F(r) \exp (i\ell \varphi + i kz)$, where $(r,\varphi,z)$ are cylindrical coordinates, $F(r)$ is the radial function which behaves as $F(r) \propto r^{|\ell |}$ for $r\to 0$, and $\ell = 0, \pm 1, \pm 2, \dots$ is the integer topological charge (azimuthal quantum number) of the vortex. 

\begin{figure*}
\centering
\includegraphics[width=\linewidth]{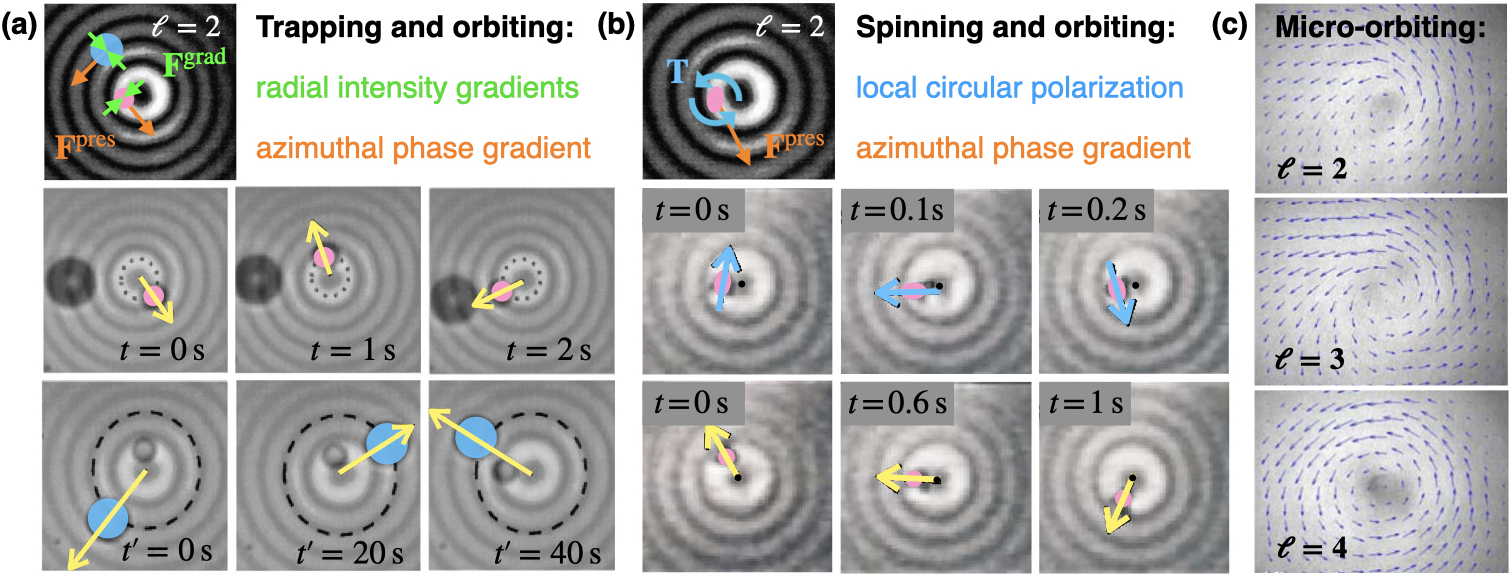}
\caption{Dynamics of particles in optical and acoustic paraxial vortex beams. (a) Radial trapping by the gradient force (either at intensity maxima or minima) and orbital motion (indicated by yellow arrows) due to the azimuthal radiation-pressure force (associated with the OAM) in an optical Bessel beam with $F(r) \propto J_{|\ell|}(\kappa r)$ ($\kappa \ll k$ is the radial wavenumber, $J_{|\ell|}$ is the Bessel function of the first kind) with $\ell =2$. Adapted from \cite{Garces-Chavez2002PRA}. (b) Similar radial trapping and orbital motion of an anisotropic particle, accompanied by its faster local spinning (indicated by blue arrows) due to the torque \eqref{eq:T_aniz} produced by the spin (circular polarization). Adapted from \cite{Garces-Chavez2003PRL}. (c) Local orbital motion of microparticles in a fluid induced by acoustic Bessel beams with different vortex charges $\ell$. These distributions correspond to the Stokes drift and visualize the transverse components of the canonical momentum density $\vb{P}^{({\bf v})}$ [see Section~\ref{sec:ponderomotive} and Fig.~\ref{fig:ponderomotive}(b)]. Adapted from \cite{Hong2015PRL}.}
\label{fig:vortex_beams}
\end{figure*}

\subsubsection{Optics}

The electromagnetic field of a paraxial optical vortex beam can be written as 
\begin{align}
\sqrt{\varepsilon}\vb{E} &\simeq \dfrac{1}{\sqrt{1+|m|^2}}\!
\left( \bar{\vb{x}} + m \bar{\vb{y}} \right) \Psi \,, \nonumber \\
\sqrt{\mu}\vb{H} &\simeq \dfrac{1}{\sqrt{1+|m|^2}}\!
\left( -m \bar{\vb{x}} + \bar{\vb{y}} \right) \Psi \,,
\label{eq:vortex_EM}
\end{align}
where $m$ is the same complex polarization parameter introduced earlier for evanescent waves, and we neglected the small longitudinal $z$-components of the fields. 
Substituting the fields (\ref{eq:vortex_EM}) into Eqs.~(\ref{eq:UEM})--(\ref{eq:Helicity}) yields the energy, momentum, and spin density distributions:
\begin{align}
U^{(\vb{E},\vb{H})} &\simeq \frac{1}{2} |F(r)|^2,~~ 
\vb{P}^{(\vb{E},\vb{H})} \simeq \frac{1}{2\omega}\! \left( k \bar{\vb{z}} + \frac{\ell}{r} \bar{\boldsymbol{\varphi}} \right)\! |F(r)|^2,
\nonumber \\ 
\vb{S}^{(\vb{E},\vb{H})} &\simeq \frac{\sigma}{2\omega}  \bar{\vb{z}}\, |F(r)|^2\,.
\label{eq:vortex_EM_UPS}
\end{align}
Here, the azimuthal component of the canonical momentum, $P_\varphi$, generates the OAM density $L_z = (\vb{r} \times \vb{P})_z = r P_\varphi \simeq (\ell/2\omega) |F(r)|^2$, which looks similar to the spin $S_z$ but with the vortex charge $\ell$ replacing the helicity parameter $\sigma$.

Substituting Eqs.~(\ref{eq:vortex_EM_UPS}) into the main force and torque Eqs.~(\ref{eq:F}), (\ref{eq:T}), and (\ref{eq:T_tot}), we find that an isotropic absorbing Rayleigh particle in a paraxial vortex beam experiences: (i) a {\it radial} gradient force, (ii) longitudinal and {\it azimuthal} radiation-pressure forces, and (iii) a longitudinal torque due to the spin. Consequently, if the particle's $z$-position is fixed (e.g., by a substrate), it can: (i) be trapped at a radial maximum or minimum of the field intensity $|F(r)|^2$ by the gradient force (depending on the sign of the real part of its dipole polarizability), (ii) orbit azimuthally along an $r=\text{const}$ circle due to the azimuthal radiation-pressure force controlled by the vortex charge $\ell$, and (iii) spin around the local $z$-axis passing through the particle due to the torque controlled by the polarization helicity $\sigma$, see Fig.~\ref{fig:vortex_beams}(a,b). The azimuthal orbiting and local spinning clearly demonstrate the qualitative difference between the orbital (extrinsic) and spin (intrinsic) angular momenta. Such dynamics of small particles in optical vortex beams was observed in \cite{Gahagan1996OL, Gahagan1999JOSAB, ONeil2002PRL, Garces-Chavez2002PRA, Volke-Sepulveda2002JOB, Garces-Chavez2003PRL, Curtis2003PRL}. Note that the $\sigma$-controlled spinning observed in these experiments originated from the particle's anisotropy and the corresponding anisotropy-induced torque \eqref{eq:T_aniz} rather than the absorption torque \eqref{eq:T_tot}. Employing anisotropy instead of absorption is typically more efficient and avoids undesirable heating.

Particles experience `negative' gradient forces and are attracted to low-intensity regions (`dark-field' trapping), when the real part of their polarizability is negative. This occurs when the refractive index of the dielectric particle is lower than the index of the surrounding medium \cite{Gahagan1999JOSAB, Garces-Chavez2002PRA} or at over-resonant frequencies near polarizability resonances (Fig.~\ref{fig:resonant-simple}) \cite{Dienerowitz2008OEO, Zemanek1998OCa, Nelson2007NP}. Such negative gradient forces enable trapping of negative-polarizability particles at the zero-field $r=0$ centers of vortex beams \cite{Sato1991EL, Simpson1998JMO, O'Neil2001OC, Gomez-Viloria2024ACSP}. 

When a particle occupies the central region of the vortex beam (e.g., when it is trapped at the center or its size exceeds the characteristic vortex radius), both the azimuthal radiation-pressure force and the spin-induced torque produce similar rotational effects with respect to the beam/particle center. In this regime, the `orbital' and `spin' torques act in similar ways \cite{He1995PRL, Simpson1997OL}, and the OAM-controlled `orbital torque' can be used for local rotational manipulation of particles \cite{Yan2013JPCL, Arzola2014OE}.

Furthermore, beyond the paraxial-field approximation (\ref{eq:vortex_EM}) and the basic gradient and radiation-pressure forces, an additional $\sigma$-dependent {\it orbital} rotation of particles can appear in vortex beams. This occurs via two mechanisms. First, in nonparaxial vortex beams, there is a $\sigma$-dependent contribution to the azimuthal canonical momentum $P_\varphi$ (i.e., OAM $L_z$) \cite{Adachi2007PRA, Zhao2007PRL, Bliokh2010PRA, Bliokh2015NP, Bliokh2015PR}. Second, a similar effect can arise due to the $\sigma$-dependent azimuthal component of the complex Poynting (kinetic) momentum density $\tilde{\Pi}_\varphi$ and the associated recoil force (\ref{eq:F_recoil}) originating from its real \cite{Angelsky2012OE, Angelsky2012OE_II, Svak2018NC} and imaginary \cite{Zhou2022PNAS, Xu2019PRL} parts.

\subsubsection{Acoustics}

In acoustics, paraxial vortex beams can be described by the scalar pressure field $\sqrt{\beta} p \simeq \Psi$ and the corresponding velocity field $\sqrt{\rho} \vb{v} \simeq \Psi \, \bar{\vb{z}}$. This longitudinal sound field produces the energy and canonical momenta densities $U^{(p,\vb{v})}$ and $\vb{P}^{(p,\vb{v})}$ similar to electromagnetic Eqs.~(\ref{eq:vortex_EM_UPS}) but without spin density: $\vb{S}^{(\vb{v})} \simeq 0$. (The transverse spin, present in both optics and acoustics, arises only in nonparaxial fields that have both transverse and longitudinal components \cite{Bliokh2015PR, Aiello2015NP, Bliokh2019b}.) As in optics, the azimuthal component of the canonical momentum is responsible for the OAM $L_z = r P_\varphi$ carried by the beam. 

Accordingly, paraxial acoustic vortex beams exert only a radial gradient force and a radiation pressure force with longitudinal and azimuthal components. A number of experiments have demonstrated acoustic `orbital torques' on Mie-sized or larger particles at the center of acoustic vortex beams, driven by the azimuthal radiation-pressure force \cite{Volke-Sepulveda2008PRL, Skeldon2008NJP, Demore2012PRL, Anhauser2012PRL, Baresch2018PRL}. The dynamical effects of acoustic vortices have found numerous applications, including trapping, levitation, and holographic tweezers \cite{Marzo2015NC, Baresch2016PRL, Hong2017SR, Marzo2018PRL, Marzo2019PNAS, Baudoin2019SA}, rotation of particles up to 6500 revolutions per second \cite{Zhang2022PRA}, manipulation of cells \cite{Baudoin2020NC, Zhang2020APL}, microbubbles \cite{Baresch2020PNAS} and objects in living bodies \cite{Ghanem2020PNAS} (see \cite{Guo2022JAP} for a review). 

Figure~\ref{fig:vortex_beams}(c) shows experimental observations \cite{Hong2015PRL} of the local orbital motion of microparticles in acoustic vortex fields. This microparticle velocity field can be associated with the local canonical momentum density or {\it Stokes drift} distribution in acoustic vortices [see Section~\ref{sec:ponderomotive} and Fig.~\ref{fig:ponderomotive}(b)].

\subsection{Trapping and tweezers}
\label{sec:tweezers}

\subsubsection{Optics}

Optical tweezers have been the most prolific implementation of optical forces. They use one or more tightly focused laser beams to trap and manipulate particles. Optical trapping was first proposed and demonstrated by Ashkin and co-workers \cite{Ashkin1986, Ashkin2000IEEE, Ashkin1987, Ashkin1987_II}. The explanation of the effect depends on the size of the trapped object, Eq.~\eqref{eq:ka_regimes}. For particles larger than several wavelengths, a ray-optics picture is appropriate. Namely, large dielectric particles act as lenses, re-directing the momentum of the incident rays, with the corresponding recoil pushing them towards the focus \cite{Ashkin1992BiophysJ, Ashkin2000IEEE}. For particles much smaller than the wavelength, the dipolar regime dominates, and the intensity-gradient force in Eq.~\eqref{eq:F} pushes the particle toward the high-intensity focus. For  particles of intermediate size, comparable to the wavelength, neither model is accurate, and one has to consider higher-order multipoles, Mie scattering calculations, or numerical calculations of forces based on the general Eqs.~\eqref{eq:force_and_torque_average}. 

Importantly, optical trapping can be applied to individual atoms \cite{Chu1998, Phillips1998, Raab1987}, Bose-Einstein condensates \cite{Gustavson2001PRL, Chikkatur2002S}, biological objects \cite{Ashkin1987, Ashkin1987_II}, and various nanostructures such as quantum dots, nanotubes, and graphene flakes \cite{Marago2013NN, Pauzauskie2006NMa, Tan2004NL, Jauffred2008NL}. Optical tweezers can also be used in nanofabrication, direct microfluidic control, and assembly \cite{Galajda2001APL, Terray2002S, Pauzauskie2006NMa}. Beyond manipulation, they enable precise measurement of tiny particle displacements, since light scattering is highly sensitive to the local intensity profile within the focus.

\begin{figure*}
\centering
\includegraphics[width=\linewidth]{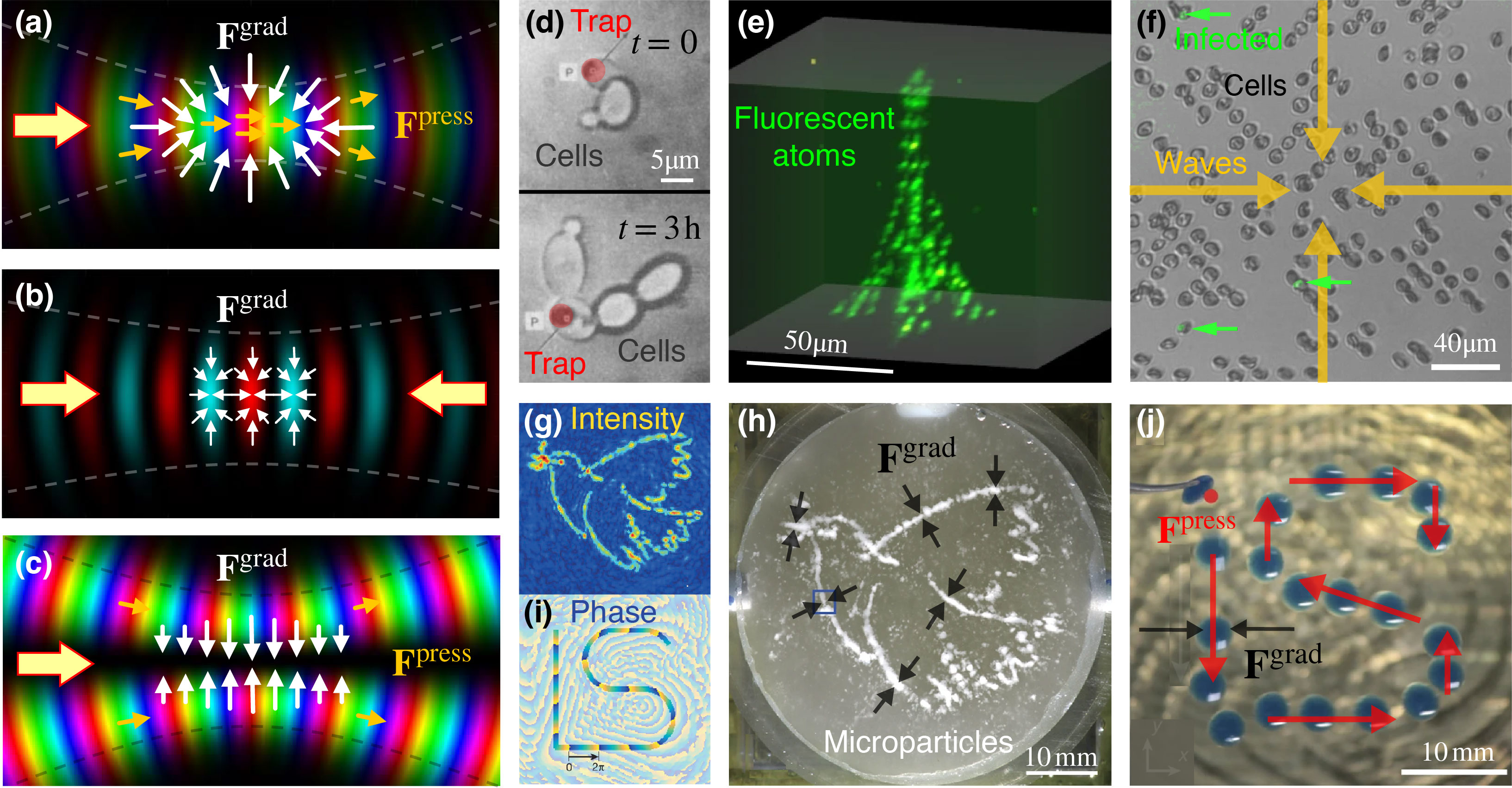}
\caption{Optical and acoustic trapping and tweezers. (a) A tightly focused Gaussian-like beam can trap particles in the focal region when the gradient force prevails over the radiation-pressure force. (b) Two counter-propagating beams form a standing-wave pattern. The radiation-pressure force vanishes, whereas the gradient force can trap particles at intensity maxima. (c) Vortex beams can provide radial `dark-field' trapping on the beam axis when the gradient force is `negative' (i.e. anti-parallel to the intensity gradient). All beams are shown here in the meridional $(z,x)$ cross-sections. (d) The first demonstration of optical trapping of biological cells with a single focused beam. Division of yeast cells was observed after three hours of trapping. Adapted from \cite{Ashkin1987_II}. (e) Holographic optical 3D trapping of 126 fluorescent Rubidium-87 atoms arranged in the shape of the Eiffel tower. Adapted from \cite{Barredo2018N}. (f) Trapping of multiple red blood cells in a 2D periodic lattice formed by two orthogonal standing surface acoustic waves. The cells are spaced by $\lambda/\sqrt{2}$. Green arrows indicate cells infected by a fluorescent parasite. Adapted from \cite{Collins2015NC}. (g-h) Holographic 2D acoustic field in which the intensity maximum forms the shape of a dove, trapping microparticles via the gradient force. (i-j) Holographic 2D acoustic field in which the intensity maximum forms the curve ``LS'', while the phase increases uniformly along the curve. Here, the gradient force traps the particle, while the radiation-pressure (phase-gradient) force propels it along the curve (a superimposed time-lapse sequence of images of one moving particle is shown). Adapted from \cite{Melde2016N}.}
\label{fig:tweezers}
\end{figure*}

Let us consider the main field configurations used in optical tweezers for small particles. In the simplest case, a single focused beam is employed. Assuming electric-dipole particles, the trapping gradient force ${\bf F}^{\rm grad}$ (for ${\rm Re} (\alpha_e) >0$) competes with destabilizing radiation-pressure and recoil forces ${\bf F}^{\rm press}$ and ${\bf F}^{\rm rec}$, which push the particle longitudinally out of the trap, Fig.~\ref{fig:tweezers}(a). According to Sections~\ref{subsec:Isotropic_Rayleigh} and ~\ref{subsec:recoil}, the gradient force depends on the tightness of focusing and on $\Re (\alpha_e )$, while the pressure force is proportional to $\Im (\alpha_e ) = k^{-1}\sigma^{\text{ext}}$. For lossless Rayleigh particles with $ka \ll 1$, we have $\Im (\alpha_e ) = k^{-1}\sigma^{\text{sc}} = g |\alpha_e|^2 \ll \Re (\alpha_e )$, Eq.~(\ref{eq:cross-sections}), so the gradient force typically dominates over the radiation-pressure one. However, both pressure and recoil forces grow with the particle size and refractive index faster than the gradient force, making it challenging to trap larger dielectric particles or high-index particles. This limitation can be overcome using very tight focusing and/or anti-reflection coatings \cite{Bormuth2009BJ, Jannasch2012NP}. In addition, for small particles, the gradient force competes with Brownian motion, which depends strongly on the surrounding medium and temperature. This can be problematic for metallic (lossy) nanoparticles that heat up under illumination \cite{Seol2006OLO}. At the same time, polarizability resonances in metallic particles can be advantageously exploited for trapping and control \cite{Dienerowitz2008, Pelton2006OLO}.

The next field configuration involves two counter-propagating focused beams with equal amplitudes, polarizations, and other parameters. 
This arrangement produces a standing wave with alternating maxima and minima of the electric energy density, separated by half a wavelength along the longitudinal axis, Fig.~\ref{fig:tweezers}(b). Importantly, in such a non-propagating field the momentum and energy flux densities vanish, and so do the pressure and recoil forces. 
Counter-propagating beams were first used in Ashkin’s pioneering work \cite{Ashkin1970, Ashkin2000IEEE} to trap large particles with sizes of several wavelengths. For such particles, the fine subwavelength standing-wave features are unimportant, and the {\it wavelength-averaged} intensity gradients provide stable trapping in the focal region. 
For subwavelength Rayleigh particles, the standing-wave structure becomes important, and such particles can be trapped at local field-intensity maxima (for ${\rm Re} (\alpha_e) >0$) \cite{Zemanek1999OLO} or minima (${\rm Re} (\alpha_e) <0$) \cite{Zemanek1998OCa}, as shown in Fig.~\ref{fig:tweezers}(b). Standing-wave trapping is particularly relevant in the context of cavity optomechanics \cite{Aspelmeyer2014RMP}.

Optical tweezers can also employ structured (essentially non-Gaussian) beams. In particular, both high-intensity (for $\Re(\alpha_e)>0$) and low-intensity (for $\Re(\alpha_e)<0$) regions have been exploited for trapping in vortex beams, see Section~\ref{subsec:vortex_beams}, Fig.~\ref{fig:vortex_beams}, and Fig.~\ref{fig:tweezers}(c). Furthermore, ``optical bottle beams'', which generate a 3D dark spot, have been used for `dark-field' trapping \cite{Arlt2000OLO, Isenhower2009OL, Xu2010OL, Barredo2020PRL, Shvedov2011OE}. Note that photophoretic forces can play an important role in such `dark-field' trapping with vortex and bottle beams \cite{Shvedov2009OE, Desyatnikov2009OE, Shvedov2010PRL, Shvedov2011OE, Smalley2018Nature}. Diffraction-free Bessel beams \cite{Durnin1987PRL} have been used to trap and manipulate particles over long distances in both single-beam \cite{Arlt2001OC} and counter-propagating two-beam \cite{Cizmar2005APL, Karasek2008PRLa, Yan2012NL} configurations. Likewise, Airy beams that exhibit `transverse acceleration' upon propagation \cite{Berry1979AJP, Siviloglou2007PRL} have also been applied in manipulation experiments \cite{Baumgartl2008NP, Baumgartl2009LC}.

A breakthrough in optical tweezers came with the introduction of multi-functional traps, often called {\it holographic tweezers} \cite{Grier2003N}. These systems generate arrays of multiple beams with independently controlled foci, enabling simultaneous multi-point manipulation of many trapped particles \cite{Dufresne1998RSI, Sasaki1991OLO, Mogensen2000OC, Sasaki1991JJAP, Dufresne2001RevSciInstrum, Reicherter1999OLOa, Liesener2000OC, Curtis2002OC, Barredo2018N, Smalley2018Nature}, see Figs.~\ref{fig:overview}(b), \ref{fig:displays}(a), and \ref{fig:tweezers}(e). 

Importantly, optical trapping can be combined with various torques, enabling controlled rotations of anisotropic or absorptive trapped particles. This has driven rapid progress in the study of optical torques and their applications \cite{Friese1998Nat, He1995PRL, Simpson1997OL, Padgett2011NP, Bustamante2021NR}. 
Remarkably, the achievable angular velocities of trapped particles span an enormous range: from quasi-static 3D alignments and slow (Hz-scale) rotations \cite{Pelton2006OLO, Tong2010NL, Wu2012NP} to MHz--GHz rotations, where material strength limits and quantum effects become relevant \cite{Arita2013NC, Kuhn2015NL, Hu2023NC, Ahn2018PRL, Reimann2018PRL, Gonzalez-Ballestero2021S}.

Optical trapping and tweezers have found wide-ranging applications across many fields. First, they 
have become an invaluable tool for manipulating and probing biomolecules and living cells \cite{Ashkin1987, Ashkin1987_II, Block1990N, Svoboda1993N, Abbondanzieri2005N, Asbury2003S, Smith1996S, Svoboda1994, Dholakia2020NRP, Zhang2008JRSI, Bustamante2021NR, Xin2020AM}, Fig.~\ref{fig:tweezers}(d). 
Second, optical trapping has played a key role in optical cooling of atoms, molecules, nanoparticles, and larger macroscopic objects \cite{Chu1998, Phillips1998, Metcalf1994PR, Chan2011N, Li2011NP, Gieseler2012PRL, Corbitt2007PRL, Tarbutt2018CP, Li2010S, Delic2020S, Schafer2021PRL, Rudolph2022PRL}, enabling advances from macroscopic quantum interference \cite{Arndt2014NP, Romero-Isart2010NJP} to gravitational-wave detection \cite{Arvanitaki2013PRL}. 
Third, optical forces near the trap centers create effective harmonic-oscillator potentials, providing a platform for both classical and quantum optomechanical studies \cite{Aspelmeyer2014RMP, Millen2020RPP, Chang2010PNAS, Gonzalez-Ballestero2021S, Jain2016PRL, Gieseler2013NP, Tebbenjohanns2021N}.
Moreover, 1D, 2D, and 3D arrays of optically trapped atoms, molecules, or particles can serve for quantum simulators and computers \cite{DeMille2002PRL, Endres2016S, Barredo2016S, Bernien2017N, Weitenberg2011PRA, Schlosser2023PRL, Ebadi2021N, Barredo2018N}, atomic clocks \cite{Young2020N}, and volumetric displays \cite{Smalley2018Nature}, see Figs.~\ref{fig:displays}(a) and \ref{fig:tweezers}(e).  
{Finally, optical trapping and tweezers play the key role in various force-microscopy techniques \cite{Neuman2008NM, Stout1997}.}

\subsubsection{Acoustics}
\label{sec:tweezers_a}

Acoustic trapping and tweezers share natural similarities with their optical counterparts, but also exhibit important peculiarities. As in optics, the gradient force ${\bf F}^{\rm grad}$, Eq.~\eqref{eq:F}, plays the central role in acoustic tweezers. However, in most optical systems the {\it electric-dipole} part of this force dominates, while the magnetic-dipole part can typically be neglected. In contrast, in acoustic systems, both the {\it monopole} and {\it dipole} parts of the gradient force are usually significant. Moreover, these parts often act in opposite directions, so the competition between the magnitudes of the monopole and dipole gradient forces largely determines the trapping properties of the acoustic field.

Consider the simplest case of a 1D standing sound wave: 
\begin{equation}
\sqrt{\beta}\, p \propto \cos (kx)\,, \quad 
\sqrt{\rho}\, {\bf v} \propto i \bar{\bf x} \sin (kx)\,.
\label{eq:standing_A}
\end{equation}
Here, the nodes of the pressure energy density $U^{(p)} \propto \cos^2 (kx)$ coincide with the anti-nodes of the velocity energy density $U^{({\bf v})} \propto \sin^2 (kx)$, and vice versa. Therefore, when the monopole and dipole polarizabilities of the particle, $\alpha_M$ and $\alpha_D$ (real-valued, for simplicity), have the same sign, the monopole and dipole parts of the gradient force act in opposite directions (see also Section~\ref{sec:sorting}). 

\begin{figure*}
\centering
\includegraphics[width=\linewidth]{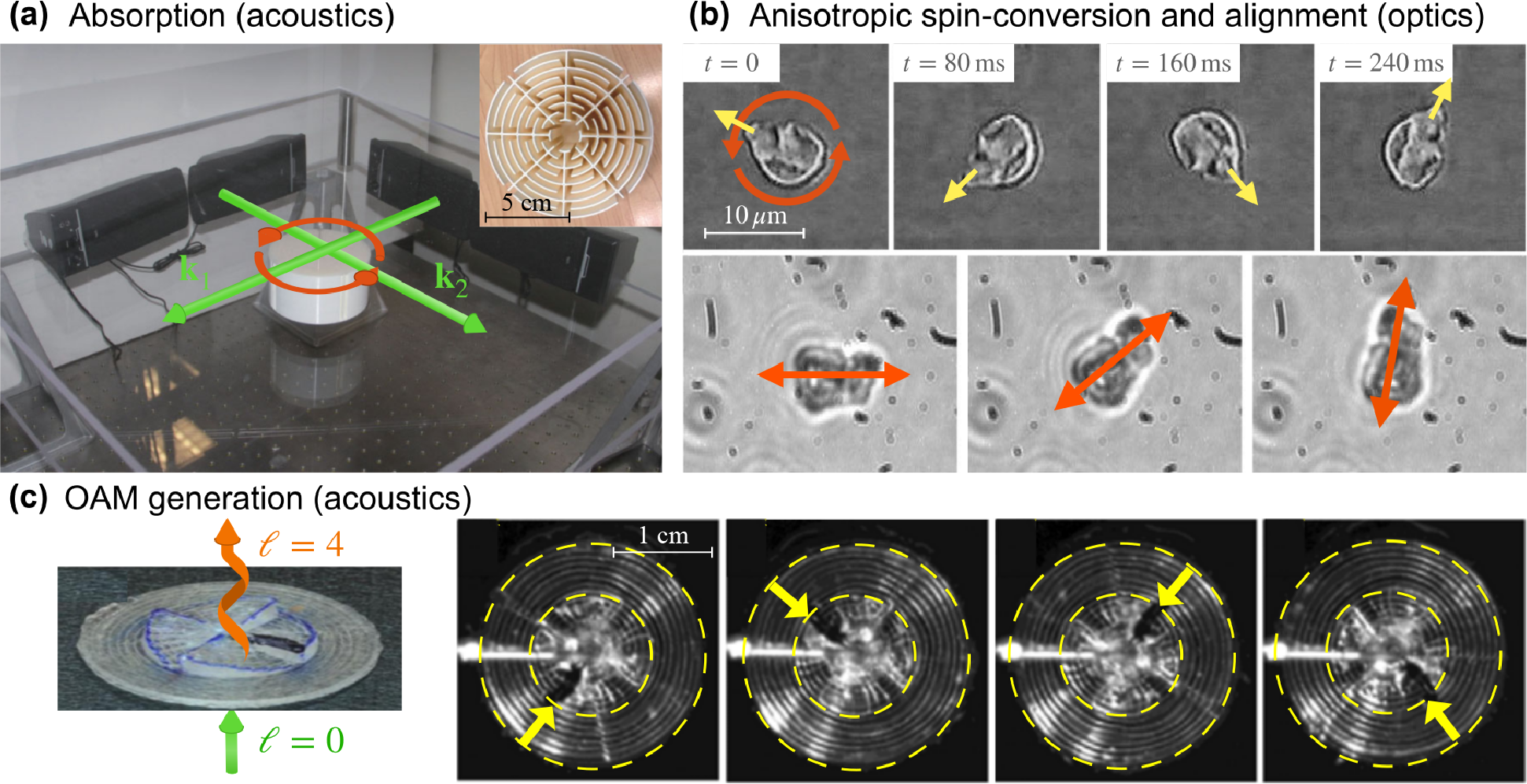}
\caption{Optical and acoustic torque experiments. (a) Acoustic torque \eqref{eq:T_tot} on a resonant-absorbing `meta-atom' generated by a non-zero vertical spin density in two interfering sound waves (the wavelength is $\lambda \simeq 40\,$cm). Adapted from \cite{Shi2019}. 
(b) Spinning (top) and alignment (bottom) of a birefringent particle in circularly and linearly polarized light (polarization is indicated by red arrows) produced by the torques \eqref{eq:T_aniz} and \eqref{eq:T_align}. Adapted from \cite{Friese1998Nat}. (c) Acoustic torque on a spiral wave plate generating OAM (vortex) in the {transmitted} sound (ultrasonic) wave. Adapted from \cite{Wunenburger2015NJP}. The torques observed in experiments (a), (b), and (c) correspond to the schematics in Figs.~\ref{fig:types_of_torque}(a), \ref{fig:types_of_torque}(b,c), and \ref{fig:types_of_torque}(e), respectively.}
\label{fig:torque_exp}
\end{figure*}

Similarly to the first optical trapping experiments \cite{Ashkin1970}, the earliest acoustic tweezers were based on a {\it standing-wave} field formed by two counter-propagating beams \cite{Wu1991JASA}. Standing waves remain one of the primary configurations for acoustic and acoustofluidic manipulation \cite{Eller1968JASA, Apfel1981JASA, Hertz1995JAP, Trinh1985RSI, Collins2015NC, Ding2012PNAS, Wu2017PNAS,  Lim2019NP, Li2015PNAS, Ozcelik2018NM, Wu2019MN, Meng2019JPD, Ding2013LC, Baudoin2020AR, Silva2019PRA}, Fig.~\ref{fig:tweezers}(f) (see also Section~\ref{sec:sorting}). Interestingly, the first historical demonstration of the mechanical action of acoustic waves was also implemented in a standing sound wave: in 1866 Kundt observed the accumulation of powder particles near the standing-wave nodes \cite{Kundt1866, Sarvazyan2010UMB}. 

More recently, single-beam acoustic tweezers have been realized \cite{Lee2009APL, Baresch2016PRL}, often employing {\it vortex beams} \cite{Baresch2016PRL, Baresch2018PRL, Baresch2020PNAS, Baudoin2020NC, Baudoin2019SA, Zhang2022PRA,  Marzo2015NC, Marzo2018PRL, Marzo2019PNAS, Baudoin2020AR}. For paraxial acoustic vortex beams (see Section~\ref{subsec:vortex_beams}), the pressure and velocity energy densities have similar distributions, $U^{(p)} \simeq U^{({\bf v})}$, with a nodal line along the beam's axis, Fig.~\ref{fig:tweezers}(c). In this case, a `negative' trapping gradient force can be achieved for relatively hard particles in air. Indeed, typical particles are heavier but less compressible than air: $\rho_p > \rho$, $\beta_p < \beta$. According to Appendix~\ref{app:polarizabilities}, this means that $\alpha_D >0$ and $\alpha_M < 0$. Thus, the dipole and monopole parts of the radial gradient force have opposite signs; trapping in the vortex center occurs when the monopole contribution prevails. Naturally, vortex-beam trapping also involves `orbital' torques induced by the vortex phase structure.

As in optics, {\it holographic} acoustic tweezers have been developed, enabling arbitrary 2D and 3D arrangements and simultaneous manipulation of multiple particles \cite{Melde2016N, Marzo2015NC, Marzo2019PNAS, Hirayama2019N}, see Figs.~\ref{fig:overview}(c), \ref{fig:displays}(b) and \ref{fig:tweezers}(g-j). Notably, holographic acoustic manipulation via phase-gradient, i.e., using the radiation-pressure force ${\bf F}^{\rm press}$, has also been achieved \cite{Melde2016N}, Fig.~\ref{fig:tweezers}(i-j).

Finally, acoustic manipulation is often accompanied by {\it acoustic streaming}: steady fluid flows induced by the dissipation of sound waves \cite{Lighthill1978}. Such flows themselves can serve as an efficient tool for acoustofluidic manipulation and acoustic-streaming tweezers \cite{Ahmed2016NC, Marmottant2004PNAS, Hashmi2012LC, Ozcelik2018NM, Meng2019JPD, Hossein2023BR, Ding2013LC, Baudoin2020AR}.   

{Numerous applications of acoustic trapping and tweezers partially parallel those of optical tweezers, including the manipulation of particles, microfluids, biological cells, and living organisms \cite{Ozcelik2018NM, Dholakia2020NRP, Evander2012LC, Lim2024RPP, Meng2019JPD}, noninvasive manipulation in living bodies \cite{Ghanem2020PNAS, Yang2023NC, Jooss2022SA}, holographic tweezers and volumetric displays \cite{Marzo2015NC, Marzo2019PNAS, Hirayama2019N}, contactless manipulation and probing of extraterrestrial materials \cite{Ferretti2024AJ}, and acoustic force spectroscopy \cite{Sitters2015NM}.}

\subsection{Torque experiments}

Here we briefly discuss the main experiments involving optical and acoustic torques of different kinds, schematized in Fig.~\ref{fig:types_of_torque}. 

\subsubsection{Optics}

The optical torque \eqref{eq:T_tot} originating from the {\it absorption} of light carrying spin AM by an isotropic particle was measured in \cite{Friese1998OL} for an absorbing particle trapped in the focus of an elliptically polarized Gaussian beam. When a particle is trapped in the center of a vortex beam carrying both spin and OAM, the torque acquires an additional orbital contribution due to the absorption of the OAM \cite{Friese1996PRA, Simpson1997OL}. Furthermore, macroscopic `Beth-type' experiments directly measuring spin-absorption and OAM-absorption torques were reported in \cite{Delannoy2005APL, Emile2018AP, Emile2014PRL}.

The {\it anisotropy}-induced optical torques \eqref{eq:T_aniz} and \eqref{eq:T_align} were first predicted theoretically and subsequently measured experimentally in early landmark works \cite{Sadowsky1899, Poynting1909, Beth1935, Beth1936, Holbourn1936} (see Section~\ref{sec:historical}). A breakthrough was achieved in \cite{Friese1998Nat}, which demonstrated and measured both the {\it aligning} and {\it continuous} (spin-conversion) torques for laser-trapped {\it birefringent} microparticles, see Fig.~\ref{fig:torque_exp}(b). Importantly, for small particles, the effective anisotropy, characterized by the anisotropic polarizability tensor $\ten{\alpha}$, may originate not only from the material birefringence but also from the {\it anisotropic shape} of the particle: {see microwave experiment \cite{Allen1966AJP} with a dipolar antenna}. Various anisotropic-shaped particles, such as micro- and nano-rods, trapped and rapidly rotated in focused laser beams, are currently attracting growing attention in optomechanics and levitodynamics \cite{Hoang2016PRL, Ahn2018PRL, Rashid2018PRL, Kuhn2015NL, Kuhn2017NC, Kuhn2017O, Gonzalez-Ballestero2021S}. In such systems, optical torques often combine several mechanisms, including absorption and anisotropy \cite{Reimann2018PRL, Monteiro2018PRA}. 

Optical torques also appear for various complex-shaped objects. In every case, these torques originate from the imbalance between the angular momenta of the incident and scattered fields. Examples include 2D chiral-shaped particles \cite{Higurashi1997JAP}, cylindrical-lens OAM converters \cite{Beijersbergen2005SPIE}, and q-plates converting spin to OAM \cite{Hakobyan2014NP, Magallanes2018NP} [see Fig.~\ref{fig:types_of_torque}(e)]. A broader overview of macroscopic torque experiments is provided in \cite{Brasselet2023AP}.

{The most recent optical-torque experiments enable full 3D angular control and rotation of optically trapped particles \cite{Wu2025NC, Zhu2023PR}.}

\subsubsection{Acoustics}

The acoustic spin-absorption torque \eqref{eq:T_tot} was described and observed only recently \cite{Shi2019, Toftul2019PRL}. This is because only in recent years was it recognized that structured sound-wave fields can exhibit nontrivial polarization and spin properties, Eq.~\eqref{eq:SA} \cite{Shi2019, Bliokh2019b, Bliokh2025CP}. The presence of such spatially varying spin density in two interfering orthogonal plane sound waves was experimentally confirmed in \cite{Shi2019} by measuring the acoustic torque on a resonant-absorbing `meta-atom', Fig.~\ref{fig:torque_exp}(a) (cf. earlier theoretical considerations in \cite{Busse1981, Zhang2014JASA}).

While it had long been believed that sound waves carry no spin, numerous experiments demonstrated the OAM of acoustic vortex beams by measuring torques on macroscopic absorbing objects \cite{Volke-Sepulveda2008PRL, Demore2012PRL, Anhauser2012PRL, Skeldon2008NJP} and on small trapped particles \cite{Baresch2018PRL, Zhang2022PRA, Marzo2018PRL} (see also theory in \cite{Zhang2011PRE_II}).

Anisotropy-induced torques \eqref{eq:T_aniz} and \eqref{eq:T_align} on anisotropic-shaped particles are expected to arise in acoustics in close analogy with optics \cite{Smagin2024PRAppl, Fan2008JASA, Lima2020PRA}. 
Although there have been experiments on acoustofluidic torques and rotations of small floating anisotropic particles \cite{Schwarz2013JASA, Schwarz2015MN, Bernard2017LC}, these results still lack a consistent theoretical interpretation based on acoustic spin. 

Finally, acoustic torques on macroscopic spiral plates that generate OAM in the scattered sound, Fig.~\ref{fig:types_of_torque}(e), were measured in \cite{Sanchez-Padilla2019PRL, Wunenburger2015NJP, Sanchez-Padilla2024APR}, see Fig.~\ref{fig:torque_exp}(c).

\subsection{Optical chiral-particle sorting}
\label{sec:sorting}

An important example involving {\it chiral} particles and forces is the separation of enantiomers, i.e., particles that are identical except for their handedness. Molecular chirality plays a key role in pharmacology, since different enantiomers often produce markedly different biological effects \cite{Challener2017, Sanganyado2017WR}. Optical chiral sorting could be a valuable alternative to the current state-of-the-art chemical separation methods, with the advantage of being independent of the specific chemical structure of the molecules. 

Experimental demonstrations of chiral manipulation have so far been achieved primarily for chiral microparticles with sizes comparable to or larger than the wavelength \cite{Tkachenko2014NC, Tkachenko2014NC_II, Tkachenko2013PRL, Kravets2019PRA, Kravets2019PRL, Zhao2017NN, Shi2020LSA}, and more recently for chiral gold nanoparticles \cite{Yamanishi2022SA}. Although such particles are beyond the range of validity of the dipole approximation used in Eq.~(\ref{eq:F_chiral}), this equation perfectly captures the main features of the observed dynamics. 

\begin{figure}
\centering
\includegraphics[width=0.9\linewidth]{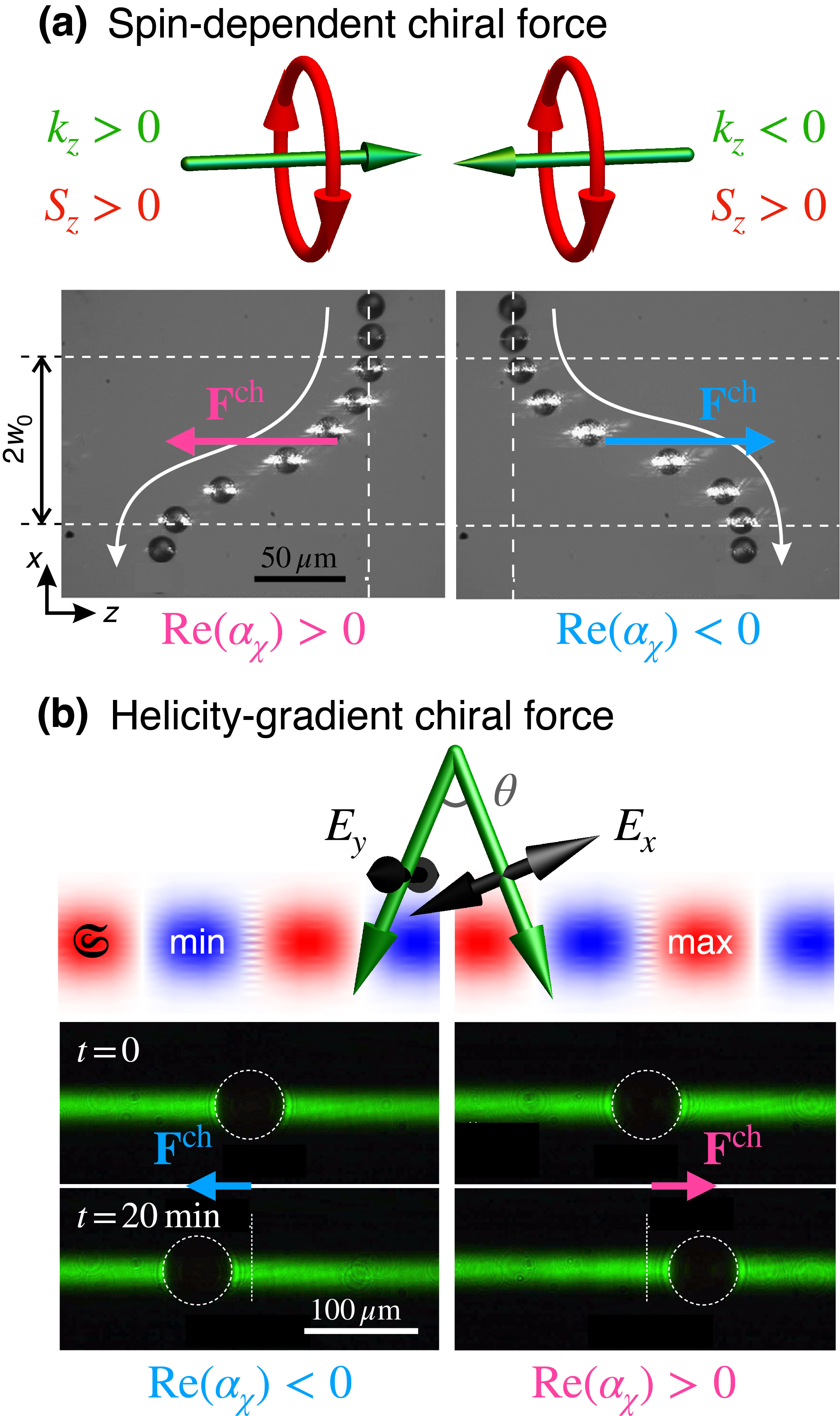}
\caption{Experiments on chiral optical forces. (a) Optofluidic sorting of chiral particles in a standing wave with zero momentum but non-zero spin density. Adapted from \cite{Tkachenko2014NC}. (b) Demonstration of the helicity-gradient chiral force using a 1D helicity-density grating formed by the interference of two slightly noncollinear laser beams with orthogonal linear polarizations. Adapted from \cite{Kravets2019PRL}.}
\label{fig:chiral}
\end{figure}

In particular, the separation in \cite{Tkachenko2014NC} employed two circularly-polarized counter-propagating beams with opposite helicities (i.e., the same spin direction), Fig.~\ref{fig:chiral}(a). In this configuration, the non-chiral radiation-pressure and recoil forces vanish, whereas the  chiral radiation-pressure and recoil forces, Eqs.~\eqref{eq:F_chiral} and \eqref{eq:P_chiral}, act along the spin density vector ${\bf S}$.
This produces opposite motions of the right- and left-handed enantiomers. {For ${\rm Im}(\alpha_\chi)\simeq {\rm Im}(\alpha_e)\simeq 0$, the direction of the chiral force is controlled by ${\rm Re}(\alpha_\chi)$.} Note that the non-chiral gradient forces, arising from the standing-wave intensity modulation, can be neglected in this case, since the particles are much larger than the standing-wave period. 
However, this method cannot be applied to much smaller particles, such as small pharmaceutical molecules, because they would be trapped by the gradient forces at the intensity maxima or minima of the standing wave \cite{Zheng2020OL}. 

Next, the experiments \cite{Kravets2019PRL, Kravets2019PRA} demonstrated the separation of chiral liquid-crystal microspheres using the chiral helicity-gradient force in Eq.~(\ref{eq:F_chiral}), see Fig.~\ref{fig:chiral}(b). Specifically, a sinusoidal helicity-density distribution $\mathfrak{S}(x) \propto \sin (2\pi x/ \Lambda)$, with $\Lambda \gg \lambda$ was generated by interfering two slightly non-collinear plane waves with orthogonal linear polarizations \cite{Cameron2014NJP}. As in the previous example, the $x$-components of non-chiral optical forces vanish in this field, while the chiral gradient force, {proportional to ${\rm Re}(\alpha_\chi)$,} drives the right- and left-handed enantiomers in opposite $x$-directions.
The helicity-gradient chiral force acting on chiral nanoparticles was also explored in \cite{Yamanishi2022SA}.

Several theoretical works have investigated transverse chiral-sorting forces acting on particles near interfaces between different media \cite{Wang2014NC, Hayat2015PNAS, Zhang2017ACS}. One such scenario was realized experimentally in \cite{Shi2020LSA} for chiral particles floating on the water surface, which essentially employed the interference between incident and reflected fields. 

More recent proposals for optical chiral sorting involve near-field configurations around optical waveguides \cite{Golat2024PRR, Martinez-Romeu2024PR, Liu2023APL}, as well as a strong effect achieved by {first orienting the molecules} using a static electric field, %
and {then sorting them with} an optical standing wave \cite{Cameron2023NJP, Cameron2024arXiv}. {The latter approach operates entirely within the electric-dipole regime and relies on nonlinear effects, making it fundamentally different from the other methods discussed here  (which are based on electric-magnetic dipole coupling $\alpha_\chi$).} 

Despite significant theoretical and experimental advances in the study of chiral radiation forces, {the} implementation of optical chiral sorting of molecules remains an open challenge. 

\subsection{Acoustofluidic sorting of biological cells}
\label{subsec:biosorting}

One of the most important applications of acoustic forces is acoustofluidic sorting of biological cells, see reviews \cite{Ozcelik2018NM, Hossein2023BR, Meng2019JPD, Laurell2007CSR, Wu2019MN, Lenshof2012LC, Ding2013LC}. 
Acoustofluidics is an interdisciplinary research area that combines acoustics with microfluidics, providing a non-destructive approach to separating biological particles, cells, and microorganisms. 

\begin{figure}
\centering
\includegraphics[width=0.95\linewidth]{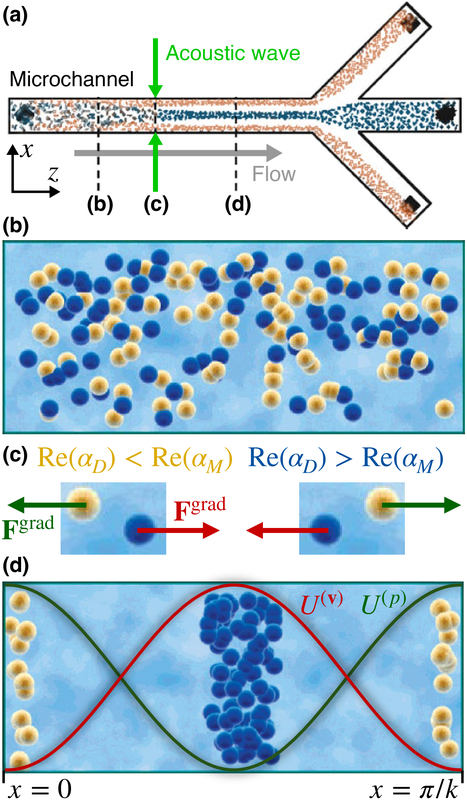}
\caption{Acoustofluidic sorting of biological cells, see also Fig.~\ref{fig:overview}(d). (a) Top view of a microfluidic channel with a flowing mixture of two types of biological cells, subjected to a transverse standing acoustic wave. Adapted from \cite{Laurell2007CSR}. (b) Cross-section of the microchannel in a region without acoustic wave. (c,d) Cross-section of the microchannel in a region with the standing acoustic wave. Panel (d) shows the intensity profiles of the pressure field $U^{(p)}(x)$ and velocity field $U^{({\bf v})}(x)$, while panel (c) indicates the corresponding acoustic gradient forces acting on two types of cells with polarizabilities $\Re(\alpha_D) > \Re(\alpha_M)$ and $\Re(\alpha_D) < \Re(\alpha_M)$. These two cell populations are separated by the gradient forces and become trapped at the nodes of the pressure and velocity fields, respectively. Adapted from \cite{Petersson2005AC}.}
\label{fig:acousto-fluidic}
\end{figure}

Acoustofluidic separation  is typically realized by driving a liquid mixture of cells through a microchannel while introducing ultrasound waves generated by piezoelectric transducers.
The main method, shown in Fig.~\ref{fig:acousto-fluidic}, relies on the excitation of a {\it standing acoustic wave} across the microchannel \cite{Weiser1984, Johnson1995SepTechnol, Yasuda1995JpnJApplPhys, Petersson2005AC,Petersson2005LC,Nilsson2004LC,Ai2013AC}. This standing wave can be either bulk or surface (i.e., evanescent along the third direction, orthogonal to both the microchannel and the wavevectors). In both cases, the $x$-dependence of the acoustic field is described by Eq.~\eqref{eq:standing_A}, with the microchannel width corresponding to half the wavelength, $x\in (0, \pi/k)$. Such a field exhibits a pressure-field node and a velocity-field maximum  at the center of the channel, as shown in Fig.~\ref{fig:acousto-fluidic}(d). 

As discussed in Section~\ref{sec:tweezers_a}, the main acoustic force in this configuration is the gradient force ${\bf F}^{\rm grad}$ in Eq.~\eqref{eq:F}, because the radiation-pressure force vanishes. This force has two contributions: from gradients of the pressure-field and velocity-field intensities, weighted by the monopole and dipole polarizabilities of the particle, respectively. Since the pressure and velocity intensity gradients have opposite directions in the standing-wave field \eqref{eq:standing_A}, the two contributions may also oppose each other. The resulting total force \eqref{eq:F} takes the form
\begin{equation}
\label{eq:F_axial}
F_x^{\rm grad} = k\, U_A \Re(\alpha_D-\alpha_M) \sin(2kx)\,,
\end{equation}
where $U_A$ is the acoustic energy density \eqref{eq:UA}. Thus, the force direction is determined by the balance between the monopole and dipole polarizabilities of the cell. The normalized parameter $\Re(\alpha_D-\alpha_M)$ divided by the cell volume is often referred to as the acoustic contrast factor $\phi$. 

If two types of cells are present in the microchannel, one with $\Re(\alpha_D) > \Re(\alpha_M)$ and the other with $\Re(\alpha_D) < \Re(\alpha_M)$, they will experience gradient forces of opposite signs.
As a result, one type is trapped near the center of the channel, while the other is pushed toward the edges, see Fig.\ref{fig:acousto-fluidic}.
After this separation, the two types of cells can be directed into different reservoirs by branching the central and side parts of the microchannel into three outlets. Since the polarizabilities $\alpha_{M,D}$ depend on multiple parameters (cell size, density, compressibility, acoustic frequency, and fluid properties, see Table~\ref{tab:polarizabilities}), this method can be tuned to achieve efficient sorting of virtually any two types of cells with distinct physical characteristics.

\begin{figure*}
\centering
\includegraphics[width=\linewidth]{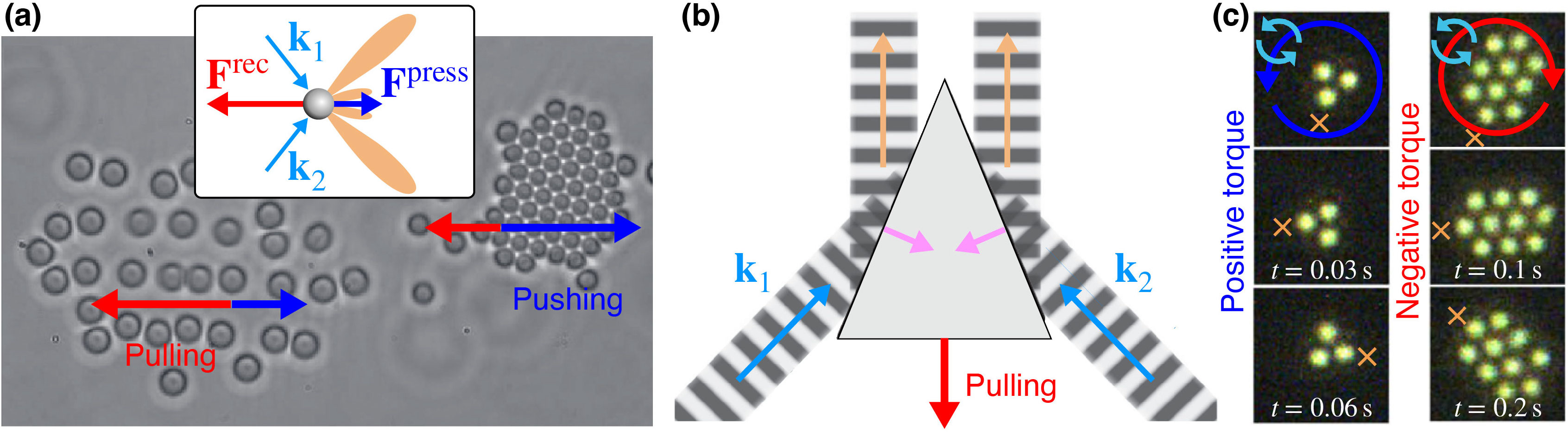}
\caption{Optical pulling forces and negative torques in optics and acoustics. (a) Optical pulling force produced by the recoil force on Mie particles (larger, of radius $1\,{\rm \mu m}$) in a two-wave interference field. For smaller particles (of radius $0.8\,{\rm \mu m}$), the pushing radiation-pressure force dominates. The inset shows schematics with the incident wavevectors, scattering diagram, radiation-pressure force and recoil force. Adapted from \cite{Brzobohaty2013NP}. (b) Acoustic pulling force of similar origin for a macroscopic triangular object. Adapted from \cite{Demore2014PRL}. (c) Observation of negative optical torque acting on clusters of nanoparticles. A circularly-polarized incident wave (polarization is shown in the upper left corners) induces opposite rotations for clusters of three particles and ten particles. In the latter case, the rotation direction is opposite to the polarization rotation. Adapted from \cite{Han2018NC}.}
\label{fig:pulling_exp}
\end{figure*}

There are other approaches to acoustofluidic sorting. For example, traveling waves which can produce significant recoil forces for Mie particles with $ka \gtrsim 2\pi$ \cite{Skowronek2013AC,Devendran2016RSCAdv,Destgeer2017RSCAdv,Ma2016AC,Destgeer2015AC,Collins2016LC,Destgeer2014LC,Destgeer2013LC}. 
There are also additional effects which must be taken into account and can be employed for manipulation of bio-particles. These include: {secondary radiation forces} arising from scattering off neighboring particles, bubbles, and surfaces; acoustic streaming; and the Stokes drag force associated with particle motion relative to the viscous fluid \cite{Ahmed2016NC, Marmottant2004PNAS, Hashmi2012LC, Ozcelik2018NM, Meng2019JPD, Hossein2023BR, Ding2013LC, Baudoin2020AR, Wu2019MN}.

\subsection{Pulling forces}
\label{sec:pulling}

One of the most intriguing examples of nontrivial radiation forces is the `{\it pulling}' or `negative' force, which draws a particle toward the source of the incident field. Wave beams that drag particles against their propagation direction are often referred to as `tractor beams'. This phenomenon has recently attracted considerable attention, in both optics \cite{Chen2011NP, Sukhov2011PRL, Novitsky2011PRL, Ruffner2012PRL_II, Brzobohaty2013NP, Kajorndejnukul2013NP, Shvedov2014NP, Canaguier-Durand2015PRA, Fernandes2015PRA, Petrov2016LPR, Zhu2018PRL, Li2019SA, Li2020PRL, Horodynski2023PRA} (for reviews, see \cite{Dogariu2013NP, Ding2019AP, Li2020AOP} and acoustics \cite{Marston2006JASA, Zhang2011PRE, Xu2012EPL, Demore2014PRL, Mitri2015EPL, Mitri2015JAP, Meng2020PRA, Wang2021PRA}. 

The most established mechanism for generating pulling forces relies on strong forward scattering, such that the backward-directed {\it recoil} force (Sec.~\ref{subsec:recoil}) dominates over the forward radiation-pressure force \cite{Brzobohaty2013NP, Chen2011NP, Novitsky2011PRL, Sukhov2011PRL, Zhang2011PRE, Demore2014PRL, Meng2020PRA}. This situation can occur in the Mie resonant regime, where forward scattering is enhanced by interference between the excited multipoles (see Figs.~\ref{fig:recoil} and \ref{fig:higher_multipoles_recoil}), as well as in the geometrical optics regime of large objects. For example, Fig.~\ref{fig:pulling_exp}(a) shows optical pulling and pushing forces depending on the size of Mie particles \cite{Brzobohaty2013NP}, while Fig.~\ref{fig:pulling_exp}(b) illustrates an acoustic pulling force acting on a macroscopic prism-like target \cite{Demore2014PRL}.  
An important limitation of this approach is that it essentially requires two or more (e.g., in Bessel beams) non-collinear plane waves in the incident field to produce a pulling force on a passive object in a homogeneous medium.

There are several other methods to generate pulling radiation forces. First, {\it active} particles with gain are characterized by a negative imaginary part of the polarizability, $\Im (\alpha) < 0$. This directly leads to negative {\it radiation-pressure} forces and torques in Eqs.~\eqref{eq:F} and \eqref{eq:T}. Such pulling forces have been considered in both optics \cite{Mizrahi2010OL, Webb2011PRE, Gao2017PRA, Alaee2018PRA} and acoustics \cite{Rajabi2016JSV, Rajabi2018U, Meng2020PRA}. 
Second, various temporal modulations of the incident beam (which make it {\it poly-chromatic}) can also produce pulling forces \cite{Ruffner2012PRL_II, Lepeshov2020O, Mitri2015JAP}. These forces can be interpreted as {\it gradient} forces arising from longitudinal intensity gradients, which may act opposite to the main wavevector \cite{Ruffner2012PRL_II}.
Third, pulling forces can occur for particles located near {\it interfaces} between two media \cite{Kajorndejnukul2013NP, Mansuripur2013NP, Qiu2015LSA, Petrov2016LPR,  Wang2021PRA, Wang2024NC}. In the dipole approximations, these forces can be described using the formalism outlined in Section~\ref{sec:interfaces}.

Other approaches to pulling forces involve complex structured media \cite{Li2020PRL, Zhu2018PRL}, nontrivial structured fields with opposite-sign projections of the energy flow and canonical momentum density \cite{Lee2010OE}, chirality \cite{Canaguier-Durand2015PRA, Fernandes2015PRA}, photophoretic forces \cite{Shvedov2014NP}, and complex wavefront shaping (Section~\ref{sec:wavefron_shaping}) \cite{Horodynski2023PRA}.

Rotational counterparts of pulling forces are {\it negative torques}, which act opposite to the spin of the incident field (at least when projected on a chosen axis) \cite{Hakobyan2014NP, Chen2014SR, Han2018NC, Diniz2019OE, Shi2022Jul, Qi2022NL}. For an isotropic sphere in a homogeneous medium, negative torque can be realized only for {\it active} particles with $\Im (\alpha) <0$. For passive particles, one must break the spherical symmetry to generate higher angular momentum in the scattered field, resulting in a negative {\it recoil} torque \cite{Hakobyan2014NP, Chen2014SR, Han2018NC}. For instance, an optical q-plate that transforms an incident circularly-polarized wave into a vortex beam with higher angular momentum experiences a corresponding negative recoil torque \cite{Hakobyan2014NP} (see Fig.~\ref{fig:types_of_torque}(e)). Another example of negative optical torque for clusters of nanoparticles \cite{Han2018NC} is shown in Fig.~\ref{fig:torque_exp}(c). Finally, for finite-size particles, torques can arise not only from the spin of the incident field but also from gradients of the radiation forces, which push different sides of the particle differently. Such `orbital' torques can compete with the spin-induced torques and may result in negative net torques. 

\subsection{{Artificial intelligence advances}}

{Modern optical and acoustic manipulation presents a number of challenges. Recent advances in artificial intelligence and machine learning have proven effective in several areas \cite{Ciarlo2024Nanophotonics}, including: reducing computational costs by predicting particle trajectories and associated forces \cite{Lenton2020MachLearn, BronteCiriza2023ACSPhot, Wu2024Sensors};
enhancing force-field calibration for optical tweezers \cite{Argun2020APR};
Improving tracking of trapped particles \cite{Helgadottir2019O, Huan2024SensActuatorsB};
real-time  control of optical tweezers \cite{Aggarwal2010RevSciInstrum};
algorithmic smart design of the environment \cite{Li2021AOM, Li2019OL, Raymond2020SR};
designing the trapping field \cite{Shen2025ACSPhot, Kondo2022JpnJApplPhys, Wu2024Sensors, Wang2023SpringerLink}
or the particle geometry \cite{Igoshin2025OL}.}
{Machine learning has also enabled several achievements that were previously out of reach. Notable examples include: the inverse design approaches mentioned above, large-scale defect-free assembly \cite{Lee2024CurrApplPhys,Lin2025PRL,Ren2024APLQuantum},
and simultaneous manipulation of many-body systems \cite{Schrage2023AdvMaterTechnol, Medany2025NatMachIntell}.}

\section{Conclusions}
\label{sec:conclusion}

In this review, we have aimed to provide a comprehensive and {\it unifying} picture of optical and acoustic radiation forces and torques. Based on numerous earlier studies, we have formulated a universal approach that naturally encompasses the main results in both optics and acoustics, while allowing straightforward application and cross-application to a wide range of problems. We have also highlighted key examples and application areas where radiation forces and torques play a central role. We believe that 
such a universal framework is illuminating and can stimulate new insights, as well as foster exchange of fruitful ideas across different fields. 

Naturally, optical and acoustic forces are central to a number of applied domains (e.g., optical tweezers, acoustofluidics, etc.), where understanding fine details is essential for addressing specific systems. 
For such details and system-specific phenomena, we refer the reader to a number of dedicated reviews mentioned in Section~\ref{sec:about}. Our goal here has been to provide the `big picture’ of the main phenomena. We have also aimed to make this review {\it pedagogical}, with technical appendices included where necessary, so that newcomers to the field can readily trace and reproduce the main results. 

Starting with the general case of an arbitrary particle in an arbitrary inhomogeneous (yet monochromatic) wave field, we have examined a number of important particle types (Rayleigh, Mie, isotropic, anisotropic, chiral, etc.) and field configurations (vortex, standing, evanescent, etc.), as well as a range of specific effects, including optical and acoustic trapping, optical chiral sorting, acoustofluidic sorting, and pulling forces.

The essence of our approach is based on three constituents: 
\begin{enumerate}
\item[(A)]  Coherent use of the fundamental dynamical characteristics of monochromatic wave fields: canonical momentum, spin angular momentum, and energy densities. These quantities share universal mathematical forms across different types of waves \cite{Bliokh2025CP}. 
\item[(B)] Characterization of particles in terms of their monopole, dipole, and higher-multipole polarizabilities, which determine the corresponding multipole moments in the scattered fields.
\item[(C)] Calculations of radiation forces and torques via general integrals of the momentum and angular-momentum fluxes. In the most important basic cases, the results are expressed directly in terms (A) and (B). 
\end{enumerate} 
As described in Section~\ref{sec:intro}, the connection between radiation forces / torques and the momentum / angular momentum of the incident field has been recognized since the earliest historical studies. Indeed, the mechanical action of wave fields on matter provides the most direct way to observe and measure the corresponding dynamical properties of waves.

Importantly, the framework provided in this review greatly {\it simplifies} the analysis and interpretation of many results in the existing literature, and will undoubtedly facilitate future research. To illustrate this, we next provide a few explicit examples.

First, analyzing the longitudinal and azimuthal components of the Poynting vector in a polarized optical Bessel beam with a vortex charge $\ell > 0$, \cite{Novitsky2007JOSA} found that these components can be {negative} in some areas and interpreted this as a ``{\it negative propagation}'' within the beam, which could lead to negative scattering (i.e., radiation-pressure, in our terminology) forces. However, our formalism clearly shows that the radiation-pressure forces are determined by the {\it canonical} momentum density rather than the Poynting vector. In such Bessel beams, the longitudinal and azimuthal components of the canonical momentum are always positive  \cite{Ghosh2024JOSA}. True ``negative propagation'', i.e., {\it negative phase gradients}, leading to negative radiation-pressure forces, can occur only in superposition of Bessel beams or different modes \cite{Bracken1994JPA, Berry2010JPA, Bliokh2013NJP_II, Eliezer2020O, Daniel2022NJP, Ghosh2023O}. 

Second, analyzing acoustic torque on isotropic particles in analogous sound Bessel beams, \cite{Zhang2018PRA} found that the longitudinal component of the torque can surprisingly be negative, despite the positive sign of the OAM number $\ell$. Our formalism makes this result more transparent: the torque on a small isotropic particle is determined by the {\it spin} density, rather than by the OAM, and the longitudinal component of the spin density in acoustic Bessel beams is indeed negative in certain regions \cite{Bliokh2019b}. Thus, this finding is no longer so surprising.

Third, the vast literature on acoustic and acoustofluidic forces and torques often employs cumbersome expressions involving the material parameters of specific particles and specific field distributions. Recasting acoustic forces in terms of universal parameters, such as the monopole and dipole polarizabilities of the particle together with fundamental dynamical field characteristics, greatly simplifies the main equations and their interpretation. For instance, the acoustic gradient force used in acoustofluidic sorting of molecules with a standing ultrasound wave (Section~\ref{subsec:biosorting}) is typically written in its final form \cite{Petersson2005AC, Hossein2023BR, Laurell2007CSR}, which provides little physical intuition and can readily adopt typographical errors \cite{Wu2019MN}. In contrast, presenting the same force in terms of particle polarizabilities and energy-density gradients, as done in our review, elucidates the physical origin of the effect and clarifies the underlying equations \cite{Meng2019JPD}.

\begin{figure}
\centering
\includegraphics[width=0.95\linewidth]{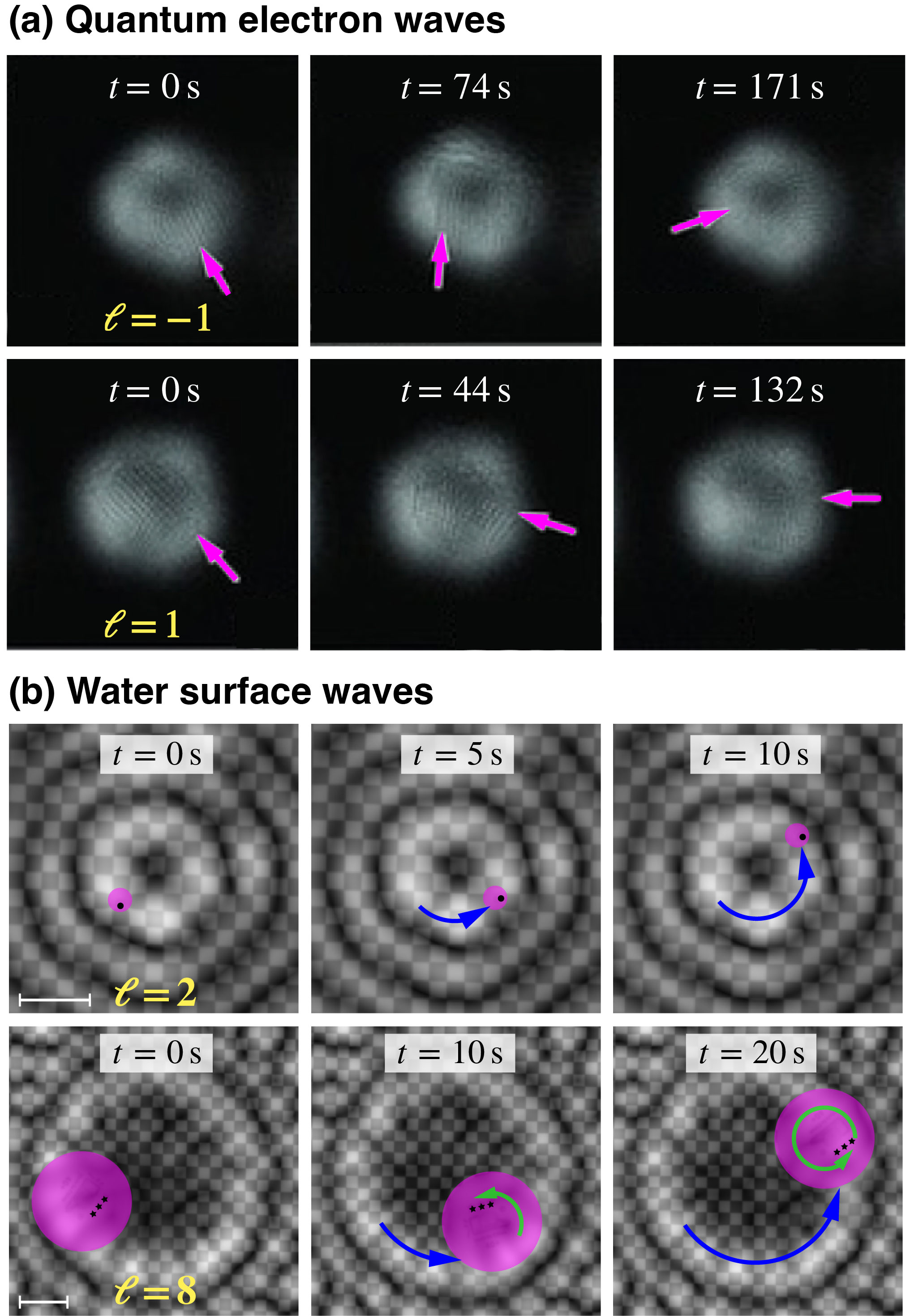}
\caption{Analogues of optical and acoustic radiation forces and torques in other types of waves. (a) Trapping and rotation of a 3~nm gold nanoparticle in quantum-electron vortex beams with $\ell=-1$ and $\ell=1$, produced in a transmission electron microscope. Adapted from \cite{Verbeeck2013AM}. (b) Radial trapping, orbital rotation, and spinning motion of floating particles in Bessel-type vortices with $\ell = 2$ and $\ell = 8$ in water-surface waves (scale bar: 2~cm). Cf. Fig.~\ref{fig:vortex_beams}. Adapted from \cite{Wang2024}.}
\label{fig:conclusion}
\end{figure}

Another advantage of the unifying approach presented in this review is its straightforward extensibility to other types of systems. For example, recent experiments have demonstrated wave-induced forces and torques acting on floating particles in structured {\it water-surface waves} \cite{Wang2024, Falkovich2005N}, Fig.~\ref{fig:conclusion}(b). Within our framework, these forces and torques can be naturally identified as generic radiation-pressure and gradient forces, together with spin-induced torques.
Moreover, similar wave-induced forces appear for nanoparticles interacting with {\it quantum matter waves} \cite{Verbeeck2013AM, Zheng2012NL, Gorlach2017PRL}, Fig.~\ref{fig:conclusion}(a). On a much larger scale, massive astronomical objects interacting with gravitational waves should also experience wave-induced forces and torques.

{Remarkably, the mechanical actions exerted by electron, optical, acoustic, and water waves nicely complement each other across wavelength ranges, and, hence, across the corresponding sizes of manipulated particles \cite{Dholakia2020NRP}, see Table~\ref{tab:sizes}. 
These different types of waves enable manipulation of objects spanning from the atomic angstrom scale to the macroscopic meter scale.
Furthermore, the water-wave regime extends to colossal wavelengths $\sim\!10,000$~km in ocean tidal waves, which can exert forces and torques on entire islands of size $\sim\!1,000$~km \cite{Domina2025}.
For any kind of wave, the typical magnitude of the radiation force can be roughly estimated as the product of the wave momentum density, the particle's geometric cross-section (i.e., its squared linear size), and the wave speed, while the typical torque magnitude is given by the same product divided by the wavenumber. These estimations follow from considering the mechanical action of a circularly polarized plane wave incident on a perfectly absorbing particle.}

\begin{table}[]
\caption{{Typical sizes of manipulated particles for different types of waves.}}
\begin{tabular}{p{0.47\linewidth}>{\centering\arraybackslash}p{0.5\linewidth}}
\hline \hline 
\white{$\dfrac{1}{1}$} Wave type  & Particle size  \\ [0.5em]  \hline
Electron waves \white{$\dfrac{1}{1}$}  
  & $\sim \SI{1}{\angstrom}$ -- $\SI{10}{\nm}$  \\ [0.5em]
Optics 
  & $\sim \SI{10}{\nm}$ -- $\SI{10}{\um}$  \\ [0.5em]
Acoustics        
  & $\sim \SI{1}{\um}$ -- $\SI{10}{\mm}$  \\ [0.5em]
Water waves      
  & $\sim \SI{1}{\mm}$ -- $\SI{1}{\m}$    \\  [0.5em]
\hline \hline
\end{tabular}
\label{tab:sizes}
\end{table}

Naturally, we could not cover {\it all} radiation force and torque phenomena. An important limitation, assumed throughout this review, is the use of {monochromatic} fields. 
Radiation forces and torques in {\it polychromatic} waves have so far been studied mainly for the simplest cases of bi-chromatic fields \cite{Grimm1990PRL, Gupta1993PRL, Williams1999PRA, Cashen2003JOSA, Partlow2004PRL, Kozyryev2018PRL, Galica2018PRA, Wenz2020PRR, Silva2006PRL,Morrell2024PRE} and fields with slow time modulation of the amplitude \cite{Ruffner2012PRL_II, Lepeshov2020O, Mitri2015JAP}. It has been shown that such fields can offer considerable advantages and new functionalities compared to the monochromatic case.
Recently, there has been a surge of interest in complex {\it space-time structured} polychromatic waves \cite{Shen2023JO, Yessenov2022AOP}, which may bring about novel opto-mechanical and acousto-mechanical phenomena. 
{We have also limited our consideration to \textit{linear} wave scattering. Nonlinear regimes can offer additional phenomena, such as nonlinearity-induced optical torque \cite{Toftul2023PRL} and bistable trapping effects \cite{Jiang2010NP,Zhang2018NL,Mirzaei-Ghormish2025PRA}.}
{Another emergent direction in optical and acoustic manipulation is the study of complex nonlinear and non-Hermitian dynamics of particles, which plays a crucial role in the stability of trapped particles, particularly in many-body systems \cite{Li2021NC, Rieser2022S, Li2024NC, Liska2024NP, Reisenbauer2024NP}.}

We hope that this review will contribute to the further development of fascinating areas of fundamental and applied science involving the mechanical action of diverse wave fields.

\begin{acknowledgments}

We thank  Etienne Brasselet, Xiao Li, Manuel Nieto-Vesperinas, and Mikhail Petrov for helpful remarks.
This work was supported by 
the Australian Research Council (Grant No. DP210101292),
the International Technology Center Indo-Pacific (ITC IPAC) via Army Research Office (contract FA520923C0023),
EIC-Pathfinder-CHIRALFORCE (101046961) which is funded by Innovate UK Horizon Europe Guarantee (UKRI project 10045438),
Ikerbasque (Basque Foundation of Science), 
Marie Skłodowska-Curie COFUND Programme of the European Commission (project HORIZON-MSCA-2022-COFUND-101126600-SmartBRAIN3), 
the International Research Agendas Programme (IRAP) of the Foundation for Polish Science co-financed by the European Union under the European Regional Development Fund and Teaming Horizon 2020 program of the European Commission [ENSEMBLE3 Project No. MAB/2020/14], 
and the project of the Ministry of Science and Higher Education (Poland) ``Support for the activities of Centers of Excellence established in Poland under the Horizon 2020 program'' [contract MEiN/2023/DIR/3797],
Nippon Telegraph and Telephone Corporation (NTT) Research,
the Japan Science and Technology Agency (JST)
[via the CREST Quantum Frontiers program Grant No. JPMJCR24I2],
the Quantum Leap Flagship Program (Q-LEAP), and the Moonshot R\&D Grant Number JPMJMS2061],
and the Office of Naval Research (ONR) Global (via Grant No. N62909-23-1-2074).

\end{acknowledgments}

\bibliography{refs.bib}

\newpage

\onecolumngrid
\newpage
\appendix

\section{Fields Produced by Monopoles and Dipoles}
\label{app:multipoles}

Due to ambiguity in the polarizability definitions in the literature, we provide explicit expressions for the radiated fields from the first multipoles: monopole/dipole in acoustics and electric/magnetic dipoles in electromagnetism.

\subsection{Electromagnetism}

The electromagnetic field radiated by a monochromatic point electric dipole $\vb{e}$ is 
\begin{align}
\sqrt{\varepsilon}\, \vb{E}_{e} &= \frac{k^3}{4\pi \sqrt{\varepsilon}} \left\{ \left(\vb{n} \times {\vb{e}}\right) \times \vb{n}  + \left[ 3 \vb{n} (\vb{n}\cdot {\vb{e}}) - {\vb{e}}\right]\! \left[ \frac{1}{(kr)^2} - \frac{i}{kr}\right] \right\} \frac{e^{i kr}}{kr},  \nonumber \\
\sqrt{\mu}\,\vb{H}_{e} &= \frac{k^3}{4\pi {\sqrt{\varepsilon}}} (\vb{n} \times {\vb{e}}) \left( 1 + \frac{i}{kr}\right) \frac{e^{ikr}}{kr} ,
\label{eq:e-dipole}
\end{align}
where ${\bf n} = {\bf r}/r$.
In the near field, $kr \ll 1$, this yields
\begin{equation}
\sqrt{\varepsilon}\, \vb{E}_{e} \simeq 
\frac{k^3}{4\pi \sqrt{\varepsilon}} 
\left[ 3 \vb{n} (\vb{n}\cdot {\vb{e}}) - {\vb{e}} \right] \frac{1}{(kr)^3}, \qquad 
\sqrt{\mu}\,\vb{H}_{e} \simeq  \frac{\iu k^3}{4\pi \sqrt{\varepsilon}} (\vb{n} \times {\vb{e}}) \frac{1}{(kr)^2}.
\label{eq:electric_dipole_NF}
\end{equation}
In the far field, $kr \gg 1$, we have
\begin{equation}
\sqrt{\varepsilon}\,\vb{E}_{e} \simeq  \sqrt{\mu}\, \vb{H}_{e} \times \vb{n}
= \frac{k^3}{4\pi \sqrt{\varepsilon}} \left[ (\vb{n} \times {\vb{e}} )\times  \vb{n} \right] \frac{e^{ikr}}{kr}, \qquad 
\sqrt{\mu}\,\vb{H}_{e}  \simeq \frac{k^3}{4\pi \sqrt{\varepsilon}} (\vb{n} \times {\vb{e}})  \frac{e^{ikr}}{kr}.
    \label{eq:electric_dipole_FF}
\end{equation}

In turn, the electromagnetic field of a point magnetic dipole $\vb{m}$ is obtained from Eqs.~\eqref{eq:e-dipole}--\eqref{eq:electric_dipole_FF} via substitution $\sqrt{\varepsilon}\vb{E} \to \sqrt{\mu} \vb{H}$, $\sqrt{\mu} \vb{H} \to -  \sqrt{\varepsilon}\vb{E}$, and $\vb{e} / \sqrt{\varepsilon} \to \vb{m}/\sqrt{\mu}$:
\begin{align}
\sqrt{\varepsilon}\,\vb{E}_{m} &= - \frac{ k^3}{4\pi \sqrt{\mu}} (\vb{n} \times  {\vb{m}}) \left( 1 + \frac{i}{kr}\right) \frac{e^{ikr}}{kr}, \nonumber \\
\sqrt{\mu}\,\vb{H}_{m} &= \frac{k^3}{4\pi \sqrt{\mu}} \left\{ (\vb{n} \times {\vb{m}}) \times \vb{n} + 
\left[3 \vb{n} (\vb{n} \cdot {\vb{m}}) - {\vb{m}}\right] \left[ \frac{1}{(kr)^2} - \frac{i}{kr}\right] \right\} \frac{e^{ikr}}{kr} .
\label{eq:m-dipole}
\end{align}
Note that our definition of the magnetic dipole moment ${\bf m}$ is related to the textbook dipole moment $\tilde{\bf m}$ \cite{jackson1998ClassicalElectrodynamics} as ${\bf m} = \mu \tilde{\bf m}$. This makes all equations nicely symmetric. 
In the near field, $kr \ll 1$:
\begin{equation}
\label{eq:magnetic_dipole_NF}
\sqrt{\varepsilon}\,\vb{E}_{m} \simeq - \frac{\iu k^3}{4\pi \sqrt{\mu}} (\vb{n} \times  {\vb{m}}) \frac{1}{(kr)^2}, \qquad 
\sqrt{\mu}\,\vb{H}_{m}  \simeq \frac{k^3}{4\pi \sqrt{\mu}}  
\left[ 3 \vb{n} (\vb{n} \cdot {\vb{m}}) - {\vb{m}}\right]  \frac{1}{(kr)^3}.
\end{equation}
In the far field, $kr \gg 1$:
\begin{equation}
\label{eq:magnetic_dipole_FF}
\sqrt{\varepsilon}\,\vb{E}_{m} \simeq - \frac{k^3}{4\pi \sqrt{\mu}} (\vb{n} \times  {\vb{m}}) \frac{e^{ikr}}{kr} , \qquad
\sqrt{\mu}\, \vb{H}_{m}  \simeq - \sqrt{\varepsilon}\, \vb{E}_{m} \times \vb{n} = 
\frac{k^3}{4\pi \sqrt{\mu}} \left[ (\vb{n} \times {\vb{m}}) \times \vb{n} \right] \frac{e^{ikr}}{kr} .
\end{equation}

The electromagnetic field produced by a combined point electric and magnetic dipole source located at $\vb{r}_0$ can also be expressed via {\it Green's tensor} \cite{Sersic2011PRB}:
\begin{equation}
    \begin{pmatrix}
        \sqrt{\varepsilon} \, \vb{E} \\ \ \\ 
        \sqrt{\mu} \, \vb{H}
    \end{pmatrix} = 
    \begin{pmatrix}
        (k^2 \ten{I}  + \grad \otimes \grad ) G_0(\vb{r}, \vb{r}_0) & - i k (\ten{I} \times \grad ) G_0(\vb{r}, \vb{r}_0) \\ \ \\ 
         i k (\ten{I} \times \grad ) G_0(\vb{r}, \vb{r}_0) & (k^2 \ten{I}  + \grad \otimes \grad ) G_0(\vb{r}, \vb{r}_0)
    \end{pmatrix} 
    \begin{pmatrix}
        \vb{e}/\sqrt{\varepsilon} \\ \ \\ 
        \vb{m} / \sqrt{\mu}
    \end{pmatrix}
    \label{eq:G_EM}
\end{equation}
where $G_0(\vb{r}, \vb{r}_0) = \eu^{\iu k \abs{\vb{r} - \vb{r}_0}} / (4\pi \abs{\vb{r} - \vb{r}_0})$ is the Green's function of the Helmholtz equation, $\left(\grad^2 + k^2\right) G_0(\vb{r}, \vb{r}_0)  = - \delta(\vb{r} - \vb{r}_0)$, with $\delta(\vb{r} - \vb{r}_0)$ being the Dirac delta function.
Using Green's tensor from Eq.~\eqref{eq:G_EM} one can derive the reciprocity restrictions on the polarizabilities in Table~\ref{tab:particles_overview}~\cite{Sersic2011PRB}.

\subsection{Acoustics}

The acoustic field radiated by a monochromatic monopole $M$ is given by
\begin{align}
\sqrt{\beta}\, p_M &= \frac{k^3}{4 \pi \sqrt{\beta}} M \frac{e^{ikr}}{kr}, \nonumber \\
\sqrt{\rho}\,\vb{v}_{M} &= \frac{k^3 }{4\pi \sqrt{\beta}} \vb{n} M \left( 1 + \frac{i}{kr}\right) \frac{e^{ikr}}{kr}.
\label{eq:M-monopole}
\end{align}
Note that for the sake of symmetry of the main expressions throughout the paper, we define the monopole moment $M$ in a slightly different way as compared to textbooks monopole moment $Q$ \cite{williams1999FourierAcousticsSound}: $Q = - {d}M/{d t} = \iu \omega M$.
In the near field, $kr \ll 1$:
\begin{equation}
    \sqrt{\beta}\, p_M \simeq \frac{k^3}{4 \pi \sqrt{\beta}}  M \frac{1}{kr}, \qquad
    \sqrt{\rho}\,\vb{v}_{M} \simeq \frac{k^3 }{4\pi \sqrt{\beta}}  \vb{n} M  \frac{i}{(kr)^2}.
\end{equation}
In the far field, $kr \gg 1$:
\begin{equation}
    \sqrt{\beta}\, p_M \simeq \frac{k^3}{4 \pi \sqrt{\beta}} M \frac{e^{ikr}}{kr}, \qquad
    \sqrt{\rho}\,\vb{v}_{M} \simeq \frac{k^3 }{4\pi \sqrt{\beta}} \vb{n} M \frac{e^{ikr}}{kr}.
\end{equation}

For a point dipole moment $\vb{D}$, we have
\begin{align}
\sqrt{\beta}\,p_{{D}} &= \frac{  k^3}{4\pi \sqrt{\rho}} (\vb{D} \cdot \vb{n}) \left( 1 + \frac{i}{kr}\right) \frac{e^{ikr}}{kr}, \nonumber \\
\sqrt{\rho}\,\vb{v}_{{D}} &= \frac{k^3}{4\pi \sqrt{\rho} } \left\{  \vb{D} \left[ - \frac{i}{kr} + \frac{1}{(kr)^2}\right] + \vb{n}(\vb{D}\cdot \vb{n}) \left[ 1 + \frac{3i}{kr} - \frac{3}{(kr)^2} \right]\right\} \frac{e^{ikr}}{kr}.
\label{eq:D-dipole}
\end{align}
Our definition of the dipole moment ${\bf D}$ is related to the textbook dipole moment $\tilde{\bf D}$ \cite{williams1999FourierAcousticsSound} as $\vb{D} = - \rho \tilde{\vb{D}}$.
In the near field, $kr \ll 1$: 
\begin{equation}
    \sqrt{\beta}\, p_{{D}}  \simeq \frac{k^3}{4\pi \sqrt{\rho}}  (\vb{D} \cdot \vb{n}) \frac{i}{(kr)^2}, \qquad 
    \sqrt{\rho}\,\vb{v}_{{D}}  \simeq - \frac{3k^3}{4\pi \sqrt{\rho}}    \vb{n}(\vb{D}\cdot \vb{n}) \frac{1}{(kr)^3}.
\end{equation}
In the far field, $kr \gg 1$: 
\begin{equation}
    \sqrt{\beta}\,p_{{D}}  \simeq \frac{  k^3}{4\pi \sqrt{\rho}} (\vb{D} \cdot \vb{n})  \frac{e^{ikr}}{kr}, \qquad
    \sqrt{\rho}\,\vb{v}_{{D}}  \simeq  \frac{k^3}{4\pi\sqrt{\rho}}   \vb{n}(\vb{D}\cdot \vb{n})  \frac{e^{ikr}}{kr}.
\end{equation}

Akin to electromagnetic Eq.~\eqref{eq:G_EM}, the generic acoustic field produced by a combined point monopole and dipole source located at $\vb{r}_0$ can be expressed via Green's function $G_0(\vb{r}, \vb{r}_0)$: 
\begin{equation}
    \begin{pmatrix}
        \sqrt{\beta}\, p \\ \ \\ 
        \sqrt{\rho}\, \vb{v}
    \end{pmatrix} = 
    \begin{pmatrix}
        k^2 G_0(\vb{r}, \vb{r}_0) &  & - \iu k \grad G_0(\vb{r}, \vb{r}_0) \\ \ \\ 
        - \iu k \grad G_0(\vb{r}, \vb{r}_0) & &  - \grad \otimes \grad  G(\vb{r}, \vb{r}_0)
    \end{pmatrix}
    \begin{pmatrix}
        M / \sqrt{\beta} \\ \ \\ 
        \vb{D} / \sqrt{\rho}
    \end{pmatrix}.
\end{equation}
Using Green's tensor in this equation, one can derive the reciprocity restrictions on the polarizabilities, Eqs.~\eqref{eq:polarizabilities_Willis} \cite{Sersic2011PRB, Quan2018PRL}.

\section{Polarizabilities and Mie Theory}
\label{app:polarizabilities}

For a spherical isotropic particle, the scattering problem is described by the electromagnetic Mie theory \cite{bohren1984AbsorbtionScatteringLight} and its acoustic counterpart \cite{blackstock2000FundamentalsPhysicalAcoustics,Yosioka1955,Toftul2019PRL}. The zero and first terms in the Mie series correspond to the monopole and dipole radiation, respectively. Then, comparing these terms with Eqs.~\eqref{eq:e-dipole}, \eqref{eq:m-dipole}, \eqref{eq:M-monopole}, and \eqref{eq:D-dipole}, one can obtain expressions for the corresponding particle's polarizabilities.  
These expressions are summarized in Table~\ref{tab:polarizabilities}. 
For small Rayleigh particles, $ka\ll 1$, one can expand the Mie coefficients in the Taylor series, where the main terms correspond to the {\it static-approximation} polarizabilities (real-valued for lossless particles), whereas the next-order terms allow one to obtain the imaginary {\it radiation corrections} to the main polarizabilities \cite{Sipe1974PRA, Nieto-Vesperinas2010OE, Simpson2010JOSA, Albaladejo2010OE, LeRu2013, Toftul2019PRL}.
Note that for lossless particles with real-valued material parameters and static polarizabilities, the radiation corrections are necessary to satisify the constraints following from Eqs.~\eqref{eq:cross-sections} and \eqref{eq:g} with $\sigma^{\rm ext} = \sigma^{\rm sc}$.

\begin{table}[ht]
\caption{\label{tab:polarizabilities} Dipole and monopole polarizabilities of spherical isotropic Rayleigh particles. Here $a$ is radius of the particle, $a_{\text{EM},n}$ and  $b_{\text{EM},n}$ are the electromagnetic Mie scattering coefficients \cite{bohren1984AbsorbtionScatteringLight}, and $a_{\text{A},n}$ are the acoustic Mie scattering coefficients \cite[SM]{Toftul2019PRL}. Factors $g$, $g_{M}$, and  $g_{D}$ are given by Eq.~\eqref{eq:g}.}
\begin{ruledtabular}
\begin{tabular}{lcccc}
\multirow{2}{*}{\bf Approximation} & \multicolumn{2}{c}{\bf Electromagnetism}                                      & \multicolumn{2}{c}{\bf Acoustics}                             \\
& \multicolumn{1}{c}{$\alpha_e$} & \multicolumn{1}{c}{$\alpha_m$} & \multicolumn{1}{c}{$\alpha_M$} & \multicolumn{1}{c}{$\alpha_D$} \\ [0.4em]  
\hline  \ \\ 
Exact $\alpha^{(\text{Mie})}$ 
& $i g^{-1} a_{\text{EM},1}$
& $i g^{-1} b_{\text{EM},1}$
& $- i g^{-1}_Ma_{\text{A},0}$
& $-i g^{-1}_D a_{\text{A},1}$ \\ [2em]
Static approximation, $\alpha^{(0)}$
& $ 4\pi a^3 \dfrac{\varepsilon_{p} - \varepsilon}{\varepsilon_{p} + 2 \varepsilon} $ 
& $ 4\pi a^3 \dfrac{\mu_{p} - \mu}{\mu_{p} + 2 \mu} $ 
& $4 \pi a^3  \dfrac{\beta_{p}-\beta}{3\beta}$
& $4 \pi  a^3 \dfrac{\rho_{p} - \rho}{2 \rho_{p} + \rho}$   \\ [2em]
With radiation corrections, $\alpha$
& $\dfrac{\alpha^{(0)}_{e}}{1 - i g \alpha^{(0)}_{e}}$
& $\dfrac{\alpha^{(0)}_{m}}{1 - i g \alpha^{(0)}_{m}}$
& $\dfrac{\alpha^{(0)}_{M}}{1 - i g_M^{} \alpha^{(0)}_{M}}$
& $\dfrac{\alpha^{(0)}_{D}}{1 - i g_D^{} \alpha^{(0)}_{D}}$  \\ [1em]
\end{tabular}
\end{ruledtabular}
\end{table}

{Explicitly, the electromagnetic and acoustic Mie coefficients for a sphere of radius $a$ are:
\begin{align}
    a_{\text{EM},n} &= \frac{\sqrt{\bar{\varepsilon}}\, \psi_n(k_{p} a) \psi^{\prime}_n(k a) - \sqrt{\bar{\mu}}\, \psi_n(k a) \psi^{\prime}_n(k_{p} a)}{\sqrt{\bar{\varepsilon}}\, \psi_n(k_{p} a) \xi^{\prime}_n(k a) - \sqrt{\bar{\mu}}\, \xi_n(k a) \psi^{\prime}_n(k_{p} a)}, 
    \qquad 
    b_{\text{EM},n}  = \frac{\sqrt{\bar{\mu}}\, \psi_n(k_{p} a) \psi^{\prime}_n(k a) - \sqrt{\bar{\varepsilon}}\, \psi_n(k a) \psi^{\prime}_n(k_{p} a)}{\sqrt{\bar{\mu}}\, \psi_n(k_{p} a) \xi^{\prime}_n(k a) - \sqrt{\bar{\varepsilon}}\, \xi_n(k a) \psi^{\prime}_n(k_{p} a)}, 
      \\
    a_{\text{A},n} &= \frac{ \sqrt{\bar{\beta}}\, j_n^{\prime} (k_p a) j_n(ka) - \sqrt{\bar{\rho}}\, j_n(k_p a) j_n^{\prime}(ka)}{ \sqrt{\bar{\rho}}\, j_n(k_{p} a) h_n^{(1)\prime} (ka) - \sqrt{\bar{\beta}}\, j_n^{\prime}(k_{p} a) h_n^{(1)}(ka)},
    \label{eq:Mie_coef}
\end{align}
where $j_n$ is the spherical Bessel function, $h_n^{(1)}$ is the spherical Hankel function of the first kind, $\psi_n(x) = x j_n(x)$ and $\xi_n(x) = x h_n^{(1)}(x)$ are the Riccati-Bessel functions, the prime denotes derivative with respect to the argument, $\bar{\varepsilon} = \varepsilon_p /\varepsilon$ and $\bar{\mu} = \mu_p /\mu$ are the relative permittivity and permeability, $\bar{\rho} = \rho_p /\rho$ and $\bar{\beta} = \beta_p /\beta$ are the relative density and compressibility, and  $k_p = \omega \sqrt{\varepsilon_p \mu_p}$ ($k_p = \omega \sqrt{\beta_p \rho_p}$) is the wavenumber inside the sphere in electromagnetism (acoustics).
The Taylor series expansions of the monopole and dipole coefficients yield:
\begin{align}
a_{\text{EM},1} & \simeq - \frac{2 i}{3} \frac{\bar{\varepsilon}  - 1}{2 + \bar{\varepsilon}} (ka)^3 -  i  \frac{4 + \bar{\varepsilon} (\bar{\varepsilon} \bar{\mu} + \bar{\varepsilon} - 6)}{5 (2 + \bar{\varepsilon})^2} (ka)^5 + \frac{4}{9} \left( \frac{\bar{\varepsilon}  - 1}{2 + \bar{\varepsilon}} \right)^2 (ka)^6 + \dots \, , \nonumber \\
b_{\text{EM},1} &\simeq  - \frac{2 i}{3} \frac{\bar{\mu}  - 1}{2 + \bar{\mu}} (ka)^3 -  i  \frac{4 + \bar{\mu} (\bar{\varepsilon} \bar{\mu}  + \bar{\mu} - 6)}{5 (2 + \bar{\mu})^2} (ka)^5 + \frac{4}{9} \left( \frac{\bar{\mu}  - 1}{2 + \bar{\mu}} \right)^2 (ka)^6 + \dots \, , \\
    a_{\text{A},0} &\simeq \frac{i}{3} \left(\bar{\beta} - 1 \right) (ka)^3 + \frac{i}{45} \left[ \bar{\beta}^2 (\bar{\rho} + 5) - 15 \bar{\beta} + 9 \right] (ka)^5 - \frac{1}{9} \left(\bar{\beta} - 1 \right)^2 (ka)^6 + \dots \, ,\nonumber \\ 
    a_{\text{A},1} &\simeq \frac{i}{3} \frac{\bar{\rho} - 1}{2 \bar{\rho} + 1} (ka)^3 + \frac{i}{5} \frac{\bar{\rho}^2 \left( \bar{\beta} - 1\right)  - \bar{\rho} + 1}{ \left( 2\bar{\rho} + 1\right)^2} (ka)^5 - \frac{1}{9} \left(\frac{\bar{\rho} - 1}{2 \bar{\rho} + 1}\right)^2 (ka)^6 + \dots \, .
\end{align}
}

\section{Derivation of the Radiation Force on a Rayleigh Particle}
\label{app:tensor_force}

Here we show derivations of the main radiation forces and torques on a Rayleigh particle presented in Table~\ref{tab:forces}. In doing so, we mostly follow the electromagnetic force derivation in \cite{Chaumet2009OE}, while the derivation of the optical torque, as well as the acoustic force and torque can be performed in a similar manner \cite[see Appendix]{Smagin2024PRAppl}.

We start with the general momentum-flux equation \eqref{eq:force_and_torque_average}, choosing the integration surface $\Sigma$ to be a sphere of radius $r$ (assuming the particle to be located at ${\bf r} = {\bf 0}$):
\begin{equation}
    \vb{F} = - \oint \limits_\Sigma \ten{\mathcal{T}} \cdot d \vb{\Sigma} =   \frac{r^2}{2} \Re  \int\limits_{4\pi} d\Omega   \left[ \varepsilon (\vb{E}\cdot \vb{n}) \vb{E}^* +  \mu (\vb{H}\cdot \vb{n}) \vb{H}^* -  \frac{\vb{n}}{2} \left( \varepsilon |\vb{E}|^2 + \mu |\vb{H}|^2 \right) \right],
    \label{eq:force_main}
\end{equation}
where $\int \limits_{4\pi} d \Omega  = \int \limits_{0}^{2\pi} d \varphi  \int \limits_{0}^{\pi} d \vartheta \sin \vartheta$, with $\vartheta$ and $\varphi$ being the spherical-coordinates angles, $\vb{n} = (\sin \vartheta \cos \varphi, \sin \vartheta \sin \varphi, \cos \vartheta)^T$ is the outer normal unit vector to the sphere $\Sigma$, and the electromagnetic field is a sum of the incident and scattered fields, $(\vb{E},\vb{H}) = (\vb{E},\vb{H})_{\rm inc} + (\vb{E},\vb{H})_{\rm sc}$.

Before starting the derivation, we list helpful identities, which will be referred to along the mathematical transformations in what follows:
\begin{equation}
\int\limits_{4\pi} d \Omega\, n_i = 0, \qquad 
 \int\limits_{4\pi} d \Omega\, n_i n_j n_k = 0 , \qquad \int\limits_{4\pi} d \Omega\, n_i n_j n_k n_{\ell} n_m = 0,
\label{eq:odd_n}
\end{equation}
\begin{equation}
\int\limits_{4\pi} d \Omega = 4\pi, \qquad 
\int\limits_{4\pi} d \Omega\, n_i n_j = \frac{4\pi}{3} \delta_{ij}, \qquad 
\int\limits_{4\pi} d \Omega\, n_i n_j n_k n_{\ell}  = \frac{4\pi}{15} \left(\delta_{ij}\delta_{k \ell} + \delta_{ik}\delta_{j \ell} + \delta_{i\ell}\delta_{jk}\right),
\label{eq:even_n}
\end{equation}
\begin{equation}
	\epsilon_{ijk} \epsilon_{imn} = \delta_{jm} \delta_{kn} - \delta_{ij} \delta_{km}\,,
	\label{eq:levi}
\end{equation}
\begin{equation}
	\div \vb{E}_{\text{inc}} = 0\,,
	\label{eq:div}
\end{equation}
\begin{equation}
	\curl \vb{E} = \iu k \sqrt{\frac{\mu}{\varepsilon}}\, \vb{H}\,,
	\label{eq:curl}
\end{equation}
where $i,j,k,\ell,m = x,y,z$, $\delta_{ij}$ is the Kronecker delta,  and $\epsilon_{ijk}$ is the Levi-Civita symbol.

We first calculate the `{\it mixed}' contribution of the incident and scattered fields to the force, Eqs.~\eqref{eq:field-split} and \eqref{eq:flux-split}:
\begin{align}
\label{eq:Fmix}
	&\vb{F}^{\text{mix}} =  - \oint \ten{\mathcal{T}}_{\text{mix}} \cdot d \vb{\Sigma} \\
 &=\frac{r^2}{2} \Re\!  \int\limits_{4\pi} d \Omega \bigg[ 
 \varepsilon (\vb{E}_{\text{sc}} \cdot \vb{n}) \vb{E}_{\text{inc}}^* + \varepsilon (\vb{E}_{\text{inc}}^* \cdot \vb{n}) \vb{E}_{\text{sc}} + 
 \mu (\vb{H}_{\text{sc}} \cdot \vb{n}) \vb{H}_{\text{inc}}^* + 
 \mu (\vb{H}_{\text{inc}}^* \cdot \vb{n}) \vb{H}_{\text{sc}} - \vb{n} \left[ \varepsilon(\vb{E}_{\text{sc}} \cdot \vb{E}^*_{\text{inc}}) + \mu (\vb{H}_{\text{sc}} \cdot \vb{H}^*_{\text{inc}}) \right] \bigg]. 
 \nonumber
\end{align}
This integral can be evaluated choosing the radius of the integration sphere to be just around the small particle, $kr \ll 1$, and using the {\it near field} approximation for the scattered field. The incident field can be expanded in the Taylor series: 
\begin{equation}
    \vb{E}_{\text{inc}}(\vb{r}) \simeq \vb{E}_{\text{inc}}(\bf{0}) + r (\vb{n} \cdot \grad) \vb{E}_{\text{inc}} (\bf{0}), \quad 
    \vb{H}_{\text{inc}}(\vb{r}) \simeq \vb{H}_{\text{inc}}(\bf{0}) + r (\vb{n} \cdot \grad) \vb{H}_{\text{inc}} (\bf{0}).
    \label{eq:NF}
\end{equation}
We first consider the {\it electric-dipole} contribution to the scattered field, $({\bf E}, {\bf H})_{\rm sc} = ({\bf E}, {\bf H})_{\bf e}$, Eq.~\eqref{eq:electric_dipole_NF}, and evaluate different terms in the integral \eqref{eq:Fmix} separately:
\begin{equation}
	\varepsilon r^2 \int\limits_{4\pi} d\Omega \, (\vb{E}_{e} \cdot \vb{n}) \vb{E}_{\text{inc}}^*
	\overset{\text{\tiny \eqref{eq:odd_n}, \eqref{eq:NF}}}= \frac{1}{4\pi}
	\int\limits_{4\pi}d\Omega\,  2 (\vb{n}\cdot \vb{e}) (\vb{n} \cdot \grad) \vb{E}_{\text{inc}}^*
	\overset{\text{\tiny \eqref{eq:even_n}}}= 
	\frac{2}{3} (\vb{e}\cdot \grad) \vb{E}_{\text{inc}}^*,
 \label{eq:term1}
\end{equation}
\begin{eqnarray}
	\varepsilon r^2\int\limits_{4\pi} d\Omega\, (\vb{E}_{\text{inc}}^* \cdot \vb{n}) \vb{E}_{{e}}
	&\overset{\text{\tiny\eqref{eq:NF}, \eqref{eq:electric_dipole_NF}}}=
 &\frac{r^2}{4\pi} \int\limits_{4\pi}d\Omega \left[ \vb{E}_{\text{inc}}^* + r (\vb{n}\cdot \grad)(\vb{E}_{\text{inc}}^* \cdot \vb{n}) \right] \frac{3 \vb{n} (\vb{n}\cdot \vb{e}) - \vb{e}}{r^3}
	\nonumber \\ 
&\overset{\text{\tiny\eqref{eq:odd_n}}}=
& \frac{1}{4\pi}
	\int\limits_{4\pi} d\Omega \left[ 3 (\vb{n} \cdot \grad)(\vb{E}_{\text{inc}}^* \cdot \vb{n})(\vb{n}\cdot \vb{e}) \vb{n} - (\vb{n} \cdot \grad) (\vb{E}_{\text{inc}}^* \cdot \vb{n}) \vb{e} \right]
	\nonumber \\ 
&\overset{\text{\tiny\eqref{eq:even_n}, \eqref{eq:div}}}=&
	\frac{1}{5} (\vb{e} \cdot \grad) \vb{E}_{\text{inc}}^* + \frac{1}{5} \vb{e} \berrydot \vb{E}_{\text{inc}}^*,
\end{eqnarray}
\begin{eqnarray}
    r^2 \mu \int d \Omega\, (\vb{H}_{e} \cdot \vb{n}) \vb{H}_{\text{inc}}^* \overset{\text{\tiny\eqref{eq:odd_n}}}= 0 
\end{eqnarray}
\begin{eqnarray}
	- \varepsilon r^2\int\limits_{4\pi} d\Omega\, (\vb{E}_{e} \cdot \vb{E}^*_{\text{inc}}) \vb{n} = - \frac{1}{5} {\bf e} \cdot (\vb{\grad}) {\bf E}_{\text{inc}}^* - \frac{1}{5} (\vb{e} \cdot \grad) \vb{E}_{\text{inc}}^* + \frac{1}{3} {\bf e}\cdot (\grad) {\bf E}_{\text{inc}}^*,
\end{eqnarray}
\begin{eqnarray}
	- \mu r^2\int\limits_{4\pi} d\Omega\, (\vb{H}_{e} \cdot \vb{H}_{\text{inc}}^*) \vb{n} = \frac{\iu k}{3}  \sqrt{\frac{\mu}{\varepsilon}}  (\vb{H}_{\text{inc}}^* \times \vb{e})
	\overset{\text{\tiny\eqref{eq:curl}, \eqref{eq:levi}}}=\frac{1}{3}  \vb{e} \berrydot \vb{E}_{\text{inc}}^* - \frac{1}{3} (\vb{e}\cdot \grad) \vb{E}_{\text{inc}}^*,
\end{eqnarray}
\begin{eqnarray}
	\mu r^2\int\limits_{4\pi} d\Omega\, (\vb{H}_{\text{inc}}^* \cdot  \vb{n}) \vb{H}_{e} = \frac{\iu k}{3}  \sqrt{\frac{\mu}{\varepsilon}} (\vb{H}_{\text{inc}}^* \times \vb{e})
	\overset{\text{\tiny\eqref{eq:curl}, \eqref{eq:levi}}}=\frac{1}{3}  {\bf e}\cdot (\grad) {\bf E}_{\text{inc}}^* - \frac{1}{3} (\vb{e}\cdot \grad) \vb{E}_{\text{inc}}^*.
 \label{eq:term6}
\end{eqnarray}
Summing up Eqs.~\eqref{eq:term1}--\eqref{eq:term6}, we arrive at
\begin{equation}
\vb{F}^{\text{mix}}_{e} = \frac{1}{2} \Re \left[ \vb{e} \berrydot \vb{E}_{\text{inc}}^* \right].
\end{equation}
In a similar way, one can derive the force on the {\it magnetic dipole} $\vb{m}$. As a result, the total mixed force, including the electric and magnetic dipole contributions, reads 
\begin{equation}
	\vb{F}^{\text{mix}} = \frac{1}{2} \Re \left[ \vb{e}^*  \berrydot  \vb{E}_{\text{inc}} +  \vb{m}^* \berrydot \vb{H}_{\text{inc}} \right].
 \label{eq:Fmix_result}
\end{equation}

Next, we calculate the {\it recoil} force originating from the pure-scattered field contribution, taking into account both the {\it electric and magnetic dipole} components, $({\bf E},{\bf H})_{\rm sc} = ({\bf E},{\bf H})_{e} + ({\bf E},{\bf H})_{m}$:
\begin{eqnarray}
	\vb{F}^{\text{rec}} = - \oint \ten{\mathcal{T}}_{\text{sc}} \cdot d \vb{\Sigma} &=& \frac{1}{2} r^2 \Re  \int\limits_{4\pi} d\Omega \bigg[ %
 \varepsilon(\vb{E}_{\text{sc}} \cdot \vb{n}) \vb{E}_{\text{sc}}^* + \mu(\vb{H}_{\text{sc}} \cdot \vb{n}) \vb{H}_{\text{sc}}^*
 -  \frac{\vb{n}}{2} \left( \varepsilon|\vb{E}_{\text{sc}}|^2 + \mu|\vb{H}_{\text{sc}}|^2 \right)\bigg]  
    \label{eq:Frec} \\
	&=& - \frac{1}{4} r^2 \int\limits_{4\pi} d\Omega \left( \varepsilon |\vb{E}_{\text{sc}}|^2 + \mu|\vb{H}_{\text{sc}}|^2 \right) \vb{n}\,.
 \label{eq:Frec2}
\end{eqnarray}
Here the terms $\varepsilon(\vb{E}_{\text{sc}} \cdot \vb{n}) \vb{E}_{\text{sc}}^* + \mu(\vb{H}_{\text{sc}} \cdot \vb{n}) \vb{H}_{\text{sc}}^*$ do not contribute to the integral \eqref{eq:Frec} due to Eq.~\eqref{eq:odd_n} and the near-field dipole expressions \eqref{eq:electric_dipole_NF} and \eqref{eq:magnetic_dipole_NF}.
Now, the integral \eqref{eq:Frec2} describes the total energy flux of the scattered field, and it is convenient to evaluate it in the \textit{far field}, $kr \gg 1$. The scattered far field is given by the sum of Eqs.~\eqref{eq:electric_dipole_FF} and \eqref{eq:magnetic_dipole_FF}:
\begin{equation}
    \sqrt{\varepsilon}\, \vb{E}_{\text{sc}} \simeq \frac{k^3}{4\pi } \left[ \left( \vb{n} \times \frac{\vb{e}}{\sqrt{\varepsilon}} \right) \times \vb{n}   -  \vb{n} \times  \frac{\vb{m}}{\sqrt{\mu}}  \right] \dfrac{e^{i kr}}{kr}, \quad 
    \sqrt{\mu}\,\vb{H}_{\text{sc}} \simeq \frac{k^3}{4\pi } \left[ \left(\vb{n} \times \frac{\vb{m}}{\sqrt{\mu}}\right) \times \vb{n}   +  \vb{n} \times  \frac{\vb{e}}{\sqrt{\varepsilon}}  \right] \dfrac{e^{i kr}}{kr}\, .
\label{eq:FFsc}
\end{equation}
Substituting Eqs.~\eqref{eq:FFsc} into Eq.~\eqref{eq:Frec2}, we derive:
\begin{eqnarray}
	\mu r^2 \int\limits_{4\pi} d\Omega\, |\vb{H}_{\text{sc}}|^2 \vb{n}
	&=& 
	\frac{k^4}{(4\pi)^2} \int\limits_{4\pi} d\Omega \left[ \left( \vb{n} \times \frac{\vb{m}}{\sqrt{\mu}} \times \vb{n} + \vb{n} \times \frac{\vb{e}}{\sqrt{\varepsilon}} \right) \cdot \left( \vb{n} \times \frac{\vb{m}^*}{\sqrt{\mu}} \times \vb{n} + \vb{n} \times \frac{\vb{e}^*}{\sqrt{\varepsilon}} \right)\right] \vb{n}
	\nonumber \\ &
 \overset{\text{\eqref{eq:odd_n}}}=
 & \frac{k^4}{(4\pi)^2} \int\limits_{4\pi} d \Omega\,  2\Re\!\left[ \left( \vb{n} \times \frac{\vb{e}}{\sqrt{\varepsilon}} \right) \cdot \left( \vb{n} \times \frac{\vb{m}^*}{\sqrt{\mu}} \times \vb{n} \right) \right] \vb{n}
 \overset{\text{\eqref{eq:even_n}}}= \frac{\omega k^3}{6\pi}    \Re \left( \vb{e} \times \vb{m}^* \right), \\
	\varepsilon r^2 \int\limits_{4\pi} d\Omega |\vb{E}_{\text{sc}}|^2 \vb{n} &=& \frac{\omega k^3}{6\pi} \Re \left( \vb{e} \times \vb{m}^* \right).
\end{eqnarray}
The resulting recoil force is  
\begin{equation}
	\vb{F}^{\text{rec}} = - \frac{\omega k^3}{12\pi} \Re \left( \vb{e}^* \times \vb{m} \right).
 \label{eq:Frec_result}
\end{equation}

Finally, summing up Eqs.~\eqref{eq:Fmix_result} and \eqref{eq:Frec_result}, we obtain the electromagnetic dipole-approximation force in the Table~\ref{tab:forces}:
\begin{equation}
	\vb{F} = \vb{F}^{\text{mix}} + \vb{F}^{\text{rec}} = \frac{1}{2} \Re \left( \vb{e}^* \berrydot  \vb{E}_{\text{inc}} +  \vb{m}^* \berrydot \vb{H}_{\text{inc}} \right)  - \frac{\omega k^3}{12\pi} \Re \left( \vb{e}^* \times \vb{m} \right).
\end{equation}

\section{Connection Between the Recoil Force and Scattering Diagram}
\label{app:recoil_and_radiation_diagram}

Here we describe the general relation between the recoil force (not restricted by the monopole-dipole approximation) and the scattering diagram, both in electromagnetism and acoustics. The recoil force is determined via the pure-scattered field contribution to the {\it momentum flux}:
\begin{equation}
\vb{F}^{\text{rec}} = - \oint \limits_{\Sigma} \ten{\mathcal{T}}_{\text{sc}} \cdot d \vb{\Sigma} \,.
\label{eq:Frec_D}
\end{equation}
In turn, the scattering diagram is determined by the {\it directivity} of the scattered field, defined via the {\it energy flux} per unit solid angle in the far field, $r\to \infty$:
\begin{equation}
D(\vartheta,\varphi) = \frac{r^2 \mathbfcal{P}_{\text{sc}} \cdot {\bf n}}{W} = \frac{r^2 c\, U_{\text{sc}}}{W} \,.
\label{eq:Directivity}
\end{equation}
Here $W = r^2 \expval{\mathbfcal{P}_{\rm sc} \cdot {\bf n}} = r^2 c \expval{U_{\rm sc}}$, where $\expval{...} =  \int_{4\pi} ... \, d \Omega$, and we used the fact that the far scattered field is an outgoing spherical wave with $\mathbfcal{P}_{\text{sc}} \parallel {\bf n}$ and the energy density $U_{\rm sc} = c^{-1} \mathbfcal{P}_{\text{sc}} \cdot {\bf n}$ decaying as $\propto 1/r^2$.
The directivity \eqref{eq:Directivity} depends on the spherical angles $(\vartheta, \varphi)$ {but is independent of the distance $r$}.

In the far (spherical-wave) field, the integral \eqref{eq:Frec_D} can be transformed into the integral of the energy flux density \cite{Gao2015LPR,Nieto-Vesperinas2010OE,Livett1956}:
\begin{equation}
\vb{F}^{\text{rec}} = 
- \frac{r^2}{c} \int \limits_{4\pi}  \mathbfcal{P}_{\text{sc}}\, d \Omega, 
    \label{eq:recoil_and_D_EM}
\end{equation}
Equations \eqref{eq:Directivity} and \eqref{eq:recoil_and_D_EM} result in
\begin{equation}
\vb{F}^{\text{rec}} = - \frac{W}{c} \expval{\vb{n} D} \,.
\end{equation}
Thus, the recoil force is directly determined by (and opposite to) the {\it averaged scattering direction} $\expval{\vb{n} D}$.

\section{Torque on a Small Anisotropic Particle from an Arbitrary Polarized Plane Wave}
\label{app:torque_anisotropic}

Here we derive the optical torque on a small anisotropic particle in a plane-wave incident field. For simplicity, we assume that the plane wave propagates along the $z$-axis, while the anisotropy axis of the particle lies in the $(x,y)$-plane. The incident field can be written as 
\begin{equation}
\sqrt{\varepsilon}\,\vb{E}_{\text{inc}} = A_0\! \begin{pmatrix}
 u_x \\ u_y \\ 0
\end{pmatrix}\! \eu^{\iu k z}, \qquad 
\sqrt{\mu}\,\vb{H}_{\text{inc}} = A_0\! \begin{pmatrix}
-u_y \\ u_x \\ 0
\end{pmatrix}\! \eu^{\iu k z},
\label{eq:arb_pol_pw}
\end{equation}
where $A_0$ is the wave amplitude, and the polarization components are normalized as $|u_x|^2 + |u_x|^2 =1$. The polarization state can be described by the normalized Stokes parameters \cite{collett2005FieldGuidePolarization}
\begin{equation}
    \tau = \abs{u_x}^2 - \abs{u_y}^2, \qquad
    \chi = 2 \Re(u_x^* u_y), \qquad 
    \sigma = 2 \Im(u_x^* u_y), 
\end{equation}
where the third Stokes parameter $\sigma$ determines the normalized helicity, i.e., here, the $z$-component of the spin density: $\sigma = \omega S_{{\rm inc}\,z} /U_{\rm inc}$. Without loss of generality, we can choose the $(x,y)$ axes such that $\chi = 0$.

Let the anisotropy axes of the particle be tilted by angle $\theta$ within the $(x,y)$ plane.
Using the $(x',y',z)$ coordinate frame, aligned with the anisotropy axes of the particle, its electric-dipole and magnetic-dipole polarizability tensors can be written in the diagonal form: 
\begin{eqnarray}
    \ten{\alpha}^{\prime}_{e} = \begin{pmatrix}
        \alpha_{e,x^{\prime}} & 0 & 0 \\
        0 & \alpha_{e,y^{\prime}} & 0 \\
        0 & 0 & \alpha_{e,z} 
    \end{pmatrix}, \qquad
    \ten{\alpha}_{m}^{\prime} = \begin{pmatrix}
        \alpha_{m,x^{\prime}} & 0 & 0 \\
        0 & \alpha_{m,y^{\prime}} & 0 \\
        0 & 0 & \alpha_{m,z} 
    \end{pmatrix}.
\end{eqnarray}
These tensors can be transformed to the $(x,y,z)$ coordinates using the rotation matrix $\ten{R}(\theta)$: 
\begin{equation}
\ten{\alpha}_{e,m} = \ten{R}(\theta) \ten{\alpha}^{\prime}_{e,m} \ten{R}^{-1}\! (\theta)\,, \qquad 
\ten{R}(\theta) = \begin{pmatrix}
        \cos \theta & - \sin\theta & 0 \\
        \sin\theta & \cos\theta & 0 \\
        0 & 0 & 1
    \end{pmatrix}.
\end{equation}

Substituting the induced electric and magnetic dipole moments of the particle, $\vb{e} =  \varepsilon  \ten{\alpha}_e \vb{E}$ and $\vb{m} =  \mu 
 \ten{\alpha}_m \vb{H}$ into the expression for the torque in Table~\ref{tab:forces}, we obtain 
\begin{eqnarray}
    \vb{T}_{EM} = \unitvec{z}\,  \frac{|A_0|^2}{2}  \bigg[\alpha^{\text{spin}} \sigma -   \alpha^{\text{align}} \tau \sin(2\theta) \bigg],
    \label{eq:torque_anis}
\end{eqnarray}
where
\begin{align}
\alpha^{\text{spin}} &= \frac{1}{2}\Im\!\left(\alpha_{e,x^{\prime}} + \alpha_{e,y^{\prime}} \right) - g \Re \!\left(\alpha^{*}_{e,x^{\prime}} \alpha_{m,y^{\prime}}\right) +  
\frac{1}{2}\Im\!\left(\alpha_{m,x^{\prime}} + \alpha_{m,y^{\prime}} \right) - g \Re \!\left(\alpha^{*}_{m,x^{\prime}} \alpha_{m,y^{\prime}}\right), 
\label{eq:alpha_spin_em}\\
\alpha^{\text{align}} &= \frac{1}{2} \Re\!\left(\alpha_{e,x^{\prime}} - \alpha_{e,y^{\prime}}\right) - g \Im \!\left(\alpha^{*}_{e,x^{\prime}} \alpha_{e,y^{\prime}} \right) - \frac{1}{2} \Re\!\left(\alpha_{m,x^{\prime}} - \alpha_{m,y^{\prime}}\right) + g \Im \!\left(\alpha^{*}_{m,x^{\prime}} \alpha_{m,y^{\prime}} \right).
\end{align}

For small {\it lossless} particles, the polarizability tensors can be written as a sum of the real {\it static} polarizability and imaginary {\it radiation correction}: $(\ten{\alpha}^{\prime}_{e,m})^{-1} = (\ten{\alpha}^{(0)\prime}_{e,m})^{-1} - i g \ten{I}$ \cite{LeRu2013} (see Table~\ref{tab:polarizabilities} for the isotropic case). Then, Eq.~\eqref{eq:alpha_spin_em} takes the form
\begin{eqnarray}
\alpha^{\text{spin}} = \frac{g}{2} \Re\! \left(\alpha^{(0)}_{e,x^{\prime}} - \alpha^{(0)}_{e,y^{\prime}} \right)^2 +  \frac{g}{2} \Re\! \left(\alpha^{(0)}_{m,x^{\prime}} - \alpha^{(0)}_{m,y^{\prime}} \right)^2.
\label{eq:alpha_spin}
\end{eqnarray}
The first term in Eq.~\eqref{eq:torque_anis} with Eq.~\eqref{eq:alpha_spin} describes the anisotropic spin-dependent torque \eqref{eq:T_aniz}, whereas the second term in Eq.~\eqref{eq:torque_anis} corresponds to the alignment torque \eqref{eq:T_align}. See also Fig.~\ref{fig:types_of_torque}(b,c) and Fig.~\ref{fig:torque_exp}(b).

\section{Resonant Polarizability}
\label{app:alpha_res} 

The near-resonance lowest-order (monopole or dipole) polarizability of a particle can be described by \cite{jackson1998ClassicalElectrodynamics}
\begin{equation}
{\alpha}(\omega)  = g^{-1}  \frac{\gamma_{\text{rad}}}{\omega_{\text{res}} - \omega - i\, (\gamma_{\text{abs}} + \gamma_{\text{rad}})}\,,
\label{eq:alpha_res2}
\end{equation}
where $\omega_{\rm res}$ is the resonant frequency, $\gamma_{\text{abs}}$ characterizes absorption in the particle, and $\gamma_{\text{rad}}$ characterizes radiation losses of particles. The total losses are described by $\gamma = \gamma_{\text{abs}} + \gamma_{\text{rad}}$ and the corresponding Q-factor $Q = \omega /(2\gamma)$. Note that for a lossless particle, $\gamma_{\text{abs}} =0$, and Eq.~\eqref{eq:alpha_res2} satisfies the constraint of Eqs.~\eqref{eq:cross-sections}:
\begin{equation}
\Im(\alpha) = g \abs{\alpha}^2\,.
\label{eq:constraint}
\end{equation}

Let us examine quantities following from Eq.~\eqref{eq:alpha_res2} and crucial for the gradient forces, radiation-pressure forces, and radiation torques on isotropic Rayleigh particles, Eqs.~\eqref{eq:F} and \eqref{eq:T_tot}. First, the real and imaginary parts of the polarizability are:
\begin{align}
\label{eq:Realpha}
\Re(\alpha) & = g^{-1} \frac{\gamma_{\text{rad}} (\omega_{\text{res}} - \omega)}{(\omega_{\text{res}} - \omega)^2 + (\gamma_{\text{rad}} + \gamma_{\text{abs}})^2}\,, \qquad \\
\Im(\alpha) & = g^{-1} \frac{\gamma_{\text{rad}} (\gamma_{\text{rad}} + \gamma_{\text{abs}})}{(\omega_{\text{res}} - \omega)^2 + (\gamma_{\text{rad}} + \gamma_{\text{abs}})^2}\, .
\label{eq:Imalpha}
\end{align}
Second, the absorption cross-section can be found using Eqs.~\eqref{eq:cross-sections} and $\sigma^{\rm abs} = \sigma^{\rm ext} - \sigma^{\rm sc}$, which yields:
\begin{eqnarray}
\sigma^{\text{abs}} = k \Im(\alpha) - k g \abs{\alpha}^2 = k g^{-1} \frac{\gamma_{\text{rad}} \gamma_{\text{abs}}}{(\omega_{\text{res}} - \omega)^2 + (\gamma_{\text{rad}} + \gamma_{\text{abs}})^2}\,.
\label{eq:sigma_abs}
\end{eqnarray}
For a near-resonant frequency $\omega \simeq \omega_{\rm res}$ in the important limiting cases for the absorption and radiation losses these quantities are simplified as:
\begin{itemize}
\item $\gamma \simeq \gamma_{\text{rad}} \gg \gamma_{\text{abs}}$:
\begin{equation}
\Re(\alpha) \simeq 2 Q\, g^{-1} \frac{\Delta \omega}{\omega_{\text{res}}} \,,\qquad 
\Im(\alpha)  \simeq g^{-1} \, \qquad
\sigma^{\text{abs}} \simeq k g^{-1} \frac{\gamma_{\text{abs}}}{\gamma_{\text{rad}}}\,,
\end{equation}
\item $\gamma \simeq \gamma_{\text{abs}} \gg \gamma_{\text{rad}}$:
\begin{equation}
\Re(\alpha) \simeq  2Q\, g^{-1} \frac{\Delta \omega}{\omega_{\text{res}}} \frac{\gamma_{\text{rad}}}{\gamma_{\text{abs}}}\,,        \qquad 
\Im(\alpha)  \simeq g^{-1} \frac{\gamma_{\text{rad}}}{\gamma_{\text{abs}}} \,, \qquad
\sigma^{\text{abs}} \simeq k g^{-1} \frac{\gamma_{\text{rad}}}{\gamma_{\text{abs}}}\,,
\end{equation}
\item Critical coupling $\gamma_{\text{rad}} = \gamma_{\text{abs}}$:
\begin{equation} 
\Re(\alpha) \simeq Q \, g^{-1} \frac{\Delta \omega}{\omega_{\text{res}}} \,, \qquad
\Im(\alpha) \simeq  \frac{1}{2}\, g^{-1}\,, \qquad
\sigma^{\text{abs}} \simeq  \frac{1}{4} k g^{-1} \,,
\end{equation}
\end{itemize}
where $\Delta \omega = \omega_{\text{res}} - \omega$. 
Here we considered the generic polarizability, without specifying its electric-dipole, magnetic-dipole, acoustic-dipole, or acoustic-monopole origin. In the latter two cases, the corresponding factors $g$ are defined in Eqs.~\eqref{eq:g}.

\end{document}